\documentclass[a4paper,twocolumn,showpacs,aps,prd,longbibliography,nofootinbib]{revtex4-1}

\usepackage{latexsym}
\usepackage{graphicx}
\usepackage{color}
\usepackage{amsmath}

\usepackage{hyperref}
\usepackage{bm}
\usepackage{acronym}
\usepackage{url}

\newcommand{\ee}[1]{\ensuremath{\!\times\!10^{#1}}}

\begin{document}

\title{Hierarchical Bayesian method for detecting continuous gravitational waves from an ensemble of pulsars}

\author{M.~Pitkin}
\email{matthew.pitkin@glasgow.ac.uk}
\affiliation{SUPA, School of Physics \& Astronomy, University of Glasgow, Glasgow G12 8QQ, United Kingdom}

\author{C.~Messenger}
\email{christopher.messenger@glasgow.ac.uk}
\affiliation{SUPA, School of Physics \& Astronomy, University of Glasgow, Glasgow G12 8QQ, United Kingdom}

\author{X.~Fan}
\email{fanxilong@outlook.com}
\affiliation{Department of Physics \& Astronomy, Hubei University of Education, 430205 Wuhan, China}

\date{\today}

\begin{abstract}
When looking for gravitational wave signals from known pulsars, targets have been treated using independent searches. Here we use a hierarchical Bayesian framework to combine observations from individual sources for two purposes: to produce a detection statistic for the whole ensemble of sources within a search, and to estimate the hyperparameters of the underlying distribution of pulsar ellipticities. Both purposes require us to assume some functional form of the ellipticity distribution, and as a proof of principle we take two toy distributions. One is an exponential distribution, defined by its mean, and the other is a half-Gaussian distribution defined by its width. We show that by incorporating a common parameterized prior ellipticity distribution we can be more efficient at detecting gravitational waves from the whole ensemble of sources than trying to combine observations with a simpler non-hierarchical method. This may allow us to detect gravitational waves from the ensemble before there is confident detection of any single source. We also apply this method using data for 92 pulsars from LIGO's sixth science run. No evidence for a signal was seen, but 90\% upper limits of $3.9\ee{-8}$ and $4.7\ee{-8}$ were set on the mean of an assumed exponential ellipticity distribution and the width of an assumed half-Gaussian ellipticity distribution, respectively.
\end{abstract}


\acrodef{GW}[GW]{gravitational wave}
\acrodef{MJD}[MJD]{modified Julian date}

\maketitle

\section{Introduction}\label{sec:intro}

So far, Advanced LIGO \cite{2015CQGra..32g4001L} and Advanced Virgo
\cite{2015CQGra..32b4001A} have detected gravitational waves from the time
varying mass quadrupole caused by the orbital motion of two compact objects:
binary systems consisting of two black holes
\cite{2016PhRvL.116f1102A,2016PhRvX...6d1015A} or two neutron stars
\cite{2017PhRvL.119p1101A}. However, individual neutron stars with deformations
can also have a time varying mass quadrupole that is not detectable as a
transient signal with a chirplike nature, but is long-lived and
quasimonochromatic. From such sources, the expected gravitational wave signal
strength at Earth is given by
\cite{2010ApJ...713..671A}
\begin{equation}\label{eq:h0}
    h_0 \approx 4.2\!\times\!10^{-24} \frac{f^2_{\rm kHz}}{D_{\rm kpc}} \left( \frac{I_{zz}}{10^{38}\,{\rm kg}\,{\rm m}^2} \right) \left( \frac{\varepsilon}{10^{-6}} \right),
\end{equation}
where $f_{\rm kHz}$ is the star's rotation frequency in kHz, $D_{\rm kpc}$ is
its distance in kpc, and the combination of the fiducial ellipticity
\cite{2013PhRvD..88d4004J}, $\varepsilon = \frac{|I_{xx}-I_{yy}|}{I_{zz}}$, and
principal moment of inertia, $I_{zz}$, are related to the mass quadrupole moment
(based on the definition of \cite{2000MNRAS.319..902U}) by
\begin{equation}\label{eq:q22}
    Q_{22} = \sqrt{\frac{15}{8\pi}} I_{zz} \varepsilon.
\end{equation}

Galactic neutron stars that are observed as pulsars are an enticing target for
such signals. The times of arrival of electromagnetic pulses from pulsars can, in
many cases, be used to fit their phase evolution to great accuracy. This leads
to high precision measurements of the spin frequency and frequency derivative, often
sub-milliarcsecond position reconstruction, and for sources in multiple systems,
estimates of their orbital parameters. Under the assumption that any pulsar's
mass quadrupole is formed by its having a triaxial moment of inertia ellipsoid
(or, more colloquially, a ``hill,'' ``mountain,'' or ``bump''), and that the
rotation is phase locked with the electromagnetic emission, then the
gravitational wave signal will be at exactly twice the observed frequency.
Therefore, the precise phase evolution of any associated gravitational wave
signal can be used as a filter for the signal in gravitational wave data. The
only unknowns in the signal model are therefore the gravitational wave amplitude
at Earth, as given in Eq.~(\ref{eq:h0}), and the orientation of the pulsar
given by the inclination angle, $\iota$, polarization angle, $\psi$, and
relative phase at the epoch of the electromagnetic timing model, $\phi_0$. As
in, e.g., \cite{2004PhRvD..69h2004A, 2005PhRvL..94r1103A, 2007PhRvD..76d2001A,
2010ApJ...713..671A, 2014ApJ...785..119A, 2017ApJ...839...12A}, following the
method of \cite{2005PhRvD..72j2002D}, Bayesian inference can be used to produce
a joint posterior probability density function on these four parameters, and the
orientation parameters can be marginalized over to give a posterior on only
$h_0$. If the distance to a pulsar is known then this can be used to make
inferences on, or set a limit on, the mass quadrupole moment, $Q_{22}$, or
alternatively, the pulsars fiducial ellipticity assuming a canonical moment of
inertia of $I_{zz} = 10^{38}\,{\rm kg}\,{\rm m}^2$ \cite{2013PhRvD..88d4004J}.\footnote{The canonical moment of inertia is roughly what you get from assuming a uniform density spherical star with a mass of 1.4\,$\rm{M}_{\odot}$ and radius of 10\,km.}

In previous searches for known pulsars in gravitational wave data from the LIGO,
Virgo and GEO600 detectors \cite{2004PhRvD..69h2004A, 2005PhRvL..94r1103A,
2007PhRvD..76d2001A, 2010ApJ...713..671A, 2014ApJ...785..119A,
2017ApJ...839...12A} no signal has been seen, but upper limits, at a 95\%
credible level, were set on $h_0$ for all included pulsars. In the latest
results \cite{2017ApJ...839...12A}, using data from the first observing run (O1)
of the advanced LIGO detectors \cite{2016PhRvL.116m1103A}, this produced limits
on $h_0$ for 200 pulsars, which in turn gave limits on $Q_{22}$ and
$\varepsilon$ using the best-fit measured distances given in the ATNF Pulsar
Catalogue \cite{2005AJ....129.1993M}. The lowest limits on $Q_{22}$ and
$\varepsilon$ found for any pulsar were $9.7\!\times\!10^{29}\,{\rm kg}\,{\rm
m}^2$ and $1.3\!\times\!10^{-8}$ respectively, for the relatively nearby ($\sim
0.2$\,kpc) pulsar PSR\,J0636+5129.

Different neutron star equations of state can allow different sizes of
deformations to be sustained in the star; stiffer equations of state allow
larger deformations than softer ones. A good review of the \emph{maximum}
allowed ellipticities is given in \cite{2013PhRvD..88d4004J}, which shows that
in very extreme cases they could be in the range of $10^{-3}$, but more
realistically, for crustal deformations, they could reach a few $\times
10^{-6}$. Internal magnetic fields can also induce an ellipticity in a neutron
star (e.g., \cite{2002PhRvD..66h4025C}), with field strengths of order
$10^{15}$\,G required to give rise to ellipticities of $\sim 10^{-6}$. For known
pulsars, assuming their rate of loss of rotational energy is all due to
gravitational wave emission, one can infer the ellipticity required to give the
observed spin-down (called the ``spin-down limit''). For the population of
millisecond pulsars, such maximum possible ellipticities are generally well
above the spin-down limit, which give values in the range $\sim
10^{-9}-10^{-8}$. So, it is fair to assume that this population of pulsars is
not strained to their maximum allowed values. The recent work in
\cite{2018arXiv180602822W} provides some tentative evidence that millisecond
pulsars may actually have a \emph{minimum} ellipticity of $\sim 10^{-9}$.

Searches for gravitational waves from pulsars have always treated each pulsar
individually. However, as we will describe further in this paper, it is possible
to combine observations from many pulsars to try and detect the ensemble of all
pulsars used in a search. The idea is that there may be several sources that are
\emph{individually} below some allowed threshold for detection, but when
combined the ensemble rises above a detection threshold. This was proposed in
\cite{2005PhRvD..72f3006C}, in which values of the $\mathcal{F}$-statistic
\cite{1998PhRvD..58f3001J}, commonly used in continuous gravitational wave
searches, for individual sources are summed. The method of
\cite{2005PhRvD..72f3006C} depends rather sensitively on the relative strengths
of the brightest few sources, and is unlikely to be able to detect an ensemble
of similar strength but individually undetectable, sources, i.e., it does not
``win'' by much over single source searches. This idea was extended in
\cite{2016PhRvD..94h4029F} to weight the individual $\mathcal{F}$-statistic
values based on the expected detectability of each source, i.e., favoring
close-by sources, which was found to always provide a more sensitive detection
statistic than \cite{2005PhRvD..72f3006C}.

In this paper we also address the task of combining an ensemble of sources, but
for two purposes: to try and estimate the functional form of the underlying
distribution of pulsar ellipticities, \emph{and} to use the combined sources as
an ensemble detection method. The former of these could allow us to make some
physical inferences about the equation of state, or internal magnetic field
strength, of the population of neutron stars. The latter, as discussed above,
could allow us to detect the ensemble before an individual source is confidently
detected.

\section{Method}

Our aim is to combine data from all pulsars within a gravitational wave signal
search to try and estimate the underlying distribution of their fiducial
ellipticities (or alternatively, $Q_{22}$). We can do this within the context of
hierarchical Bayesian inference, whereby we use the data to infer parameters of
a probability distribution that represents the prior on the ellipticities (in
this context these parameters are often called \emph{hyper}parameters). This is
essentially the same underlying method as described in
\cite{2018PhRvX...8b1019S}, where in that context the main use is for detecting
a stochastic background of compact binaries coalescences that are individually
unresolvable.

In previous searches, each pulsar has been treated individually, and a posterior
distribution on the gravitational wave amplitude, $h_0$, has been estimated, via
\begin{equation}
p(h_0|\mathbf{x}_i,I) \propto \int^{\bm{\theta}_i} p(\mathbf{x}_i|h_0,\bm{\theta}_i,I) p(h_0|I) p(\bm{\theta}_i|I) {\rm d}\bm{\theta}_i,
\end{equation}
where $\mathbf{x}_i$ is the data for the $i$th pulsar, $\bm{\theta}_i =
\{\cos{\iota},\phi_0, \psi\}_i$ are the pulsar's orientation parameters,
$p(\mathbf{x}_i|h_0,\bm{\theta}_i,I)$ is the likelihood of $\mathbf{x}_i$ given
the signal model and particular parameter values, $p(\bm{\theta}_i|I)$ is the
prior probability on the $\bm{\theta}_i$ parameters (which is flat over the
allowed parameter ranges, see Sec.~3.2 of \cite{2015MNRAS.453.4399P}), and
$p(h_0|I)$ is the prior on $h_0$. More recently, in \cite{2017ApJ...839...12A},
the marginal likelihood (or Bayesian evidence) for each pulsar's data being
consistent with the expected signal model has also been calculated, via
\begin{equation}
p(\mathbf{x}_i|I) = \int^{\bm{\theta}_i}\int^{h_0} p(\mathbf{x}_i|h_0,\bm{\theta}_i,I) p(h_0|I) p(\bm{\theta}_i|I) {\rm d}\bm{\theta}_i{\rm d}h_0.
\end{equation}
The prior on $h_0$ has often been flat between zero and some hard upper cutoff,
or having a Fermi-Dirac-like distribution (see, e.g., Sec.~2.3.5 of
\cite{2017arXiv170508978P}) with a flat section followed by an exponential-like
decay at some predefined value.

Alternatively, we can rearrange Eq.~(\ref{eq:h0}), and for each pulsar we can
estimate the joint posterior on $\varepsilon_i$ and the pulsar's distance, $D_i$,
\begin{align}\label{eq:eDposterior}
p(\varepsilon_i, D_i|\mathbf{x}_i,I) = \frac{p(\varepsilon_i|I)}{Z_i} \int^{\bm{\theta}_i} & p(\mathbf{x}_i|\varepsilon_i,D_i,\bm{\theta}_i,I)  \nonumber \\
 & \times p(D_i|I) p(\bm{\theta}_i|I) {\rm d}\bm{\theta}_i,
\end{align}
where $Z_i \equiv p(\mathbf{x}_i|I)$ is the evidence of the data (we will use
the term \emph{evidence}, rather than \emph{marginal likelihood} in the rest of
this paper as it is more compact). To produce a posterior on $\varepsilon_i$ we
need to set the priors for $D_i$ and $\varepsilon_i$. We can use a Gaussian
prior on $D_i$ defined by a mean value $\mu_{D_i}$, taken as the best-fit
estimate from the ATNF Pulsar Catalogue \cite{2005AJ....129.1993M} (which is
generally a dispersion measure-derived distance based on the galactic electron
density model of \cite{2017ApJ...835...29Y}), and a standard deviation,
$\sigma_{D_i}$,\footnote{Throughout this work we will take the distance
uncertainty to be 20\% of the $\mu_{D_i}$ value, which is roughly what Figure~12
in \cite{2002astro.ph..7156C} suggests for dispersion measure-based distance
estimates. For some pulsars better distance estimates are available (using, for example, parallax measurements obtained using very long baseline interferometry), while for
others the uncertainties can be considerably worse, so in a detailed analysis
more reliable uncertainties could be used for individual pulsars.}
\begin{equation}\label{eq:Dprior}
p(D_i|\mu_{D_i}, \sigma_{D_i},I) = \frac{1}{\sqrt{2\pi}\sigma_{D_i}}\exp{\left( -\frac{(D_i-\mu_{D_i})^2}{2\sigma_{D_i}^2} \right)}.
\end{equation}
For each pulsar the distance is independent, and as the distances are not of
interest to us, we can marginalize over them in Eq.~(\ref{eq:eDposterior}) as
nuisance parameters.\footnote{For pulsars in the same globular cluster the
distance measurements and error would essentially be the same, or at least very
highly correlated, but we will ignore that fact in this study. Of the pulsars
currently listed in the ATNF Pulsar Catalogue \cite{2005AJ....129.1993M} with
rotation frequencies above 10\,Hz, 139 of the 451 pulsars are in globular
clusters. The clusters with the largest numbers of pulsars are 47 Tucanae, with
25, and Terzan 5, with 35.}

The prior on $\varepsilon_i$ can be a function that represents the underlying
distribution of pulsar ellipticities. We will say that the function is defined
by a set of parameters, which in this context are called the hyperparameters,
$\bm{\Theta}$ that are intrinsic to the pulsar population. We can combine
likelihoods for all $N$ pulsars in a search, where for each we marginalize over
the parameters $\bm{\theta}_i$, $D_i$, and $\varepsilon_i$, leaving a likelihood
for $\bm{\Theta}$,
\begin{widetext}
\begin{equation}\label{eq:elllike}
p(\mathbf{X}|\bm{\Theta},I) = \prod_i^N \int^{\varepsilon_i} \int^{\bm{\theta}_i} \int^{D_i} p(\mathbf{x}_i|\varepsilon_i,\bm{\theta}_i,D_i,I) p(\bm{\theta}_i|I) p(\varepsilon_i|\bm{\Theta}, I) p(D_i|\mu_{D_i},\sigma_{D_i},I) {\rm d}\varepsilon_i {\rm d} \bm{\theta}_i {\rm d}D_i.
\end{equation}
\end{widetext}
where $\mathbf{X} \equiv \{\mathbf{x}_i\}$ means the combined data from all
pulsars.\footnote{It is worth emphasizing that even though the same raw data is
used for all pulsars, the precise signal and the Gaussian component of the
colored noise spectrum for each pulsar will be entirely independent. Tracking
the signal's precise phase over a long (weeks, months, or years) observation
time means that the phase templates for different sources (even if there were
billions of sources!) are highly orthogonal.} We are interested in estimating
the parameters of the underlying ellipticity distribution, $\bm{\Theta}$, and
also calculating the evidence for the data given our particular chosen
ellipticity distribution. To do this we further need to define a prior on
$\bm{\Theta}$ and apply Bayes theorem,
\begin{equation}\label{eq:thetapost}
    p(\bm{\Theta}|\mathbf{X},I) = \frac{p(\bm{\Theta}|I)p(\mathbf{X}|\bm{\Theta},I)}{p(\mathbf{X}|I)},
\end{equation}
where the evidence is
\begin{equation}\label{eq:totmarglike}
    p(\mathbf{X}|I) = \int^{\bm{\Theta}} p(\mathbf{X}|\bm{\Theta},I) p(\bm{\Theta}|I) {\rm d}\bm{\Theta}. 
\end{equation}

The posterior on $\bm{\Theta}$ allows us to estimate the distribution of
$\varepsilon$, while we can use the evidence, $p(\mathbf{X}|I)$, for the
hypothesis that the data is consistent with that particular distribution to
perform model selection. For example, we could calculate the ratio of this
evidence to one where the hypothesis is that the data for all pulsars consists
purely of noise, i.e.\ the Bayes factor or odds (assuming prior odds of unity),
and use this as a detection statistic for the ensemble of sources.

\subsection{Ellipticity distribution priors}\label{sec:ellpriors}

We study two different toy models for the underlying ellipticity distribution:
an exponential distribution defined only by its mean, $\mu_{\varepsilon}$,
\begin{equation}\label{eq:exponential}
 p(\varepsilon|\mu_{\varepsilon},I) = \frac{1}{\mu_{\varepsilon}}e^{-\varepsilon/\mu_{\varepsilon}}
\end{equation}
such that $\bm{\Theta} \equiv \mu_{\varepsilon}$; and, a half-Gaussian
distribution peaking at zero, and defined by its width, $\sigma_{\varepsilon}$,
\begin{equation}\label{eq:halfgaussian}
     p(\varepsilon|\sigma_{\varepsilon},I) = \frac{2}{\sqrt{2\pi}\sigma_{\varepsilon}}e^{-\varepsilon^2/2\sigma_{\varepsilon}^2},
\end{equation}
such that $\bm{\Theta} \equiv \sigma_{\varepsilon}$. We note that more
attention is given to the exponential model in most of our examples.
 
We also need to set a prior on $\bm{\Theta}$ for both cases. For both models we
chose a prior on the hyperparameter that is uniform in log-space between some
lower and upper bounds, e.g., for the exponential distribution we have
\begin{equation}
 p(\mu_{\varepsilon}|I) = \left[\ln{\left({\mu_{\varepsilon}}_{\rm max}/{\mu_{\varepsilon}}_{\rm min}\right)}\right]^{-1} \frac{1}{\mu_{\varepsilon}},
\end{equation}
and equivalently for $\sigma_{\varepsilon}$. For the analyses described in
Sec.~\ref{sec:analysis} we use a lower bound of $10^{-10}$ and an upper bound
of $10^{-5}$.
 
In our analyses we assume that all pulsar ellipticities are drawn from these
toy distributions. However, in reality they are likely to be too simplistic to
describe the true $\varepsilon$ distribution. For example, the population of
young pulsars and old recycled millisecond pulsars, have undergone different
evolutions, with the latter having most likely gone through a stage of
accretion (see, e.g., the review in \cite{2008LRR....11....8L}) that spins them
up and reduces their external dipole magnetic field.\footnote{Current searches for gravitational waves from pulsars \cite{2017ApJ...839...12A} consider all pulsars with rotation frequencies above 10\,Hz. Within this subset of $\sim 450$ pulsars, as given by the ATNF Pulsar Catalogue \cite{2005AJ....129.1993M}, about 50 of them have high spin-down rates and as such would be considered ``young'', while the rest are old recycled pulsars.} This could mean that the
distribution of ellipticities for these two populations may be quite different,
and therefore a bimodal distribution, or two independent exponential
distributions with different means, could be more appropriate. Alternatively,
it may be more appropriate to use a quasinonparametric approach, such as the
histogram binning in \cite{2010ApJ...725.2166H}, an infinite Gaussian mixture
model (see, e.g., \cite{2018arXiv180108009D} for a recent example of this in
the context of gravitational waves), or a Gaussian process to model the
ellipticity distribution function space. All these approaches have larger
$\bm{\Theta}$ parameter spaces to marginalize over and are therefore
computationally more challenging. We leave the exploration of these ideas to
future work.

\subsection{Using posterior samples}\label{sec:usepost}

If we have many pulsars, e.g., the 200 used in \cite{2017ApJ...839...12A}, and
wanted to directly calculate Eqs.~(\ref{eq:thetapost}) and
(\ref{eq:totmarglike}) then it would require integrals over $200 \times 4$
independent parameters ($\bm{\theta}_i = \{\cos{\iota},\phi_0,\psi\}_i$ and
$D_i$ for each pulsar), and a large data set consisting of the data for each
pulsar. However, as those parameters and the noise in each pulsar's
gravitational wave data are independent, we can calculate the likelihood over
$\varepsilon_i$ for each pulsar individually, as is already done for current
targeted pulsar searches \cite{2017ApJ...839...12A}. This simplifies
Eq.~\ref{eq:elllike} to be
\begin{equation}\label{eq:likeX}
    p(\mathbf{X}|\bm{\Theta},I) = \prod_i^N \int^{{\varepsilon}_i} p(\mathbf{x}_i|\varepsilon_i,I) p(\varepsilon_i|\bm{\Theta}, I) {\rm d}\varepsilon_i.
\end{equation}

In reality those searches do not output a functional form of the likelihood for
each pulsar, but they use nested sampling \cite{Skilling:2006} to compute the
evidence that the data contains a signal (coherently combined over multiple
detectors), $p(\mathbf{x}_i|I)$, and produce samples drawn from the marginalized
{\it posterior} distribution $p(\varepsilon_i|\mathbf{x}_i,I)$. The posterior
samples therefore need to be converted into a functional form of the likelihood
for use in Eq.~(\ref{eq:thetapost}).\footnote{In reality the software that
performs the nested sampling \cite{2017arXiv170508978P} and outputs the
posterior samples uses $Q_{22}$ rather than $\varepsilon$, but we can easily
convert between the two using Eq.~(\ref{eq:q22}) and assuming a principal
moment of inertia of $10^{38}\,{\rm kg}\,{\rm m}^2$.}

First, to convert these samples into a smooth functional form we can use
kernel density estimation (KDE), with a Gaussian kernel. To remove edge effects
for samples that rail against the lower bound of zero we reflect all the samples
about zero and concatenate these with the original samples before performing the
KDE. When evaluating the resulting KDE only at allowed positive values of
$\varepsilon$ it must then be multiplied by two to get the correct probability
density.

Next, rearranging Bayes theorem shows how to turn a posterior for an individual
pulsar into a likelihood
\begin{equation}\label{eq:bayesrearrange}
    p(\mathbf{x}_i|\varepsilon_i,I) = \frac{p(\varepsilon_i|\mathbf{x}_i,I)}{p(\varepsilon_i|I)}p(\mathbf{x}_i|I),
\end{equation}
where in this case $\varepsilon_i$ is the prior on $\varepsilon$ used for the
individual pulsar. If the prior $p(\varepsilon_i|I)$ is uniform (i.e.\ a
constant) in some range within which the likelihood goes to zero then this is
simple to calculate. In all our examples this is the case, with $\varepsilon_i$
being defined between zero and an upper range $\varepsilon_{\rm max}$, such that
\begin{equation}\label{eq:ellindprior}
p(\varepsilon_i|I) = \varepsilon_{\rm max}^{-1}.
\end{equation}
To be explicit, we are undoing an effective uniform prior that was used when
calculating the original pulsar posteriors, so that we can then reapply our new
ellipticity distribution prior for the whole ensemble. We can therefore
substitute
\begin{equation}\label{eq:bayesrearrange2}
    p(\mathbf{x}_i|\varepsilon_,I) = \varepsilon_{\rm max} p(\varepsilon_i|\mathbf{x}_i,I)p(\mathbf{x}_i|I),
\end{equation}
into Eq.~(\ref{eq:elllike}). We could also work with samples in $h_0$ rather
than ellipticity, and numerically marginalize over distance errors, which we
show in Appendix~\ref{ap:h0samples}, but we will not use $h_0$ samples in this
paper.

We have found that in our analysis, when we have many posteriors that peak at,
or close to, zero, and a prior function that increases rapidly as it approaches
zero, we cannot use the method as used in, e.g., \cite{2010ApJ...725.2166H}.
That method approximates the integral over the likelihood multiplied by the
prior with the expectation value of the prior evaluated at each of the samples.
We find that the accuracy of this approach is significantly reduced for small
values of prior parameters like $\mu_{\varepsilon}$ (see
Sec.~\ref{sec:ellpriors} above) when there are no samples with roughly
equivalent values.

\subsection{Spin-down limits}\label{sec:sdlims}

We could use the electromagnetically-derived spin-down limits described briefly
in Sec.~\ref{sec:intro} to infer the hyperparameters of our ellipticity
distribution. For each pulsar we have a spin-down limit on $\varepsilon$ based
on its observed electromagnetic spin-down, independent of the pulsar's distance
(combining Eqs. (5) and (6) of \cite{2014ApJ...785..119A})
\begin{equation}\label{eq:ellspindown}
\varepsilon^{\rm sd} = \frac{1.9\!\times\!10^{-8}}{f_{\rm kHz}^{5/2}} \left( \frac{|\dot{f}_{\rm rot}|}{10^{-11}\,{\rm Hz}\,{\rm s}^{-1}} \right)^{1/2} \left( \frac{10^{38}\,{\rm kg}\,{\rm m}^2}{I_{zz}} \right)^{1/2}.
\end{equation}
The observed frequency derivatives of pulsars are not necessarily their true
values, as they will be contaminated by proper motion effects (the Shklovskii
effect \cite{1970SvA....13..562S}), differential galactic rotation
\cite{1991ApJ...366..501D}, and, for pulsars in globular clusters, local
accelerations. Indeed, these effects lead to some pulsars being observed to
spin-up. Therefore in any inference using spin-down limits one would have to
either correct for these effect, or exclude pulsars for which the effects cannot
be estimated.

We can use these spin-down limit derived values to provide a likelihood on
$\varepsilon_i$ for each pulsar when estimating the underlying distribution's
hyperparameters. For an individual pulsar we can (simplistically) say that the
spin-down limit gives us a flat likelihood on $\varepsilon$
\begin{equation}\label{eq:sdprior}
p({d_{\rm EM}}_i|\varepsilon_i,I) = \begin{cases} 1/\varepsilon_i^{\rm sd} & \text{if } 0 \leq \varepsilon_i \leq \varepsilon^{\rm sd}_i, \\
0 & \text{otherwise} \end{cases}
\end{equation}
where ${d_{\rm EM}}_i$ is the electromagnetically derived information, i.e.,
from observed pulse time of arrivals, that gives the spin-down limit for a given
pulsar.

These likelihoods for each pulsar can be combined into a joint likelihood, and
form a likelihood on the distribution's hyperparameters, $\Theta$,
\begin{equation}
p(\{\mathbf{d}_{\rm EM}\}|\Theta,I) = \prod_i^N \left(  \int p({d_{\rm EM}}_i|\varepsilon_i,I) p(\varepsilon_i|\Theta,I){\rm d}\varepsilon_i \right).
\end{equation}

If we assume the underlying distribution is exponential (so $\Theta \equiv
\mu_\varepsilon$) we have
\begin{equation}
    p(\{\mathbf{d}_{\rm EM}\}|\mu_\varepsilon,I) = \prod_i^N \left(\frac{1}{\varepsilon_i^{\rm sd}} \right) \int_0^{\varepsilon_i^{\rm sd}} \frac{1}{\mu_\varepsilon}e^{-\varepsilon_i/\mu_\varepsilon} {\rm d}\varepsilon_i,
\end{equation}
where we can see that the integral is just the cumulative distribution function
for an exponential distribution, which is given by ${\rm CDF}(\varepsilon_i^{\rm
sd},\mu_\varepsilon) = 1-e^{-\varepsilon_i^{\rm sd}/\mu_\varepsilon}$. So, we
have
\begin{equation}
    p(\{\mathbf{d}_{\rm EM}\}|\mu_\varepsilon,I) \propto \prod_i^N 1-e^{-\varepsilon_i^{\rm sd}/\mu_\varepsilon}.
\end{equation}
A similar result can be found for the half-Gaussian distribution using its CDF.

\begin{figure}[!htbp]
\includegraphics[width=\columnwidth]{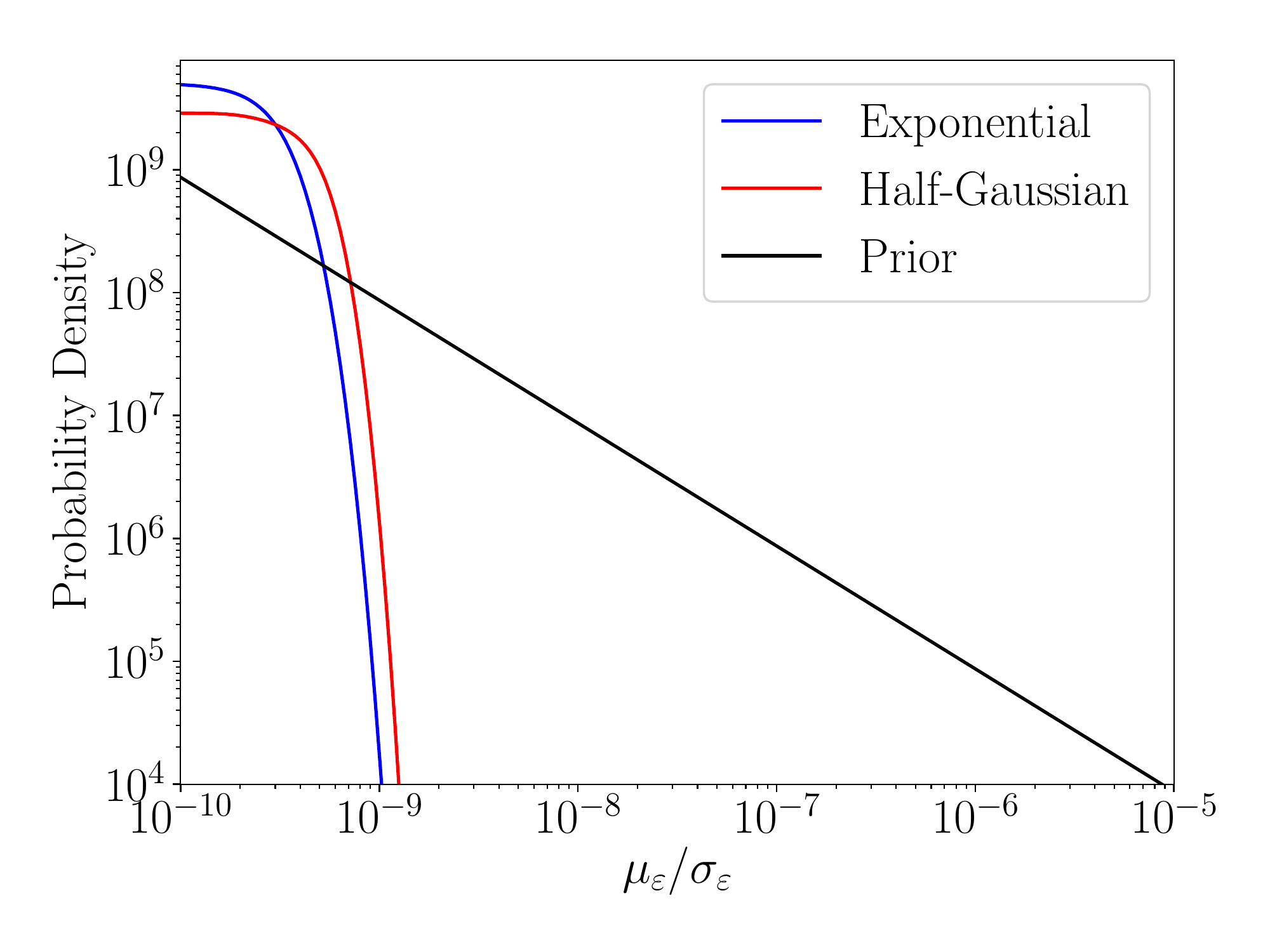}
\caption{\label{fig:spindownlimit}
The posterior probability distributions on the ellipticity distribution prior
hyperparameters $\mu_\varepsilon$ and $\sigma_\varepsilon$ for the exponential
and half-Gaussian distributions when using electromagnetically-derived spin-down
limits.}
\end{figure}

Using the priors given in Sec.~\ref{sec:ellpriors} we can calculate the
posteriors $p(\mu_\varepsilon|\{\mathbf{d}_{\rm EM}\},I)$ and
$p(\sigma_\varepsilon|\{\mathbf{d}_{\rm EM}\},I)$ for the exponential and
half-Gaussian distributions, respectively. To do this we use all pulsars in the
ATNF Pulsar Catalogue \cite{2005AJ....129.1993M}, excluding those in globular
clusters and with observed spin-ups, estimate their intrinsic period
derivatives (by calculating the combined Shklovskii and galactic rotation
effects using best-fit distances from the catalogue), and calculate
their spin-down limits using Eq.~(\ref{eq:ellspindown}). For this simple
analysis we ignore uncertainties in each pulsar's moment of inertia. The
posterior probability distributions on $\mu_\varepsilon$ and
$\sigma_\varepsilon$ for the exponential and half-Gaussian are shown in
Fig.~\ref{fig:spindownlimit}. We find 90\% credible upper bounds on these two
hyperparameters of $\mu_\varepsilon^{90\%} \le 3.0\!\times\!10^{-10}$ and
$\sigma_\varepsilon^{90\%} \le 4.1\!\times\!10^{-10}$.

In the rest of this paper we will purely assume the use of gravitational wave
data alone for making inferences about the ellipticity distribution. However, an
interesting extension to this work could be combining the gravitational wave data
for pulsars, where that places a more stringent limit than the spin-down limit,
with spin-down limits for the pulsars where it does not.

\subsection{Bayes factor}\label{sec:bayesfactor}

We saw in Eq.~(\ref{eq:totmarglike}) that we can calculate the evidence that
the data is consistent with a particular parameterized ellipticity distribution
(i.e., the exponential or half-Gaussian distributions). However, to make this
number informative we need to be able to compare it to another model. We can
form the odds between this evidence and one given by a different hypothesis,
such as the data from all pulsars being entirely consistent with Gaussian noise,
e.g.\
\begin{equation}\label{eq:odds}
\mathcal{O}^{\varepsilon}_{\rm n} = \frac{p(\mathbf{X}|H_{\varepsilon},I)}{p(\mathbf{X}|H_{n},I)} \frac{p(H_{\varepsilon}|I)}{p(H_n|I)},
\end{equation}
where $H_{\varepsilon}$ explicitly denotes the hypothesis of our data containing
signals drawn from the particular underlying ellipticity distribution, and $H_n$
denotes the noise-only hypothesis. The first term on the right-hand side, the
ratio of evidences, is often called the Bayes factor, while the second term is
the prior odds of the two hypotheses, which we will generally take as unity
unless otherwise stated.

\subsubsection{Noise hypothesis}

The simplest way to form the evidence for the noise hypothesis is just as the
product of the null likelihoods for all pulsars, where the null likelihood is
the likelihood with the signal model explicitly set of zero (see, e.g.,
Sec.~2.2.3 of \cite{2017arXiv170508978P} for the null likelihoods used in the
analysis presented here). For a particular pulsar $i$ and detector $j$ the null
likelihood for the hypothesis that the data is consistent with noise alone,
$H_{\rm n}$, is given by
\begin{equation}
    p(\mathbf{x}_i^j|H_{\rm n}^j, I) \equiv p(\mathbf{x}_i^{j}|\varepsilon_i=0,I),
\end{equation}
i.e., it is just the likelihood [as seen in, e.g.,
Eq.~(\ref{eq:eDposterior})] evaluated with the signal model set to zero.
Therefore, with $\mathbf{x}_i \equiv \{\mathbf{x}_i^j\}$, the multidetector
($N_{\rm det}$) noise hypothesis for a single pulsar is
\begin{equation}
    p(\mathbf{x}_i|H_{\rm n}, I) = \prod_{j}^{N_{\rm det}} p(\mathbf{x}_i^j|H_{\rm n}^j, I),
\end{equation}
and evidence for the noise hypothesis for all pulsars is
\begin{equation}
    p(\mathbf{X}|H_{\rm n}, I) = \prod_{i} p(\mathbf{x}_i|H_{\rm n}, I).
\end{equation}

In reality such a noise hypothesis is not robust against instrumental spectral
disturbances and lines in the data (see, e.g., \cite{2014PhRvD..89f4023K}),
which can contaminate the frequencies near, or at, those of an expected
astrophysical pulsar signal. If data from more than one detector is available, a
more robust noise hypothesis is to allow it to incorporate incoherent signals
\emph{or} noise between detectors. When using multiple detectors, the individual
pulsar signal hypothesis enforces coherence between them, so we can more
explicitly call the hypothesis the \emph{coherent} hypothesis, $H_{\rm coh}$,
which is what the evidence term $p(\mathbf{x}_i|I)\equiv
p(\{\mathbf{x}_i^j\}|H_{\rm coh},I)$ in Eqs.~(\ref{eq:eDposterior}),
(\ref{eq:bayesrearrange}) and (\ref{eq:bayesrearrange2}) truly represents.

The evidence for the new noise hypothesis for each pulsar, which we will call
the \emph{incoherent} hypothesis, is described in more detail in Sec.~2.6 of
\cite{2017arXiv170508978P}, or the Appendix of \cite{2017ApJ...839...12A}. In
all the analyses presented here we use two detectors (the LIGO Hanford and
Livingston detectors, which we will denote as H1 and L1 respectively from here
onwards), so for each pulsar the evidence for the incoherent hypothesis, $H_{\rm
in}$ is
\begin{equation}\label{eq:incohind}
p(\mathbf{x}_i|H_{\rm in}) = \prod_{j \in \{{\rm H1},{\rm L1}\}} \left[ p(\mathbf{x}_i^j|H_{\rm s}^j,I) + p(\mathbf{x}_i^j|H_{\rm n}^j,I) \right],
\end{equation}
where $H_{\rm s}^j$ is the evidence for the hypothesis that the data is
consistent with a signal for a single detector. This is generalizable to any
number of detectors. Again, we can take the product of this to give the
incoherent hypothesis for all $N$ pulsars
\begin{equation}\label{eq:incoh}
    p(\mathbf{X}|H_{\rm in}) = \prod_{i}^N p(\mathbf{x}_i|H_{\rm in}).
\end{equation}
This implies that in Eq.~(\ref{eq:odds}) we would substitute $H_{\rm n}
\equiv H_{\rm in}$, which is the case for all results presented here.

\subsubsection{Nonhierarchical statistic}

Without using any of the machinery in defining an underlying ellipticity
distribution, and hyperparameters, described previously in this section, we can
form a more na\"ive odds. This assumes all the pulsars have mass quadrupoles
(not ellipticities) drawn from the same flat prior distribution with fixed
bounds (between zero and a large value at which all likelihoods have approached
zero). To differentiate it from the above odds formed using an ellipticity
distribution with unknown hyperparameters, we call this our
\emph{nonhierarchical} statistic.

As described above, for each pulsar we have the evidence for the coherent signal
model $p(\{\mathbf{x}_i^j\}|H_{\rm coh},I)$ and the null likelihood
$p(\mathbf{x}_i|H_{\rm n},I)$. Following a similar route to Eqs.~(49--52) of
\cite{2017PhRvD..96d2001I} we can form the probability (\emph{not} the evidence
here as we are explicitly including priors on the hypotheses) for a compound
hypothesis of any combination of individual signal and noise hypotheses
\begin{align}\label{eq:compound}
p(H_{\rm com}|\mathbf{X},I) \propto &
    \begin{aligned}[t]\prod_{i}^{N} \Big[&p(\mathbf{x}_i|{H_{\rm coh}}_i,I)p({H_{\rm coh}}_i|I) + \\
    &p(\mathbf{x}_i|{H_{\rm in}}_i,I)p({H_{\rm in}}_i|I) \Big] - \end{aligned} \nonumber \\
 & \prod_{i}^{N} p(\mathbf{x}_i|{H_{\rm in}}_i,I)p({H_{\rm in}}_i|I),
\end{align}
where $p({H_{\rm coh}}_i|I)$ and $p({H_{\rm in}}_i|I)$ are the priors for each
hypothesis, which here we explicitly state for each pulsar $i$. The second
product term on the right-hand side of Eq.~(\ref{eq:compound}) is also the
probability for the all-pulsar noise hypothesis, so we can produce an odds
between the two probabilities (noting that they would both have the same
proportionality coefficient and it would thus cancel out) as
\begin{align}\label{eq:independentodds}
    \mathcal{O} \equiv & \frac{p(H_{\rm
com}|\mathbf{X},I)}{p(H_{\rm in}|\mathbf{X},I)} = \frac{p(\mathbf{X}|H_{\rm
com},I)}{p(\mathbf{X}|H_{\rm in},I)}\frac{p(H_{\rm
com}|I)}{p(H_{\rm in}|I)} \nonumber \\
=&  \begin{aligned}[t]\Bigg(\prod_{i}^{N} \Big[&p(\mathbf{x}_i|{H_{\rm coh}}_i,I)p({H_{\rm coh}}_i|I) + \\
&p(\mathbf{x}_i|{H_{\rm in}}_i,I)p({H_{\rm in}}_i|I) \Big]\Bigg)\times\end{aligned} \nonumber \\
& \left(\prod_{i}^{N} p(\mathbf{x}_i|{H_{\rm in}}_i,I)p({H_{\rm in}}_i|I)\right)^{-1} - 1.
\end{align}
If we state that we have equal \emph{a priori} probability for every combination of the coherent signal and incoherent hypotheses, e.g.,
\begin{align}
    \prod_{i=1}^N p({H_{\rm in}}_i|I) &= \prod_{i=2}^N p({H_{\rm in}}_i|I) \prod_{j=1}^1 p({H_{\rm coh}}_j|I) \nonumber \\
    &= \prod_{i=3}^N p({H_{\rm in}}_i|I) \prod_{j=1}^3 p({H_{\rm coh}}_j|I) \nonumber \\
    & = \ldots,
\end{align}
then Eq.~(\ref{eq:independentodds}) becomes
\begin{equation}\label{eq:independentodds2}
    \mathcal{O} = 
\frac{\prod_{i}^{N} \left[p(\mathbf{x}_i|{H_{\rm coh}}_i,I) + 
p(\mathbf{x}_i|{H_{\rm in}}_i,I) \right]}{\prod_{i}^{N} p(\mathbf{x}_i|{H_{\rm in}}_i,I)} - 1,
\end{equation}
where implicitly $p(H_{\rm com}|I) = (2^N-1)p(H_{\rm in}|I)$. Therefore, to
assign equal prior odds for the compound hypothesis that data for any pulsar
contains a signal one must use
\begin{equation}\label{eq:Oind}
    \mathcal{O}^{\rm NH}_{\rm n} = \frac{\mathcal{O}}{2^N-1}.
\end{equation}
We will use this as a comparison to Eq.~(\ref{eq:odds}) in the subsequent
analyses.

\subsubsection{$\mathcal{B}$-statisticlike comparison}

It is interesting to compare the above statistics to that proposed in
\cite{2005PhRvD..72f3006C}. Due to how we have constructed our data sets, it is
not simple to calculate the $\mathcal{F}$-statistic (which is the log-likelihood
ratio maximized over the parameter space) for our simulated data for each
pulsar. However, the natural logarithm of our signal versus noise evidence
ratios for each pulsar are essentially the $\mathcal{B}$-statistic of
\cite{2009CQGra..26t4013P} (although we additionally marginalize over pulsar
distance) including the line robust incoherent noise denominator
\cite{2014PhRvD..89f4023K} in our Bayes factors. Therefore, as a comparison
using an ensemble of pulsars, we can simply sum these $\mathcal{B}$-statistics.
In our notation, where we are not implicitly in log-space, we would have a
product rather than a sum of
\begin{equation}\label{eq:bstat}
\mathcal{O}^{\mathcal{B}}_{\rm n} = \prod_{i}^N \frac{p(\mathbf{x}_i|{H_{\rm coh}}_i, I)}{p(\mathbf{x}_i|{H_{\rm in}}_i,I)}.
\end{equation}
As was shown in \cite{2009CQGra..26t4013P} the $\mathcal{B}$-statistic is a
slightly more efficient discriminator between signal and noise than the
$\mathcal{F}$-statistic, so we would expect this to produce a comparable, but
slightly more efficient statistic than that in~\cite{2005PhRvD..72f3006C}.

On a final note, all the above likelihood, prior, posterior, and odds
calculations are in practice computed entirely using the natural logarithm of
the values to avoid numerical underflow and overflow.


\section{Analysis}\label{sec:analysis}

Here we will discuss analyses that have been performed on simulated data to
assess the performance of the three odds given in Eqs.~(\ref{eq:odds}),
(\ref{eq:independentodds2}) and (\ref{eq:bstat}) for detecting an ensemble of
gravitational wave signals from pulsars. In the former case this assumes that
all pulsars have ellipticities drawn from some underlying distribution defined
by unknown hyperparameters, while the latter two assume no unknown
hyperparameters for the distribution. In the former case we also assess how well
we can recover the hyperparameters defining the ellipticity distribution.

\subsection{Simulations}\label{sec:simulations}

To make these assessments we have produced a series of simulated data sets to
account for different ellipticity distributions, hyperparameter values, and for
different realizations of noise. In all cases we take the 200 pulsars searched
for in the analysis of LIGO Observing Run 1 (O1) \cite{2017ApJ...839...12A} as
our sample of sources, with their sky positions and best-fit distances obtained
from the ATNF Pulsar Catalogue \cite{2005AJ....129.1993M}. We also create data
from two detectors, the LIGO Hanford (H1) and Livingston (L1) observatories,
assuming that they are operating at their advanced design sensitivities
\cite{aLIGOdesign,2016LRR....19....1A} over one year with a 100\% duty cycle.
For each pulsar a complex time series is simulated at a sample rate of once per
1800\,s, to replicate the data that would be produced in a real targeted pulsar
search (e.g., \cite{2017ApJ...839...12A}) following the application of a
heterodyne procedure \cite{2005PhRvD..72j2002D} to remove the rapidly varying
signal phase evolution.

Firstly, we will discuss generating sources from the exponential ellipticity
distribution defined in Eq.~(\ref{eq:exponential}). There is a single
hyperparameter that defines the distribution, $\mu_\varepsilon$, the mean of the
distribution. We take 15 values of $\mu_\varepsilon$ spaced uniformly in
log-space between $5\ee{-10}$ and $5\ee{-8}$. For an exponential distribution
defined by a particular $\mu_\varepsilon$ value, we randomly draw $\varepsilon$
values for each pulsar, which (assuming $I_{zz} = 10^{38}\,{\rm kg}\,{\rm m}^2$)
we convert into that pulsar's equivalent $h_0$ via Eq.~(\ref{eq:h0}). We also
randomly generate values of $\cos{\iota}$, $\psi$ and $\phi_0$ for each pulsar,
drawn uniformly from the ranges $[-1, 1]$, $[0, \pi/2]$ and $[0, \pi]$,
respectively. Using these values, the time series' for the two detectors are
generated via the signal model defined in Equation~13 of
\cite{2005PhRvD..72j2002D} with additive Gaussian noise with zero mean and
standard deviation derived from the Advanced LIGO (aLIGO) design curve
sensitivity \cite{aLIGOdesign} at the appropriate gravitational wave frequency
(twice the pulsar's rotation frequency). For each pulsar we also calculate the
signal-to-noise ratio it would have via Eq.~(2) of \cite{2011MNRAS.415.1849P}
for the two-detector fully coherent analysis.

For $\mu_\varepsilon$ values below $\sim 9.6\ee{-10}$ and above $4.5\ee{-9}$ we
regenerate the ensemble of sources 10 times using different random seeds, while
for values between that range we regenerate the ensemble 100 times to provide a
better statistical sample. So, to summarize, assuming an underlying exponential
distribution of ellipticities, we have $(9 \times 10) + (6 \times 100) = 690$
realizations of data for two detectors containing signals from an ensemble of
200 pulsars.

We perform exactly the same steps for the half-Gaussian distribution defined in
Eq.~(\ref{eq:halfgaussian}), where we take 15 values of $\sigma_\varepsilon$
spaced uniformly in log-space over the same range as above.\footnote{The mean of
a half-Gaussian distribution, which may be more directly comparable to the
exponential distribution mean, is given by $(\sqrt{2/\pi})\sigma_\varepsilon$.}
In this case we only create 10 ensembles of pulsars for each value of
$\sigma_\varepsilon$. So, we have $15\times10 = 150$ realizations of data for
two detectors containing signals from an ensemble of 200 pulsars.

It is very useful to have ``background'' data sets with which to compare odds
values from ensembles containing signals. So, for this purpose we have also
generated 400 realizations of our data sets of 200 pulsars, with each pulsar's
data containing simulated Gaussian noise derived from the aLIGO design
sensitivity as above, but containing no signal.

\subsection{Processing the data}\label{sec:processing}

For each pulsar in each ensemble realization we run the parameter estimation and
evidence evaluation code \cite{2017arXiv170508978P} used for real known pulsar
searches such as \cite{2017ApJ...839...12A}. For each pulsar the code is run
individually for the two detectors, and with data from both detectors combined
coherently. The code uses a nested sampling algorithm
\cite{Skilling:2006,Veitch:2010,2015PhRvD..91d2003V} to evaluate the model
evidence given by the integral in Eq.~(\ref{eq:eDposterior}), and also
outputs the null likelihood in each case. For all pulsars, the code set up was
identical and 512 ``live,'' or ``active,'' points were used to initialize the nested
sampling algorithm. For the signal variables $Q_{22}$, $\cos{\iota}$, $\phi_0$,
and $\psi$ we defined uniform priors in the ranges $[0, 10^{37}]\,{\rm kg}\,{\rm
m}^2$, $[-1, 1]$, $[0, \pi]$\,rad, and $[0, \pi/2]$\,rad, respectively. The
distance was also included as a variable and assigned a Gaussian prior with a
mean given by each pulsar's best fit distance, and a standard deviation of 20\%
of that value, and a hard cutoff at zero. Note that we did not sample in
$\varepsilon$, but the $Q_{22}$ samples were easily converted into $\varepsilon$
samples via the relation in Eq.~(\ref{eq:q22}).

Nested sampling outputs a chain of samples, where each sample is a vector
containing particular values of each of the variable parameters
$\{\varepsilon_j, \cos{\iota}_j, {\phi_0}_j, \psi_j, D_j\}$. These are then
resampled to provide draws from the posterior distribution using the method
described in \cite{2015PhRvD..91d2003V}. As described in
Sec.~\ref{sec:usepost}, for Eq.~(\ref{eq:bayesrearrange2}), which must be
inserted into Eq.~(\ref{eq:likeX}), we cannot use posterior samples, but
instead need a functional form of the posterior. We use Gaussian kernel density
estimation (KDE), in particular the method implemented in the {\sc Scikit-learn}
Python package \cite{scikit-learn}, to convert the $\varepsilon$ samples into a
function that can be evaluated. In practice, when samples are cut-off at a hard
boundary, such as not being allowed to be negative for the $\varepsilon$ value,
it can lead to boundary artifacts in the KDE. So, to avoid such artifacts we
produce a copy of the samples, reflect them about zero, and concatenate them to
the original samples before performing the KDE. KDEs are also dependent on the
bandwidth chosen for the Gaussian kernels. We use the Scott rule-of-thumb
\cite{scott} to estimate the bandwidth of the kernel, but only using the
original samples rather than the concatenated version to avoid the kernel being
too broad. The use of a finite number of posterior samples, and their conversion
into a KDE, means that there will be some associated uncertainties that will
propagate through the analysis, which are discussed briefly in
Appendix~\ref{ap:evidence}.

For each ensemble of pulsars the outputs from the above processing for the
coherent two-detector analyses are inserted into Eq.~(\ref{eq:likeX}) to form
the likelihood for a particular ellipticity distribution model. This is then
used in Eq.~(\ref{eq:totmarglike}), along with the hyperparameter priors in
Sec.~\ref{sec:ellpriors}, to evaluate the evidence for that distribution (in
practice trapezoidal integration is used over the hyperparameter range). This
then gives the evidence for the numerator of Eq.~(\ref{eq:odds}). We also
obtain the posterior distribution on the hyperparameter via
Eq.~(\ref{eq:thetapost}). For both the simulated distributions (the
exponential and half-Gaussian) we calculate Eq.~(\ref{eq:totmarglike}) twice,
once with the actual distribution used for the simulations and once assuming the
alternative distribution, i.e., for the ensembles containing sources with
ellipticities drawn from an exponential distribution we calculate the evidence
that the distribution was exponential \emph{and} the evidence that the
distribution was half-Gaussian. This enables us to do model comparison between
the two distributions.

Likewise the outputs from the individual detector analyses can be combined via
Eqs.~(\ref{eq:incohind}) and (\ref{eq:incoh}) to form the denominator of
Eq.~(\ref{eq:odds}). This means that for each ensemble of pulsars drawn from
a given distribution we have two values of $\mathcal{O}^{\varepsilon}_{\rm n}$.
The first is $\mathcal{O}^{\varepsilon_{\rm exp}}_{\rm n}$ for the assumption
that the distribution is exponential, and the second is
$\mathcal{O}^{\varepsilon_{\rm hg}}_{\rm n}$ for the assumption that it is
half-Gaussian.

Similarly, we use all the individual pulsar signal and noise evidences to
produce values for $\mathcal{O}^{\rm NH}_{\rm n}$ via
Eqs.~(\ref{eq:independentodds2}) and (\ref{eq:Oind}), and values for
$\mathcal{O}^{\mathcal{B}}_{\rm n}$ via Eq.~(\ref{eq:bstat}), for each
ensemble.


\section{Results}

\subsection{Odds values}

\begin{figure*}[!htbp]
    \begin{tabular}{c}
    \includegraphics[width=\textwidth]{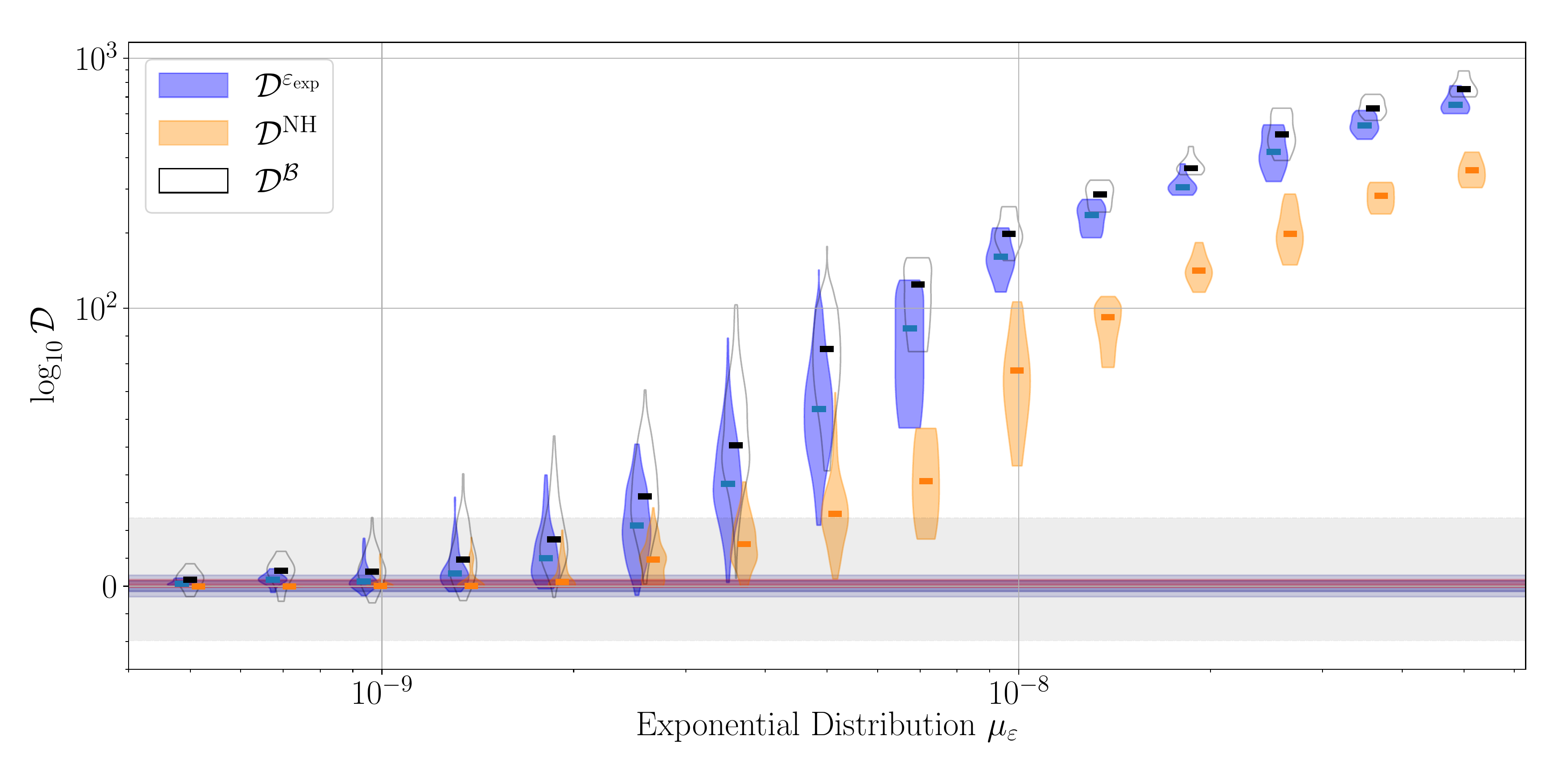} \\
    \includegraphics[width=\textwidth]{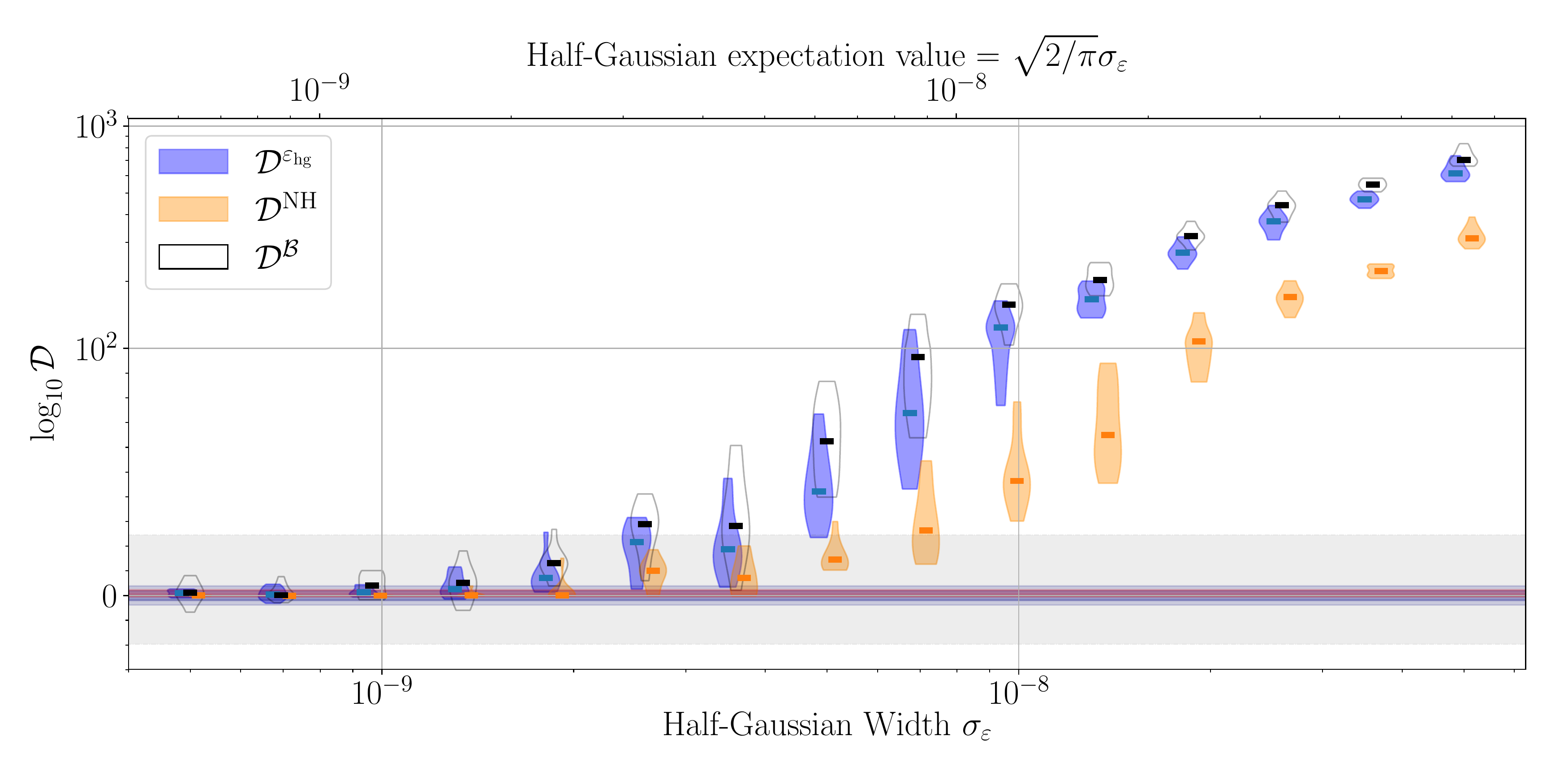}
    \end{tabular}
    \caption{\label{fig:ensemble_exp} The distributions of odds values for the
    ensembles of pulsars with ellipticities drawn from, in the upper panel,
    exponential distributions with a range of mean values, $\mu_\varepsilon$,
    and in the lower panel, half-Gaussian distributions with a range of widths,
    $\sigma_\varepsilon$ (the mean of the half-Gaussian distributions,
    $(\sqrt{2/\pi})\sigma_\varepsilon$, is displayed on the top axis). The
    distributions for the statistic assuming there is a common distribution
    $\mathcal{D}^{\varepsilon_{{\rm exp}/{\rm hg}}}$, using the nonhierarchical
    method $\mathcal{D}^{\rm NH}$, and using the $\mathcal{B}$-statistic-based
    method $\mathcal{D}^{\mathcal{B}}$, are shown. The shaded bands show the
    $5\sigma$ extent of the background distributions in all cases (for clarity
    these are shown in greater detail in Fig.~\ref{fig:ensemble_exp_snr}).
    Note that the y-axis is linear below $10^2$ and logarithmic above it.}
\end{figure*}

For each ensemble we have calculated the value of the odds for a given
$\varepsilon$ distribution: $\mathcal{O}^{\varepsilon_{\rm exp}}_{\rm n}$ and
$\mathcal{O}^{\varepsilon_{\rm hg}}_{\rm n}$. We have also calculated the
non-hierarchical odds assuming no unknown hyperparameters $\mathcal{O}^{\rm
ind}_{\rm n}$, and $\mathcal{O}^{\mathcal{B}}_{\rm n}$. This has been done for
the 400 background ensembles in which the data for each pulsar purely contains
Gaussian noise. For reasons we discuss in Appendix~\ref{ap:absodds} the true
scaling of odds values may not be reliable. However, as we are able to calculate
the odds for a background distribution in an identical way to those containing
signals, we can remove any scaling dependent effects by looking at ratios with
respect to the background. So, we produce a detection statistic from the ratio
between the observed odds for each ensemble containing signals and the
\emph{mean} odds from the background ensembles
\begin{equation}\label{eq:dstat}
\mathcal{D} = \frac{\mathcal{O}}{\langle \mathcal{O}_{\rm background} \rangle},
\end{equation}
where we might have, e.g.,
\begin{equation}\label{eq:dstatexp}
    \mathcal{D}^{\varepsilon_{\rm exp}} = \frac{\mathcal{O}^{\varepsilon_{\rm exp}}_{\rm n}}{\langle {\mathcal{O}^{\varepsilon_{\rm exp}}_{\rm n}}_{\rm background} \rangle},
\end{equation}
or equivalent values of $\mathcal{D}^{\rm NH}$ for the nonhierarchical odds
$\mathcal{O}^{\rm NH}_{\rm n}$, and $\mathcal{D}^{\mathcal{B}}$ for the
$\mathcal{B}$-statistic based odds from Eq.~(\ref{eq:bstat}). The base-10
logarithm of these three ratios is what is shown in
Fig.~\ref{fig:ensemble_exp}, and subsequent figures.

In the upper panel of Fig.~\ref{fig:ensemble_exp} we see that the statistic
$\mathcal{D}^{\varepsilon_{\rm exp}}$ appears to become completely disjoint from
the background distribution at values of $\mu_\varepsilon \approx 5\ee{-9}$,
while the nonhierarchical statistic $\mathcal{D}^{\rm NH}$ becomes disjoint a
little higher at $\mu_\varepsilon \approx 7\ee{-9}$. We see that the
$\mathcal{B}$-statistic-based distributions, $\mathcal{D}^{\mathcal{B}}$, have
generally larger values, but have a much broader background distribution than
the other two statistics (we see how this translates into efficiency in
Sec.~\ref{sec:efficiency}). We see a rather similar situation in the lower
panel of Fig.~\ref{fig:ensemble_exp} for $\mathcal{D}^{\varepsilon_{\rm hg}}$
with the distributions becoming completely disjoint from the background at
$\sigma_\varepsilon \approx 5\ee{-9}$, or $(\sqrt{2/\pi})\sigma_\varepsilon
\approx 6\ee{-9}$.

\begin{figure*}[!htbp]
    \begin{tabular}{cc}
    \includegraphics[width=\columnwidth]{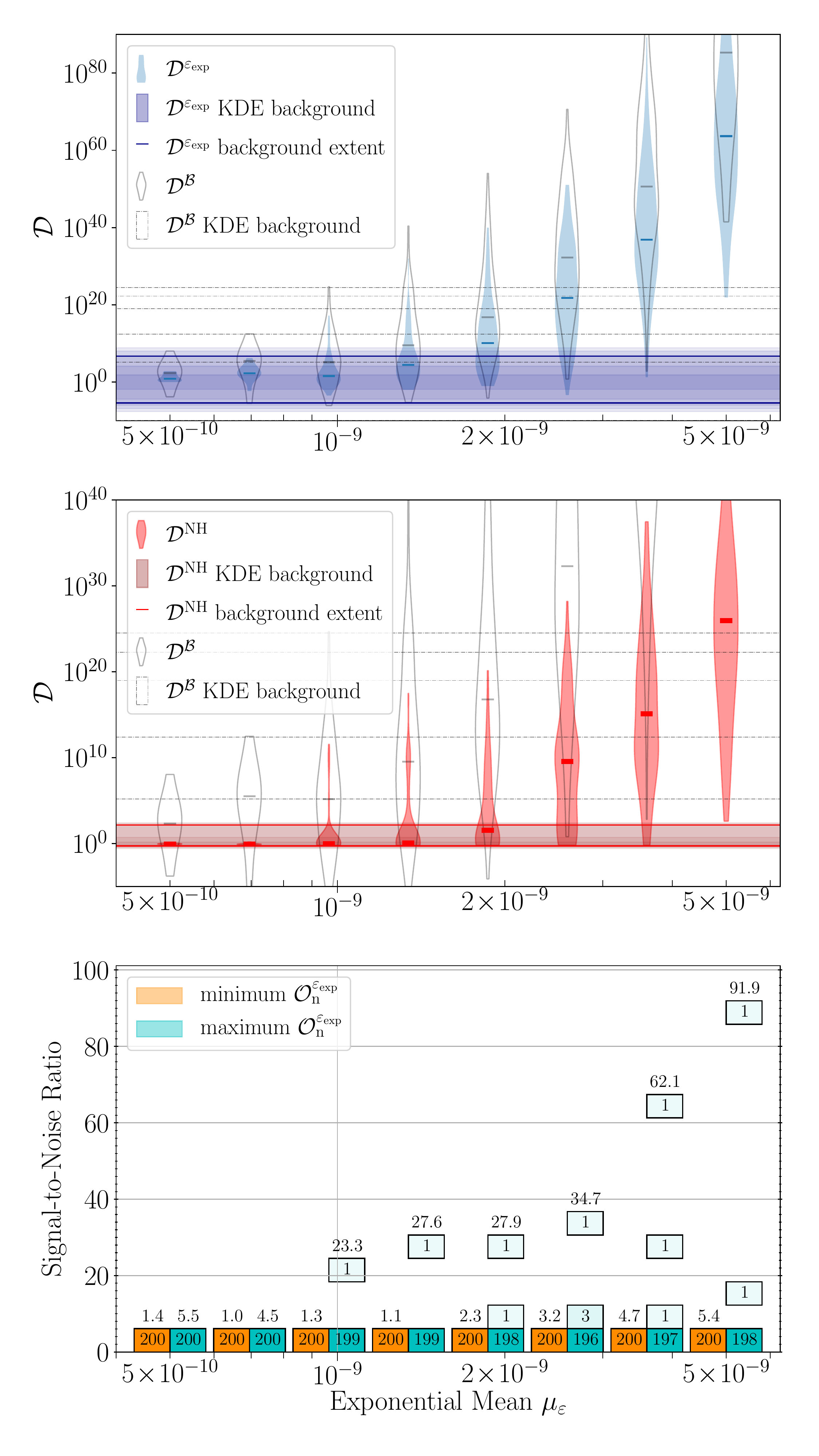}
    &
    \includegraphics[width=\columnwidth]{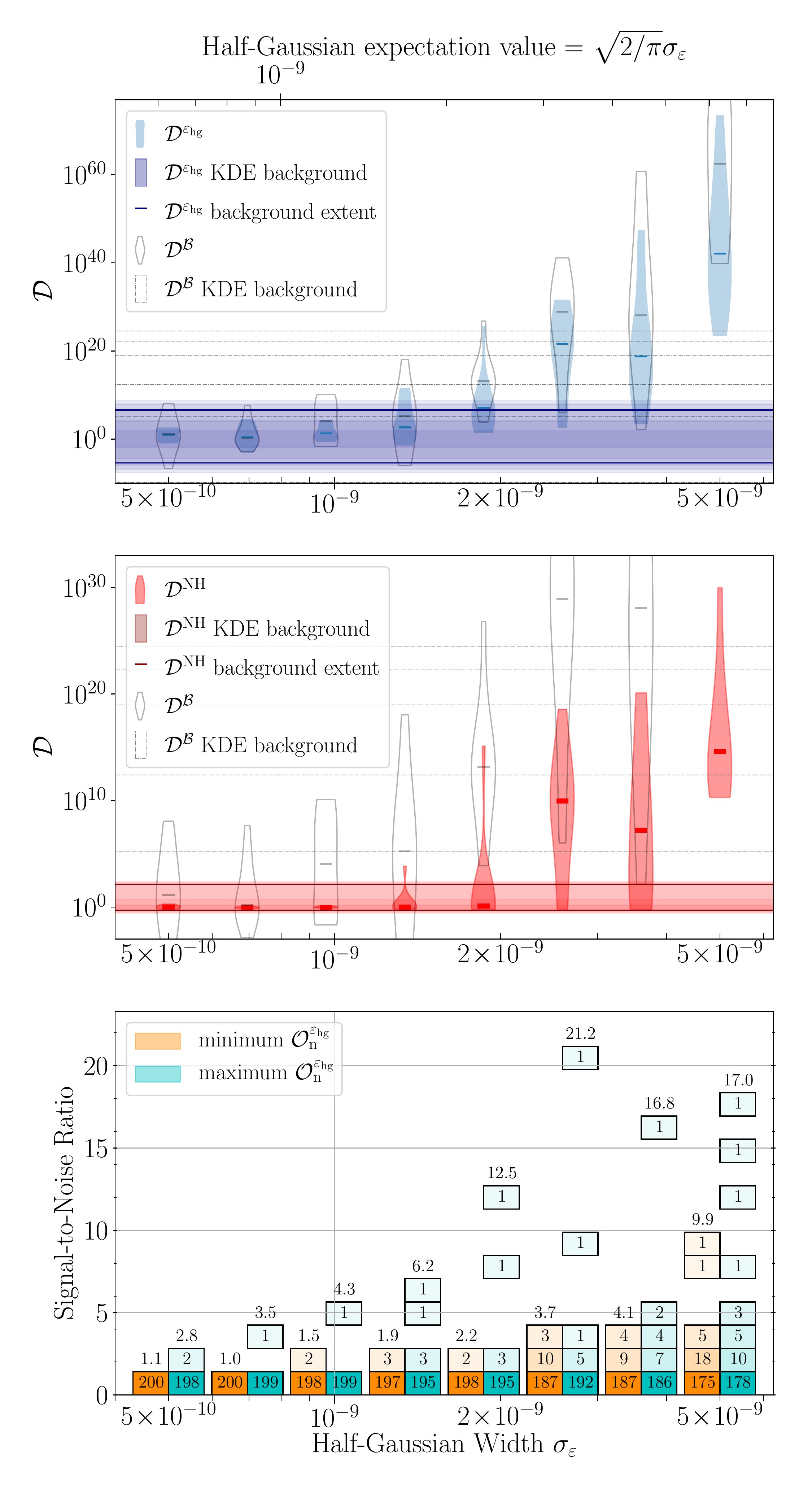}
    \end{tabular}
        \caption{\label{fig:ensemble_exp_snr} The left panel shows results for
        the exponential distribution simulations and the right panel shows
        results for the half-Gaussian distribution simulations. On each side the
        top panel shows the distribution of detection statistics
        $\mathcal{D}^{\varepsilon_{\rm exp}}$
        (left)/$\mathcal{D}^{\varepsilon_{\rm hg}}$ (right) for ensembles with
        distributions defined by eight different means/widths, $\mu_\varepsilon$
        (left)/$\sigma_\varepsilon$ (right), with the equivalent
        $\mathcal{D}^{\mathcal{B}}$ distribution shown faintly for comparison.
        The middle panel shows the distributions of $\mathcal{D}_{\rm NH}$
        (again with $\mathcal{D}^{\mathcal{B}}$ shown for comparison). In both
        these panels the horizontal shaded regions show the distribution for the
        background (noise-only) ensembles (see text), with the solid horizontal
        lines showing the maximum and minimum extent of the distribution. The
        background distributions for the $\mathcal{D}^{\mathcal{B}}$ values are
        shown as the faint dashed lines. The bottom panel shows the distribution
        of signal-to-noise ratios for the ensemble with the minimum (left box of
        each pair) and maximum (right box of each pair)
        $\mathcal{D}^{\varepsilon_{\rm exp}}$
        (left)/$\mathcal{D}^{\varepsilon_{\rm hg}}$ (right) value, with numbers
        inside the boxes showing the histogram count, and numbers above the
        boxes showing the maximum signal-to-noise ratio for that ensemble.}
\end{figure*}

In Fig.~\ref{fig:ensemble_exp_snr} we show a zoomed in version of
Fig.~\ref{fig:ensemble_exp} for both the exponential distribution (left) and
half-Gaussian (right), with the results for $\mathcal{D}^{\varepsilon_{\rm
exp}}$/$\mathcal{D}^{\varepsilon_{\rm hg}}$ and $\mathcal{D}^{\rm NH}$ split
into separate panels (top and middle respectively). The background distribution
is shown in two ways: horizontal solid lines show the upper and lower extent of
the 400 background realizations; shaded bands show the bounds from $1\sigma$ to
$5\sigma$ on the distribution. For comparison the distributions of
$\mathcal{D}^{\mathcal{B}}$ and its background, also with bounds from $1\sigma$
to $5\sigma$, are shown faintly. For $\mathcal{D}^{\varepsilon_{\rm
exp}}$/$\mathcal{D}^{\varepsilon_{\rm hg}}$ the background distribution is
roughly symmetrical and Gaussian, while for $\mathcal{D}^{\rm NH}$ the
distribution is not symmetric and is more similar to a Gamma or
$\chi^2$-distribution. However, to reflect the true distribution as best we can
we form a KDE from them, and use that to estimate the $\sigma$-bounds by finding
the intervals that bound the probabilities defining the $1-5\sigma$ bounds of a
Gaussian distribution.

The bottom panels of Fig.~\ref{fig:ensemble_exp_snr} show binned distributions
of signal-to-noise ratios---the numbers within the boxes show the number of
pulsars in that bin for the ensemble, whilst the number \emph{above} the boxes
shows the maximum signal-to-noise ratio of all pulsars within that ensemble. For
each adjacent pair the left set of bins show the signal-to-noise ratio
distribution for the ensemble with the \emph{smallest}
$\mathcal{O}^{\varepsilon_{\rm exp}}_{\rm n}$/$\mathcal{O}^{\varepsilon_{\rm
hg}}_{\rm n}$ (which will be those that gave rise to the lower extent of the
distributions shown above); and the right set of bins show the signal-to-noise
ratio distribution for the ensemble with the \emph{largest}
$\mathcal{O}^{\varepsilon_{\rm exp}}_{\rm n}$/$\mathcal{O}^{\varepsilon_{\rm
hg}}_{\rm n}$ (which will be those that gave rise to the upper extent of the
distributions shown above). If we look at the rightmost value of
$\mu_\varepsilon$ in the left panel of Fig.~\ref{fig:ensemble_exp_snr} we see
that the lowest value of $\mathcal{O}^{\varepsilon_{\rm exp}}_{\rm n}$ came from
an ensemble in which all pulsars had signal-to-noise ratios of $\le 5.4$. For an
individual pulsar analysis, such a signal-to-noise ratio could easily not be
enough to confidently assign it as detected (the probable noise outlier in
Fig.~2 of \cite{2017ApJ...839...12A} had a similar signal-to-noise ratio, as
did the rejected outliers in \cite{2017PhRvD..96l2006A}), but this gave rise to
a $\mathcal{D}^{\varepsilon_{\rm exp}}$ value well outside the background
distribution as seen in the top panel. However, from the middle panel it is
interesting to note that this ensemble is just on the edge of the background
distribution for the nonhierarchical statistic $\mathcal{D}^{\rm NH}$. In the
bottom panels on both sides we see what would be expected from the two different
distributions; there is a longer tail giving rise to larger outlier
signal-to-noise ratios for the exponential case than for the half-Gaussian case.

We will see what that means in terms of detection efficiency for the exponential
distribution in Sec.~\ref{sec:efficiency} and Fig.~\ref{fig:efficiency}.

\subsubsection{Detection efficiency}\label{sec:efficiency}

\begin{figure*}[!htbp]
    \includegraphics[width=\textwidth]{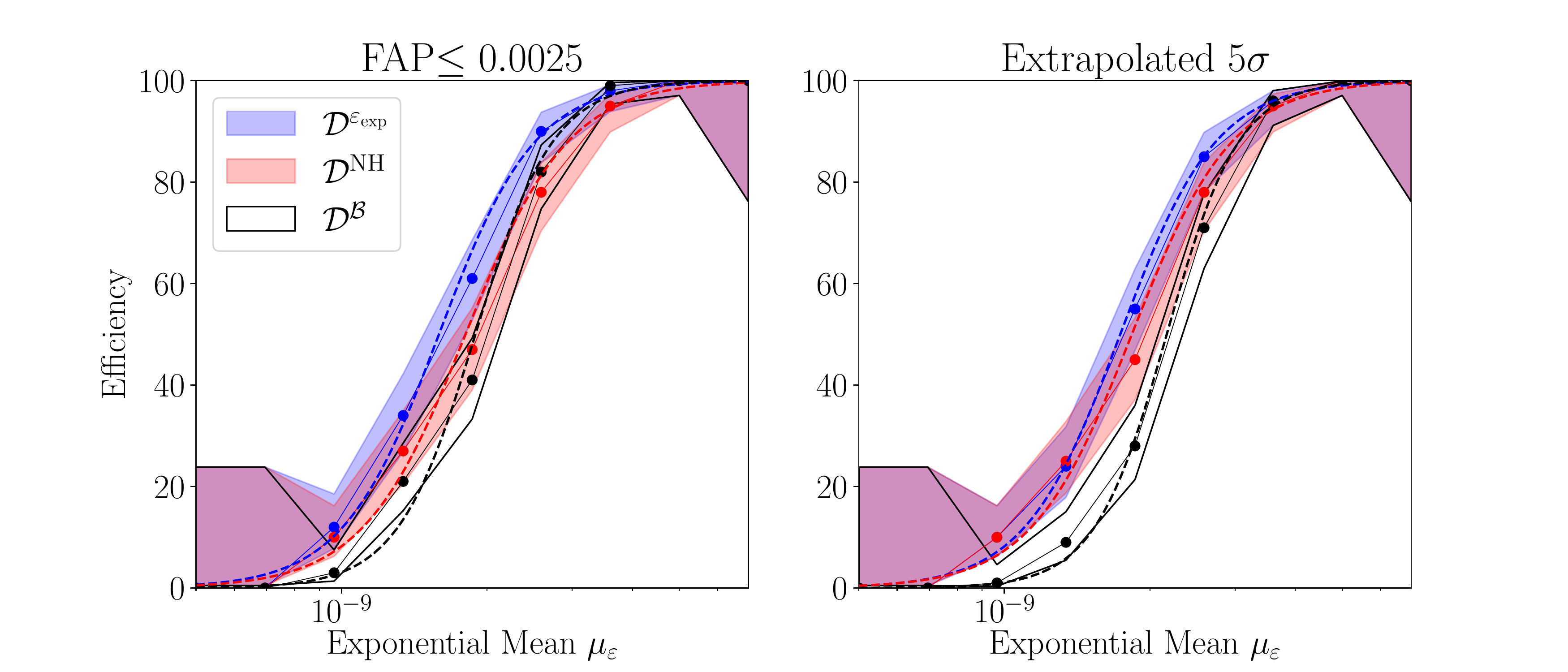}
    \caption{\label{fig:efficiency} Efficiency curves for the $\mathcal{D}^{\varepsilon_{\rm exp}}$, $\mathcal{D}^{\rm NH}$ and $\mathcal{D}^{\mathcal{B}}$ detection statistics for two different false alarm probabilities: a false alarm probability of 0.0025 based on the number of background realizations, and an equivalent $5\sigma$ false alarm probability extrapolated from KDEs of the background distributions. The dashed curves show sigmoid fits to the measured efficiencies.}
\end{figure*}

Using our distributions of statistics from the background analysis we can define
a threshold in $\mathcal{D}$ at which to claim detection with a given false
alarm probability (FAP). Here we do this in two ways, noting that we only do
this for the exponential distribution as we have enough simulations to get
reasonable statistics unlike for the half-Gaussian. The simplest threshold is
one based on the maximum background value of $\mathcal{D}$, which, given our 400
background realizations, will represent a FAP of $1/400 = 0.0025$. The other way
that we use is via extrapolation from the KDEs of the background distributions
out to values of $\mathcal{D}$ that yield equivalent cumulative probabilities to
that of a Gaussian distribution at $5\sigma$. To compute efficiencies we find
the number of ensembles for each $\mu_\varepsilon$ value that are above the two
FAP $\mathcal{D}$ thresholds for $\mathcal{D}^{\varepsilon_{\rm exp}}$,
$\mathcal{D}^{\rm NH}$, and $\mathcal{D}^{\mathcal{B}}$, compared to the total
number of simulations at that value. These efficiency curves are shown in
Fig.~\ref{fig:efficiency}, which also shows (as shaded regions) the 90\%
credible regions for each curve, and sigmoid fits to the efficiencies as dashed
lines.

We see that the false alarm probability based on the number of background
realizations $\mathcal{D}^{\varepsilon_{\rm exp}}$ clearly outperforms both the
$\mathcal{D}^{\rm NH}$ and $\mathcal{D}^{\mathcal{B}}$ statistics. For the false
alarm probability extrapolated from the background out to $5\sigma$ the
efficiencies of the two statistics, $\mathcal{D}^{\varepsilon_{\rm exp}}$ and
$\mathcal{D}^{\rm NH}$, are far more comparable. It is worth noting that the
efficiency curves for $\mathcal{D}^{\rm NH}$ for both FAPs are very similar.
This is because the actual threshold value in both cases is very similar due to
there being one large outlier that dominates the first FAP, but which only
contributes a small amount to the extrapolated threshold (which will also depend
on the KDE kernel width used for the background). Therefore, this KDE based
threshold extrapolated to $5\sigma$ for $\mathcal{D}^{\rm NH}$ is probably
unreliable and if more background realizations were performed larger outliers
may be found. So, we expect the performance of the two statistics, as observed
in the left panel of Fig.~\ref{fig:efficiency} to be more reliable, and better
represent the true gain by including the common ellipticity distribution. Both
$\mathcal{D}^{\varepsilon_{\rm exp}}$ and $\mathcal{D}^{\rm NH}$ considerably
outperform $\mathcal{D}^{\mathcal{B}}$ for the smallest distributions, i.e.,
when the contribution from the single strongest source is not overwhelming.

\subsection{Parameter estimation}

One of the purposes of this paper has been to develop a method to estimate the
hyperparameters defining a common ellipticity distribution from which the
ellipticities of pulsars are drawn. As described in Sec.~\ref{sec:ellpriors}
we have taken two simple distributions with which to test this: an exponential
distribution and a half-Gaussian.

\begin{figure}[!htbp]
    \begin{tabular}{c}
    \includegraphics[width=\columnwidth]{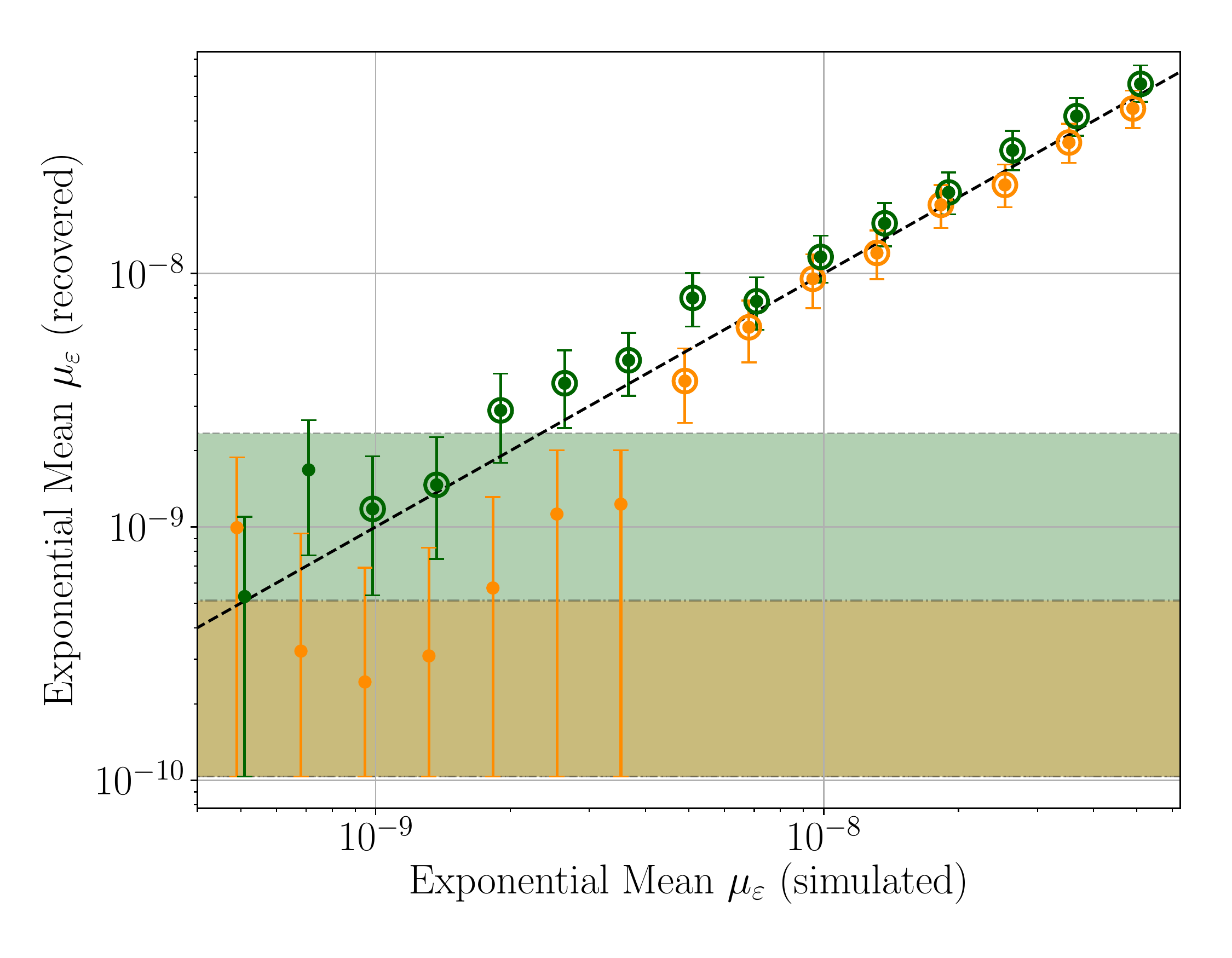} \\
    \includegraphics[width=\columnwidth]{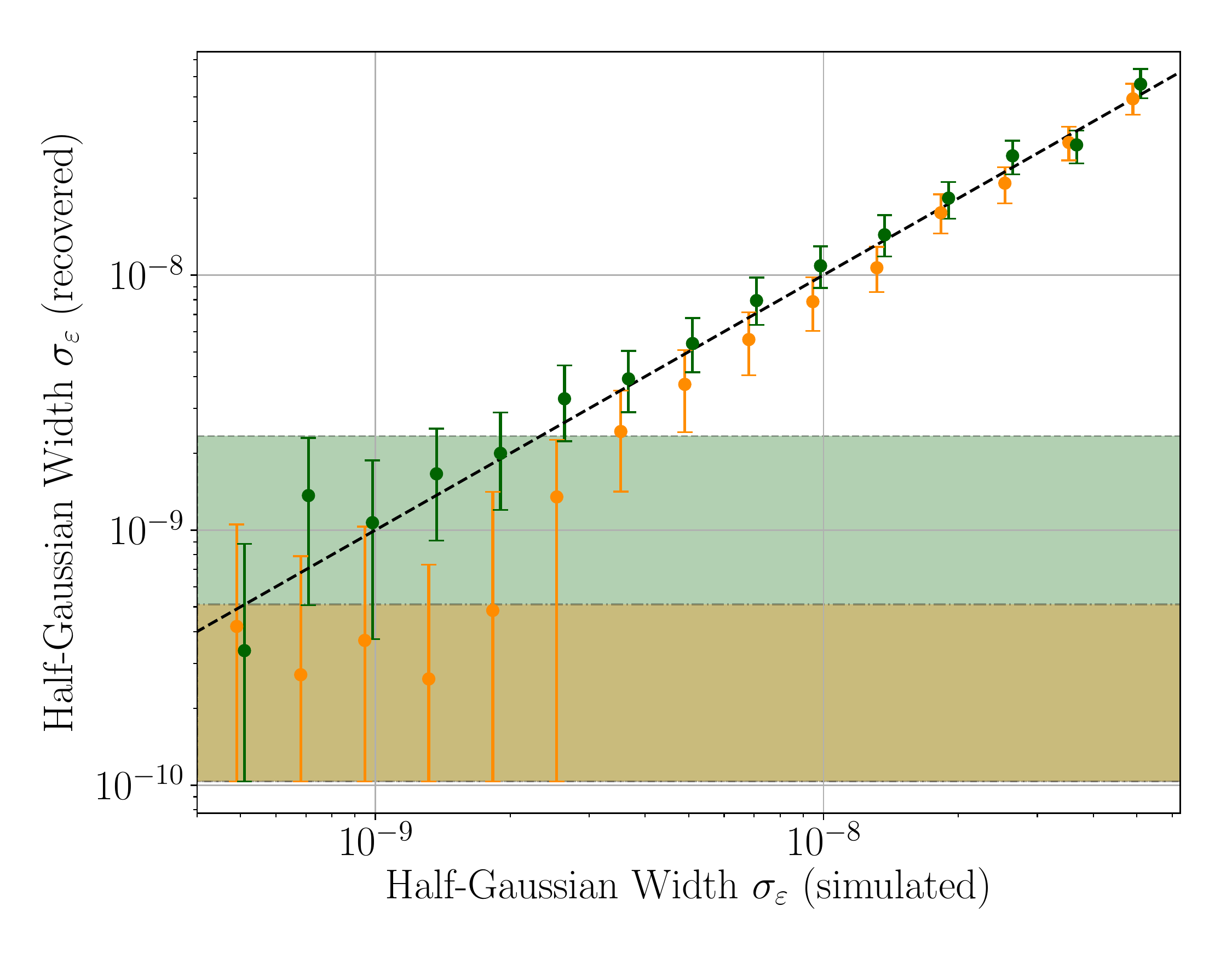}
    \end{tabular}
\caption{\label{fig:muepsilon}
Credible intervals (90\%) about the median value from the recovered posterior
probability distributions on the exponential ellipticity distribution
hyperparameter $\mu_\varepsilon$, across the range of simulated
$\mu_\varepsilon$ values (top), and the half-Gaussian ellipticity distribution
hyperparameter $\sigma_\varepsilon$, across a range of simulated
$\sigma_\varepsilon$ values (bottom). The dark green error bars represent the
credible interval for the ensemble with the largest odds, while the orange error
bars are for the ensemble with the lowest odds. These, therefore, are indicative
of the general range of possible results. In the top plot error bars with open
circles around the median values are those that would be considered ``detected''
above the $\mathcal{D}^{\varepsilon_{\rm exp}}$ false alarm probability (see
Sec.~\ref{sec:efficiency}) of 0.0025. The shaded bands show the intervals
with the largest and smallest upper bound for ensembles consisting purely of
noise.}
\end{figure}

For each of the ensembles of pulsars described in Sec.~\ref{sec:simulations}
we have estimated the posterior for the distribution means, $\mu_\varepsilon$,
for the exponential distribution simulations, and for the distribution widths,
$\sigma_\varepsilon$, for the half-Gaussian distribution simulations. These can
be seen as a function of the simulated values used to construct the ensembles in
Fig.~\ref{fig:muepsilon} (top panel for the exponential distribution and
bottom panel for the half-Gaussian). The plot shows the 90\% credible interval
on the posterior for $\mu_\varepsilon$ (top)/$\sigma_\varepsilon$ (bottom) for
the ensemble with the largest (dark error bars) and smallest (light error bars)
odds, $\mathcal{O}^{\varepsilon_{\rm exp/hg}}_{\rm n}$, for each simulated
value. We see that the true distribution parameters are recovered accurately
and, in all bar one case for the exponential distribution simulations, when no
signal is detected (based on the FAP of 0.0025) the posteriors include the lower
prior boundary on $\mu_\varepsilon$ (and thus in these case the maximum bound
represents an upper limit).

\subsection{Model comparison}

\begin{figure}[!htbp]
    \begin{tabular}{c}
    \includegraphics[width=\columnwidth]{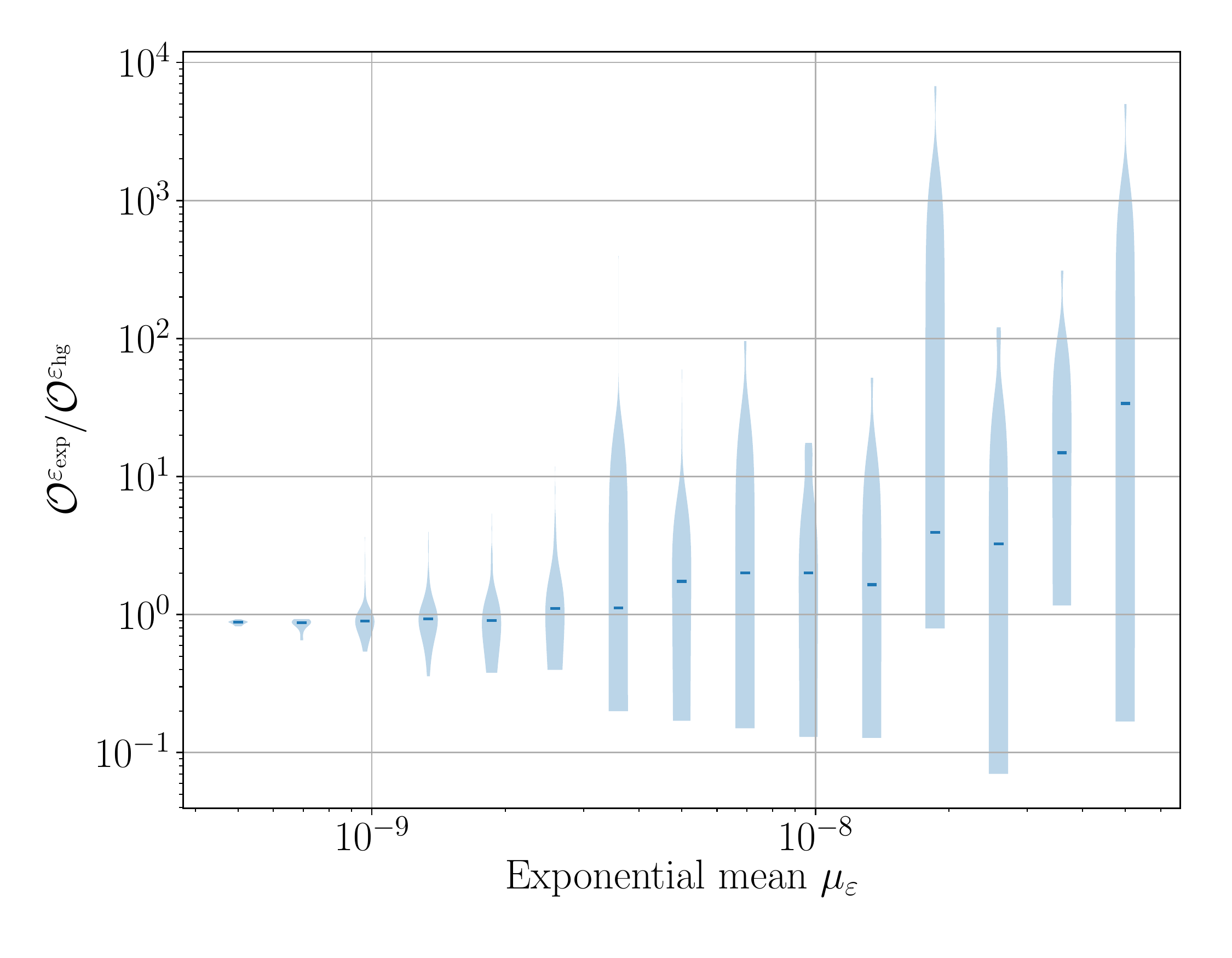} \\
    \includegraphics[width=\columnwidth]{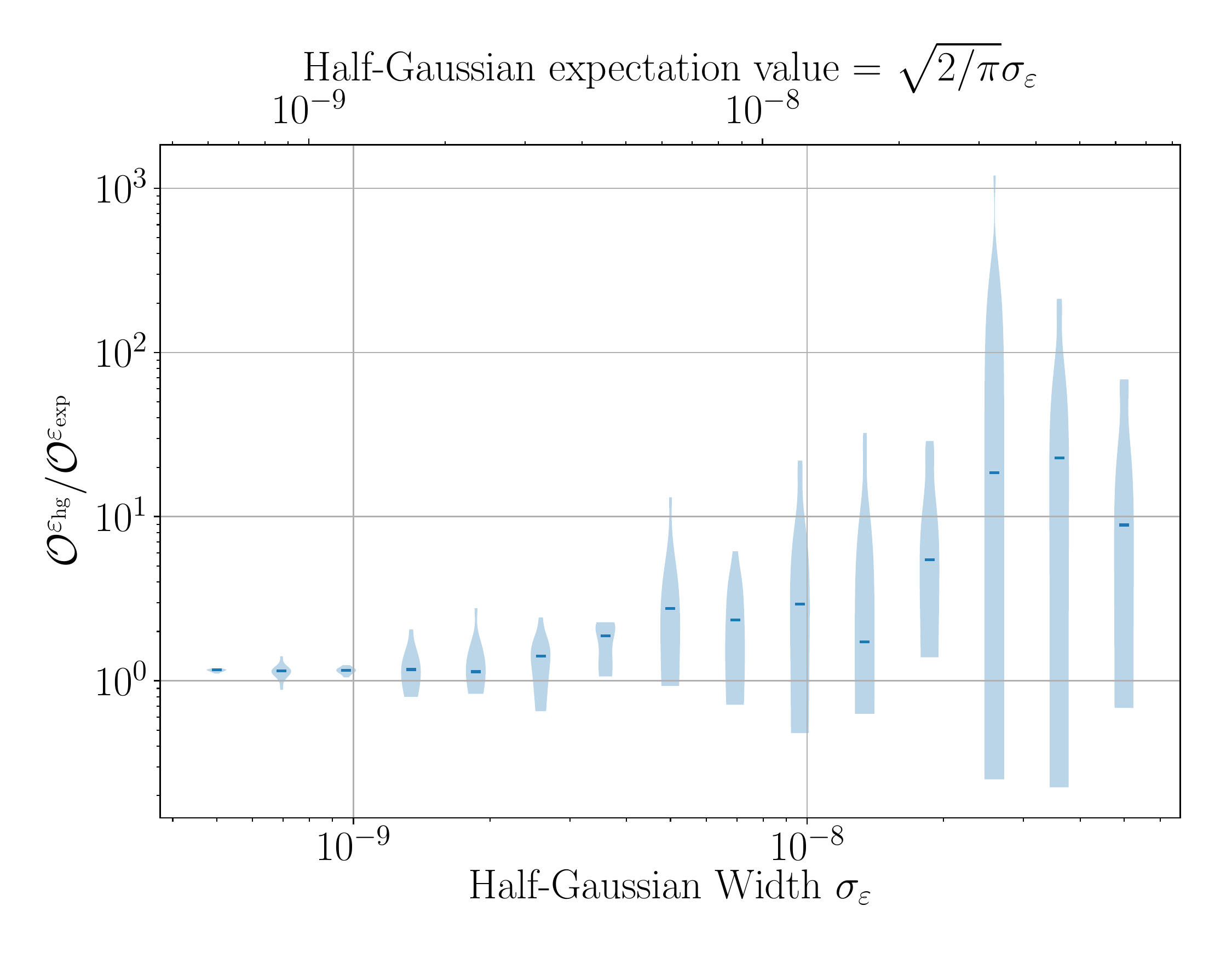}
    \end{tabular}
    \caption{\label{fig:comparison_ell} The distributions of ratio $\mathcal{O}^{\varepsilon_{\rm exp}}_{\rm n}/\mathcal{O}^{\varepsilon_{\rm hg}}_{\rm n}$ (top) for sources with ellipticities drawn from exponential distributions with given means $\mu_\varepsilon$, and $\mathcal{O}^{\varepsilon_{\rm hg}}_{\rm n}/\mathcal{O}^{\varepsilon_{\rm exp}}_{\rm n}$ (bottom) for sources with ellipticities drawn from half-Gaussian distributions with given widths $\sigma_\varepsilon$.}
\end{figure}

We can compare the evidences for the two different models of the ellipticity
distribution. As discussed earlier we have specified two toys distributions (the
exponential and half-Gaussian distributions) and have calculated the evidence
for both for all our ensembles of pulsars. We can take the ratio of these
evidences and see which distribution is preferred in each case.
Figure~\ref{fig:comparison_ell} shows this ratio for the ensembles created from
sources with ellipticities drawn from the exponential distribution (top) and the
half-Gaussian distribution (bottom). In the top panel, while there is a definite
trend towards strongly favoring the correct distribution there are still
ensembles with the largest exponential mean value that favor the half-Gaussian
model. At the smallest values of $\mu_\varepsilon$ the half-Gaussian is always
favored, probably due to it having a more sharply falling off tail and
therefore smaller prior volume. We find that the cases where the true
exponential distribution is most highly favored are when there are a few
outliers with large signal-to-noise ratios compared to the bulk of the
distributions, which would be allowed by the longer tail exponential
distribution, but not by the half-Gaussian. In the bottom panel of
Fig.~\ref{fig:comparison_ell} we find that the true half-Gaussian distribution
is generally favoured in the majority of cases, although we have fewer
simulations with which to truly probe the tails of the distribution.

\section{S6 results}

Rather than purely working with simulated data with have also run on real data
from the LIGO detectors. Here we present the results of this analysis using data
from LIGO's sixth science run (S6)
\cite{2009RPPh...72g6901A,2012arXiv1203.2674T}.\footnote{The full LIGO S6 data
set is publicly available through the LIGO Open Science Center
\url{https://losc.ligo.org/S6/} \cite{2015JPhCS.610a2021V}.} The run had data
from the two LIGO detectors at Hanford and Livingston operating in enhanced
configuration between 8 July 2009 and 20 October 2010. We use 92 of the pulsars
included in the analysis of \cite{2014ApJ...785..119A}.\footnote{For this
analysis there was a selection cut meaning not all the pulsars in
\cite{2014ApJ...785..119A} were used. To enable the relatively quick production
of background realizations of the data we made use of the more efficient
spectral interpolation algorithm \cite{2017CQGra..34a5010D}, rather than the
standard heterodyne method \cite{2005PhRvD..72j2002D}, to process the S6 data.
This method makes use of Fourier transforms of 1800\,s chunks of data, which had
been created for the analyses in \cite{2015ApJ...813...39A,2016PhRvD..94j2002A}
and included a low frequency cut-off at 40\,Hz, so pulsars with gravitational
wave frequencies below this cut-off were excluded. The spectral interpolation
method also has limitations for high spin-down pulsars or those in tight
binaries, so many where excluded on these criteria.} For each pulsar and for
each individual detector, and for both detectors combined, the evidence and
posterior samples (marginalized over orientation and pulsar distance) as
required in Eq.~(\ref{eq:bayesrearrange2}) were produced. As with the
simulated data sets described in Sec.~\ref{sec:simulations}, using these and
the noise evidence values we have calculated $\mathcal{O}^{\varepsilon_{\rm
exp}}_{\rm n}$ and $\mathcal{O}^{\varepsilon_{\rm hg}}_{\rm n}$ for the ensemble
of pulsars. As before, we require a background distribution of these values to
compare our ``foreground'' to. To create a background we required data that
shares the same noise characteristics as the foreground, but in which an
astrophysical signal from a given pulsar would not be present. To achieve this
we use the method described briefly in Section~V.A.\ of
\cite{2017PhRvD..96l2006A} (also see \cite{maxpaper}), in which for each pulsar
we reprocess (using the spectral interpolation method from
\cite{2017CQGra..34a5010D}) the data with the pulsar's sky location changed to a
different randomly selected sky location. We performed this background
generation 100 times to find the distribution of values.

\begin{figure*}[!htbp]
    \includegraphics[width=\textwidth]{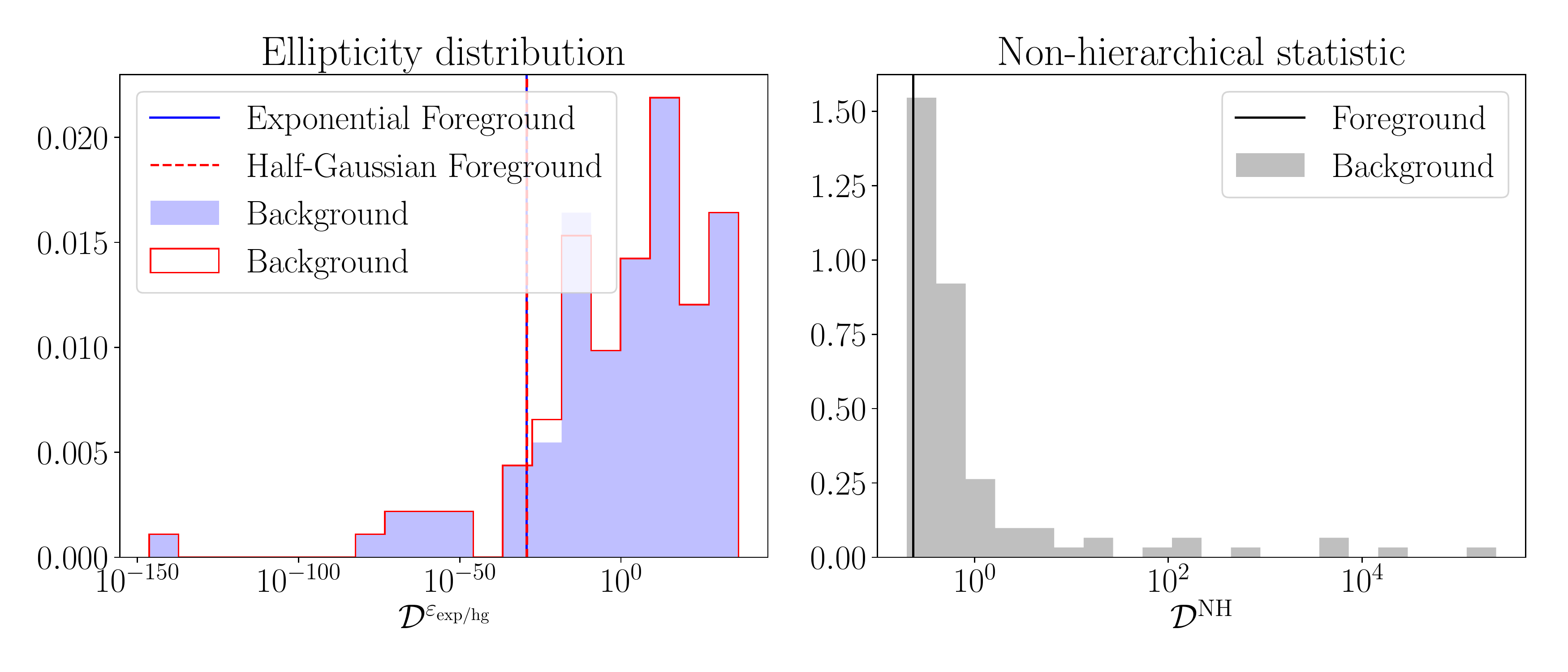}
    \caption{\label{fig:S6odds} The left panel shows the distributions of $\mathcal{D}^{\varepsilon_{\rm exp}}$ and $\mathcal{D}^{\varepsilon_{\rm hg}}$ for the foreground and background distributions from LIGO S6 data. The right panel shows $\mathcal{D}^{\rm NH}$ for the same data.}
\end{figure*}

The foreground and background distribution of $\mathcal{D}^{\varepsilon_{\rm
exp}}$, $\mathcal{D}^{\varepsilon_{\rm hg}}$, and $\mathcal{D}^{\rm NH}$ are
shown in Fig.~\ref{fig:S6odds}. We see that the foreground for both
ellipticity distributions are well within the background distribution, so we see
no evidence for gravitational wave emission from the ensemble of pulsars. 

\begin{figure}[!htbp]
    \includegraphics[width=\columnwidth]{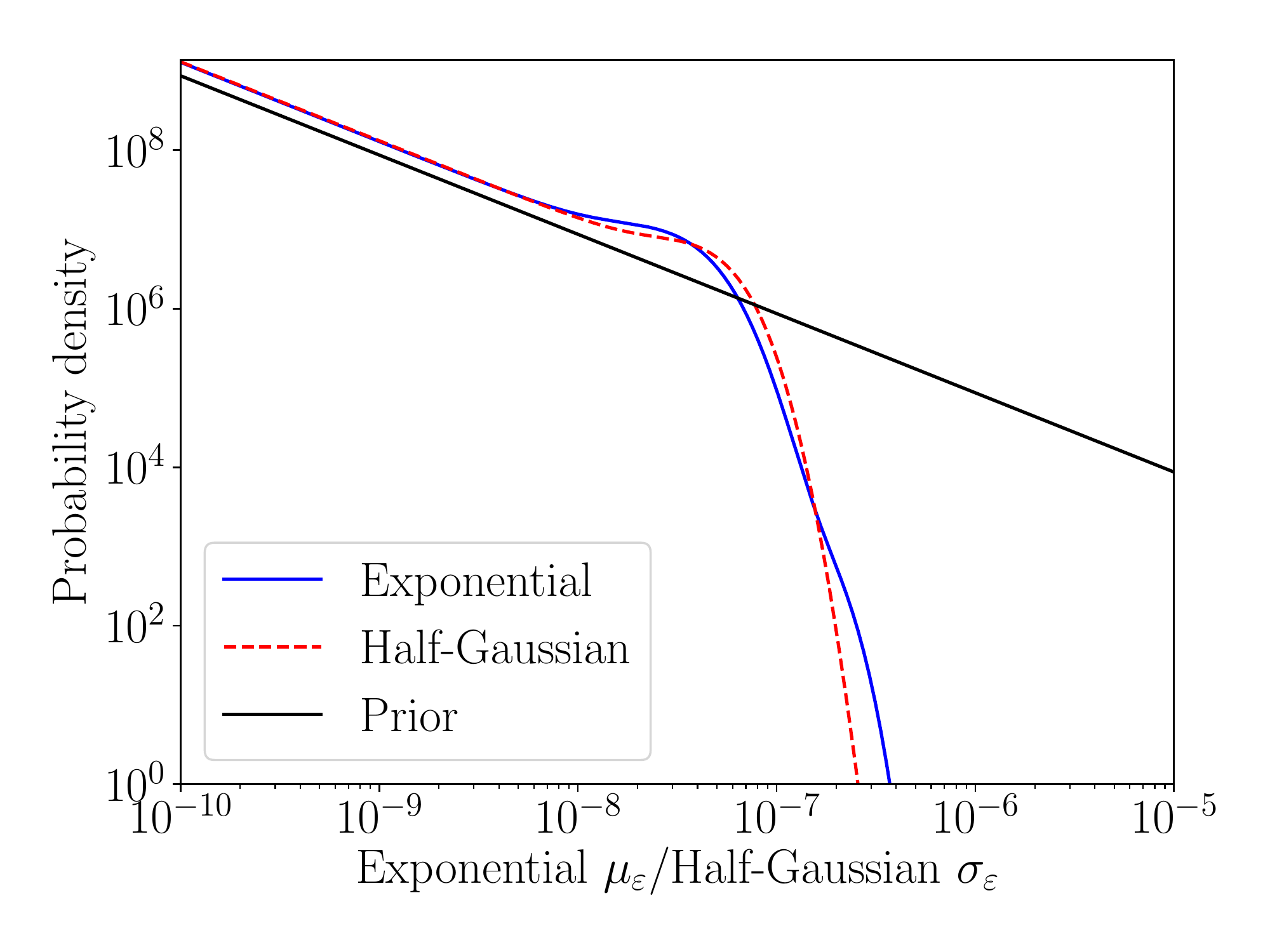}
    \caption{\label{fig:S6posteriors} The posterior probability distributions 
    for the hyperparameters $\mu_\varepsilon$ and $\sigma_\varepsilon$ defining 
    the exponential and half-Gaussian distributions respectively.}
\end{figure}

However, we can set upper limits on the ellipticity distribution
hyperparameters. Figure~\ref{fig:S6posteriors} shows the posteriors for
$\mu_\varepsilon$ and $\sigma_\varepsilon$ for the exponential and half-Gaussian
distributions, respectively. From these we find 90\% credible upper limits of
$\mu_\varepsilon^{90\%} \le 3.9\ee{-8}$ and $\sigma_\varepsilon^{90\%} \le
4.7\ee{-8}$. These are about two orders of magnitude less constraining than the
purely spin-down limit based limits discussed in Sec.~\ref{sec:sdlims},
although they are the first such limits to be set based purely on gravitational
wave observations.

\section{Conclusions}

In this work we have described a Bayesian hierarchical method for combining
gravitational wave observations from an ensemble of known pulsars for two
purposes: to create a detection statistic for identifying a signal from the
ensemble, and to estimate the parameters of the distribution of pulsars'
fiducial ellipticities.\footnote{Here we have worked with fiducial ellipticities
as they are a convenient and relatable quantity (i.e., they express the relative
deformation of the star). However, the analysis actually estimates the mass
quadrupole moment of the stars and converts that into the ellipticity given the
canonical moment of inertia of $10^{38}\,{\rm kg}\,{\rm m}^2$. So, one could
convert back to the moment of inertia independent mass quadrupole if required.
We should note that simulated signals were drawn from the fiducial ellipticity
$\varepsilon$ parameter, and we converted to mass quadrupoles using the
canonical moment of inertia, so they do not incorporate a realistic
equation-of-state dependent spread of moments of inertia.} For two toy
ellipticity distributions, an exponential and a half-Gaussian, we have used
simulations to find that incorporating this distribution as a common prior on
the ellipticity of stars, with an unknown hyperparameter, can produce a more
efficient detection statistic than combining the data for the ensemble of
pulsars in a nonhierarchical way. We also show that it is more efficient than a
statistic derived in a similar way to that in \cite{2005PhRvD..72f3006C}. We
find that the detection of the ensemble could even be seen in cases where
individual sources may not be individually detectable with high confidence.
However, we should note that the efficiency may not be improved if the true
distribution does not well match our assumed prior form.

For ensembles for which gravitational wave emission would be considered detected
we have shown in Fig.~\ref{fig:muepsilon} that we can correctly constrain the 
hyperparameters of the simulated ellipticity distribution. If no signal is seen we can 
also set upper limits on these. However, as shown in Fig.~\ref{fig:comparison_ell}, we 
have also found that it is difficult to distinguish between our two toy distributions as 
they are broadly similar.

We have performed the analyses using real data for 92 pulsars from the LIGO S6
science run, with the assumption of the same two ellipticity distributions: an
exponential and a half-Gaussian. We saw no evidence of a signal from the
ensemble, but set upper limits on the two distributions hyperparameters of
$\mu_\varepsilon^{90\%} \le 3.9\ee{-8}$ and $\sigma_\varepsilon^{90\%} \le
4.7\ee{-8}$. These upper limits are $\sim 2$ orders of magnitude less
constraining than those that can be produced using the electromagnetically
derived pulsar spin-down limits. However, they are the first such limits to be
produced purely from gravitational wave observations.

We note that the exponential and half-Gaussian distributions used are rather
simple. They were chosen as simple toy models that were easy to use due to being
defined by a single hyperparameter. However, they are not necessarily physically
realistic distributions. Observationally, we know that there are different
populations of pulsars, like the old recycled millisecond pulsars and the young
pulsars. The fact that the former are most likely to have undergone an accretion
phase, which could alter the structure of their crust and magnetic field
strength compared to non-recycled pulsars, meaning they could well have a
different distribution. So, it could be that the two populations should be
treated independently, or a more complex distribution that allows separation of
the two distributions should be used (for a simple case it could be a bimodal
Gaussian).

It should also be noted that in this work we assume a 20\% uncertainty of the
distance to all pulsars, but in reality there are range of distance
uncertainties from a few percent, or hundreds of percent. In a more thorough
analysis the actual measurement uncertainties for each pulsar should be
included, although we do not imagine it would lead to a particularly significant
change in the results.

As discussed in Sec.~\ref{sec:sdlims} the electromagnetic-observation-derived
spin-down limits could be incorporated more fully into the ellipticity
distribution analysis. The simplest way to do this would be use the spin-down
limits as a priors on the ellipticity (or mass quadrupole) for each pulsar. In
this way, for pulsars for which the gravitational wave data alone is not
particularly informative the spin-down limit-based prior would dominate.

Finally, it is worth highlighting that this type of analysis would only
constrain the underlying distribution of known pulsars, but not necessarily the
entire neutron star population. There could, for example, be a different
distribution for accreting stars, or stars that are purely gravitars (i.e.\
neutron stars that are purely spinning down due to gravitational wave emission.)

\begin{acknowledgments}

M.P.\ is funded by the UK Science \& Technology Facilities Council (STFC) under
Grant No.\ ST/N005422/1. X.F.\ was also supported by the National Natural Science
Foundation of China under Grants No.\ 11673008 and Newton International
Fellowship Alumni Follow-on Funding. We are grateful for computational resources
provided by Cardiff University, and funded by an STFC Grant No.\ (ST/I006285/1)
supporting UK Involvement in the Operation of Advanced LIGO, and also those of
the Atlas cluster at the Max-Planck Institut f\"{u}r Gravitationsphysik/Leibniz
Universit\"{a}t Hannover. We would like to thank members of the LIGO Scientific
Collaboration and Virgo Collaboration, in particular members of the Continuous
Waves (CW) working group, for useful discussions that have helped develop these
ideas. We specifically thank Max Isi for his very useful comments on draft
versions of the paper. We would also like to thank the CW working group for the
creation of the short Fourier transform data set of LIGO S6 data
\cite{2015ApJ...813...39A,2016PhRvD..94j2002A} that we have used in this
analysis. M.P.\ thanks the Institute for Nuclear Theory at the University of
Washington, the Department of Energy, and the organizers and attendees of the
``Astro-Solids, Dense Matter, and Gravitational Waves'' workshop for their
hospitality and for interesting discussion that enhanced this work.

This document has been given LIGO Document Number LIGO-P1800171 and INT preprint
number INT-PUB-18-037.

This work has been made possible thanks to a variety of software packages. The
main analyses for each pulsar used software available in {\sc LALSuite}
\cite{LALSuite}. The hierarchical analysis used {\sc Python} and functions written in
{\sc Cython} \cite{behnel2010cython}. The post-processing was performed using Jupyter
notebooks \cite{kluyver2016jupyter} and all plots have been produced using
{\sc Matplotlib} \citep{Hunter:2007,michael_droettboom_2017_248351}. Pulsar data from
the ATNF Pulsar Catalogue \cite{2005AJ....129.1993M} has been accessed using
{\sc PSRQpy} \cite{psrqpy}.

\end{acknowledgments}

\appendix

\section{$h_0$ posteriors}\label{ap:h0samples}

If an analysis produced posterior samples on the observed gravitational wave
amplitude, $h_0$, rather than on $\varepsilon$ or $Q_{22}$ (as has been the case
for previous known pulsar searches), the analysis described in this paper could
still be performed, although the marginalizations over the each pulsar's
distance would have to be explicitly computed. Assuming that a kernel density
estimate has been used to turn sets of $h_0$ posterior samples into probability
densities, then for each pulsar the likelihood on $\varepsilon_i$ (assuming a
uniform prior on $h_0$ that is only nonzero between zero and some maximum
${h_0}_{\rm max}$) is
\begin{align}
p(\mathbf{x}_i|\varepsilon_i,I) = \int^{D_i} & p(\mathbf{x}_i|h_0=\frac{\kappa_i\varepsilon_i}{D_i},D_i,I) \frac{p(\mathbf{x}_i|I)}{p({h_0}_i|I)} {\rm d}D_i \nonumber \\
 = \int^{D_i} & p({h_0}_i=\frac{\kappa_i\varepsilon_i}{D_i}|\mathbf{x}_i,D,I) {h_0}_{\rm max} \times \nonumber \\
 & p(\mathbf{x}_i|I)  p(D_i|I) {\rm d}D_i,
\end{align}
where
\begin{equation}
    \kappa_i = \frac{16\pi^2 G {f_{\rm rot}}_i^2}{c^4}I_{zz}.
\end{equation}

\section{Evidence evaluation}\label{ap:evidence}

Here we describe the various features/issues that we have found regarding
evaluating the evidences. These relate to the systematic biases on evidences
from nested sampling (see, e.g., \cite{2017arXiv170508978P}), and statistical
uncertainties on the final evidence due to a combination of uncertainties from
the KDE posterior estimates and statistical uncertainties on the individual
pulsar evidences.

\subsection{Statistical uncertainties on evidence values}

To produce our final evidence values given by Eq.~(\ref{eq:totmarglike}) we
rely on the output of a code that uses a stochastic sampler to perform the
required integrals. We also rely on a finite number of posterior samples from
each pulsar's estimate of $\varepsilon$ to form a kernel density estimate of the
true posterior. Both of these mean that even on identical initial data (with
identical noise realizations) there will be some stochastic variation in the
results.

To estimate these variations we have performed the processing described in
Sec.~\ref{sec:processing} 10 times on the same ensemble set, which in this
case is one containing no pulsar signals. A different random seed is used for
the sampler in each case, otherwise the results really should be identical. To
estimate the variations in the final evidence caused by the finite number of
samples used for the KDE of the $\varepsilon$ posteriors, we use
Eq.~(\ref{eq:bayesrearrange2}) but with $p(\mathbf{x}_i|I) = 1$ for each
pulsar, so the only variation is from differences between
$p(\varepsilon_i|\mathbf{x}_i,I)$ in each analysis. We find that there is a
standard deviation on the base-10 logarithm of the final evidence from
Eq.~(\ref{eq:totmarglike}) of $\sim 0.6$, which comes from the variation in
the estimates of the individual pulsar $\varepsilon$ posteriors. Each KDE is
estimated using $\sim 1000$ posterior samples, so we can check if this is
roughly the variation you might expect. We can draw 1000 samples from 200
half-Gaussian distributions with known standard deviations, for each
distribution produce a KDE and evaluate it at a range of points, sum the
logarithms of these KDEs, and then numerically integrate it over the range of
points. Doing this multiple times, but with the same set of standard deviations
for each half-Gaussian, we find the variation in the final integral is of the
same order as that which we see for our analysis using simulated pulsar data.
This uncertainty can be reduced by increasing the number of posterior samples
used for the KDE estimate. To get more samples in our actual analysis we would
need to use a greater number of nested sampling live points, which increases the
run time. However, in future real analysis it may be worth doing this to cut
down the uncertainties.

To see the uncertainty in the final evidence cause by both the stochastic
variation in the individual pulsar evidences \emph{and} the KDE, we repeat the
above with the actual estimated $p(\mathbf{x}_i|I)$ for each pulsar. We find a
standard deviation on in the base-10 logarithm of the final evidence from
Eq.~(\ref{eq:totmarglike}) of $\sim 1.3$. This suggests that the stochastic
variations in the individual pulsar evidences and KDEs contribute roughly
equally to the overall uncertainty. Again, this could be reduced by using a
larger number of live points for the nested sampling algorithm. However, this
variation is smaller than the distribution of background values, so will not be
of great significance.

\subsection{Systematic uncertainties on evidence values}\label{ap:absodds}

When creating the odds for signals being drawn from a particular distribution
compared to the noise distribution, we find that there can be a severe bias on
the odds value. This comes from individually small systematic biases on the
signal evidence values calculated for individual pulsars by the nested sampling
algorithm (see, for example, the study on biases in nested sampling in
\cite{Buchner2016}). For example, over the 200 pulsars used, if the average
natural logarithm of the signal evidence was 0.05 smaller than the (unknowable)
truth, then when the likelihoods for individual pulsars are combined there will
be a $0.05 \times 200 = 10$ offset compared to what the value should be. In
tests performed on identical realizations of a population of 200 pulsars, all
containing only noise, where the only difference in the analysis is the number
of live points used for the nested sampling, we find, for example, a mean
difference in the pulsar's signal natural logarithm evidence values of -0.08
between using 512 live points compared to 2048. While this difference is fairly
insignificant for a single pulsar, it adds up to a $\sim 16$ offset between the
final odds values. We do not have a handle on whether using more live points
produces evidences that are systematically closer to the truth (although
Fig.~11 of \cite{2017arXiv170508978P} suggests that systematic errors are
present even for large numbers of live points in similar cases when the true
evidence can be calculated), as we do not have access to the truth for this
specific case, it means that we cannot use the actual value of the odds as a
trustworthy indicator of the true odds. Therefore, to gauge whether our model
comparison favours one model over the other we have to resort to comparing it to
a {\it background} distributions of odds, and see at what point our simulations
diverge from the background.

We believe that these individually small systematic errors in evidences may be
important for other detection statistics based on hierarchical analysis that
combine evidences from multiple independent analyses, such as that proposed in
\cite{2018PhRvX...8b1019S}. This deserves more study (like that in
\cite{Buchner2016}) to see how if similar features are observed in nested
sampling algorithms that differ from that used here
\cite{2015PhRvD..91d2003V,2017arXiv170508978P}.

\bibliography{multipulsar}

\begin{thebibliography}{57}%
\makeatletter
\providecommand \@ifxundefined [1]{%
 \@ifx{#1\undefined}
}%
\providecommand \@ifnum [1]{%
 \ifnum #1\expandafter \@firstoftwo
 \else \expandafter \@secondoftwo
 \fi
}%
\providecommand \@ifx [1]{%
 \ifx #1\expandafter \@firstoftwo
 \else \expandafter \@secondoftwo
 \fi
}%
\providecommand \natexlab [1]{#1}%
\providecommand \enquote  [1]{``#1''}%
\providecommand \bibnamefont  [1]{#1}%
\providecommand \bibfnamefont [1]{#1}%
\providecommand \citenamefont [1]{#1}%
\providecommand \href@noop [0]{\@secondoftwo}%
\providecommand \href [0]{\begingroup \@sanitize@url \@href}%
\providecommand \@href[1]{\@@startlink{#1}\@@href}%
\providecommand \@@href[1]{\endgroup#1\@@endlink}%
\providecommand \@sanitize@url [0]{\catcode `\\12\catcode `\$12\catcode
  `\&12\catcode `\#12\catcode `\^12\catcode `\_12\catcode `\%12\relax}%
\providecommand \@@startlink[1]{}%
\providecommand \@@endlink[0]{}%
\providecommand \url  [0]{\begingroup\@sanitize@url \@url }%
\providecommand \@url [1]{\endgroup\@href {#1}{\urlprefix }}%
\providecommand \urlprefix  [0]{URL }%
\providecommand \Eprint [0]{\href }%
\providecommand \doibase [0]{http://dx.doi.org/}%
\providecommand \selectlanguage [0]{\@gobble}%
\providecommand \bibinfo  [0]{\@secondoftwo}%
\providecommand \bibfield  [0]{\@secondoftwo}%
\providecommand \translation [1]{[#1]}%
\providecommand \BibitemOpen [0]{}%
\providecommand \bibitemStop [0]{}%
\providecommand \bibitemNoStop [0]{.\EOS\space}%
\providecommand \EOS [0]{\spacefactor3000\relax}%
\providecommand \BibitemShut  [1]{\csname bibitem#1\endcsname}%
\let\auto@bib@innerbib\@empty
\bibitem [{\citenamefont {{Aasi}}\ \emph
  {et~al.}(2015{\natexlab{a}})\citenamefont {{Aasi}}, \citenamefont {{Abbott}},
  \citenamefont {{Abbott}}, \citenamefont {{Abbott}}, \citenamefont
  {{Abernathy}}, \citenamefont {{Ackley}}, \citenamefont {{Adams}},
  \citenamefont {{Adams}}, \citenamefont {{Addesso}} \emph
  {et~al.}}]{2015CQGra..32g4001L}%
  \BibitemOpen
  \bibfield  {author} {\bibinfo {author} {\bibfnamefont {J.}~\bibnamefont
  {{Aasi}}}, \bibinfo {author} {\bibfnamefont {B.~P.}\ \bibnamefont
  {{Abbott}}}, \bibinfo {author} {\bibfnamefont {R.}~\bibnamefont {{Abbott}}},
  \bibinfo {author} {\bibfnamefont {T.}~\bibnamefont {{Abbott}}}, \bibinfo
  {author} {\bibfnamefont {M.~R.}\ \bibnamefont {{Abernathy}}}, \bibinfo
  {author} {\bibfnamefont {K.}~\bibnamefont {{Ackley}}}, \bibinfo {author}
  {\bibfnamefont {C.}~\bibnamefont {{Adams}}}, \bibinfo {author} {\bibfnamefont
  {T.}~\bibnamefont {{Adams}}}, \bibinfo {author} {\bibfnamefont
  {P.}~\bibnamefont {{Addesso}}},  \emph {et~al.},\ }\bibfield  {title}
  {\enquote {\bibinfo {title} {{Advanced LIGO}},}\ }\href {\doibase
  10.1088/0264-9381/32/7/074001} {\bibfield  {journal} {\bibinfo  {journal}
  {Classical Quantum Gravity}\ }\textbf {\bibinfo {volume} {32}},\ \bibinfo
  {eid} {074001} (\bibinfo {year} {2015}{\natexlab{a}})},\ \Eprint
  {http://arxiv.org/abs/1411.4547} {arXiv:1411.4547} \BibitemShut {NoStop}%
\bibitem [{\citenamefont {{Acernese}}\ \emph {et~al.}(2015)\citenamefont
  {{Acernese}}, \citenamefont {{Agathos}}, \citenamefont {{Agatsuma}},
  \citenamefont {{Aisa}}, \citenamefont {{Allemandou}}, \citenamefont
  {{Allocca}}, \citenamefont {{Amarni}}, \citenamefont {{Astone}},
  \citenamefont {{Balestri}}, \citenamefont {{Ballardin}} \emph
  {et~al.}}]{2015CQGra..32b4001A}%
  \BibitemOpen
  \bibfield  {author} {\bibinfo {author} {\bibfnamefont {F.}~\bibnamefont
  {{Acernese}}}, \bibinfo {author} {\bibfnamefont {M.}~\bibnamefont
  {{Agathos}}}, \bibinfo {author} {\bibfnamefont {K.}~\bibnamefont
  {{Agatsuma}}}, \bibinfo {author} {\bibfnamefont {D.}~\bibnamefont {{Aisa}}},
  \bibinfo {author} {\bibfnamefont {N.}~\bibnamefont {{Allemandou}}}, \bibinfo
  {author} {\bibfnamefont {A.}~\bibnamefont {{Allocca}}}, \bibinfo {author}
  {\bibfnamefont {J.}~\bibnamefont {{Amarni}}}, \bibinfo {author}
  {\bibfnamefont {P.}~\bibnamefont {{Astone}}}, \bibinfo {author}
  {\bibfnamefont {G.}~\bibnamefont {{Balestri}}}, \bibinfo {author}
  {\bibfnamefont {G.}~\bibnamefont {{Ballardin}}},  \emph {et~al.},\ }\bibfield
   {title} {\enquote {\bibinfo {title} {{Advanced Virgo: a second-generation
  interferometric gravitational wave detector}},}\ }\href {\doibase
  10.1088/0264-9381/32/2/024001} {\bibfield  {journal} {\bibinfo  {journal}
  {Classical Quantum Gravity}\ }\textbf {\bibinfo {volume} {32}},\ \bibinfo
  {eid} {024001} (\bibinfo {year} {2015})},\ \Eprint
  {http://arxiv.org/abs/1408.3978} {arXiv:1408.3978} \BibitemShut {NoStop}%
\bibitem [{\citenamefont {{Abbott}}\ \emph
  {et~al.}(2016{\natexlab{a}})\citenamefont {{Abbott}}, \citenamefont
  {{Abbott}}, \citenamefont {{Abbott}}, \citenamefont {{Abernathy}},
  \citenamefont {{Acernese}}, \citenamefont {{Ackley}}, \citenamefont
  {{Adams}}, \citenamefont {{Adams}}, \citenamefont {{Addesso}}, \citenamefont
  {{Adhikari}} \emph {et~al.}}]{2016PhRvL.116f1102A}%
  \BibitemOpen
  \bibfield  {author} {\bibinfo {author} {\bibfnamefont {B.~P.}\ \bibnamefont
  {{Abbott}}}, \bibinfo {author} {\bibfnamefont {R.}~\bibnamefont {{Abbott}}},
  \bibinfo {author} {\bibfnamefont {T.~D.}\ \bibnamefont {{Abbott}}}, \bibinfo
  {author} {\bibfnamefont {M.~R.}\ \bibnamefont {{Abernathy}}}, \bibinfo
  {author} {\bibfnamefont {F.}~\bibnamefont {{Acernese}}}, \bibinfo {author}
  {\bibfnamefont {K.}~\bibnamefont {{Ackley}}}, \bibinfo {author}
  {\bibfnamefont {C.}~\bibnamefont {{Adams}}}, \bibinfo {author} {\bibfnamefont
  {T.}~\bibnamefont {{Adams}}}, \bibinfo {author} {\bibfnamefont
  {P.}~\bibnamefont {{Addesso}}}, \bibinfo {author} {\bibfnamefont {R.~X.}\
  \bibnamefont {{Adhikari}}},  \emph {et~al.},\ }\bibfield  {title} {\enquote
  {\bibinfo {title} {{Observation of Gravitational Waves from a Binary Black
  Hole Merger}},}\ }\href {\doibase 10.1103/PhysRevLett.116.061102} {\bibfield
  {journal} {\bibinfo  {journal} {\prl}\ }\textbf {\bibinfo {volume} {116}},\
  \bibinfo {eid} {061102} (\bibinfo {year} {2016}{\natexlab{a}})},\ \Eprint
  {http://arxiv.org/abs/1602.03837} {arXiv:1602.03837} \BibitemShut {NoStop}%
\bibitem [{\citenamefont {{Abbott}}\ \emph
  {et~al.}(2016{\natexlab{b}})\citenamefont {{Abbott}}, \citenamefont
  {{Abbott}}, \citenamefont {{Abbott}}, \citenamefont {{Abernathy}},
  \citenamefont {{Acernese}}, \citenamefont {{Ackley}}, \citenamefont
  {{Adams}}, \citenamefont {{Adams}}, \citenamefont {{Addesso}}, \citenamefont
  {{Adhikari}} \emph {et~al.}}]{2016PhRvX...6d1015A}%
  \BibitemOpen
  \bibfield  {author} {\bibinfo {author} {\bibfnamefont {B.~P.}\ \bibnamefont
  {{Abbott}}}, \bibinfo {author} {\bibfnamefont {R.}~\bibnamefont {{Abbott}}},
  \bibinfo {author} {\bibfnamefont {T.~D.}\ \bibnamefont {{Abbott}}}, \bibinfo
  {author} {\bibfnamefont {M.~R.}\ \bibnamefont {{Abernathy}}}, \bibinfo
  {author} {\bibfnamefont {F.}~\bibnamefont {{Acernese}}}, \bibinfo {author}
  {\bibfnamefont {K.}~\bibnamefont {{Ackley}}}, \bibinfo {author}
  {\bibfnamefont {C.}~\bibnamefont {{Adams}}}, \bibinfo {author} {\bibfnamefont
  {T.}~\bibnamefont {{Adams}}}, \bibinfo {author} {\bibfnamefont
  {P.}~\bibnamefont {{Addesso}}}, \bibinfo {author} {\bibfnamefont {R.~X.}\
  \bibnamefont {{Adhikari}}},  \emph {et~al.},\ }\bibfield  {title} {\enquote
  {\bibinfo {title} {{Binary Black Hole Mergers in the First Advanced LIGO
  Observing Run}},}\ }\href {\doibase 10.1103/PhysRevX.6.041015} {\bibfield
  {journal} {\bibinfo  {journal} {Phys. Rev. X}\ }\textbf {\bibinfo {volume}
  {6}},\ \bibinfo {eid} {041015} (\bibinfo {year} {2016}{\natexlab{b}})},\
  \Eprint {http://arxiv.org/abs/1606.04856} {arXiv:1606.04856} \BibitemShut
  {NoStop}%
\bibitem [{\citenamefont {{Abbott}}\ \emph
  {et~al.}(2017{\natexlab{a}})\citenamefont {{Abbott}}, \citenamefont
  {{Abbott}}, \citenamefont {{Abbott}}, \citenamefont {{Acernese}},
  \citenamefont {{Ackley}}, \citenamefont {{Adams}}, \citenamefont {{Adams}},
  \citenamefont {{Addesso}}, \citenamefont {{Adhikari}}, \citenamefont {{Adya}}
  \emph {et~al.}}]{2017PhRvL.119p1101A}%
  \BibitemOpen
  \bibfield  {author} {\bibinfo {author} {\bibfnamefont {B.~P.}\ \bibnamefont
  {{Abbott}}}, \bibinfo {author} {\bibfnamefont {R.}~\bibnamefont {{Abbott}}},
  \bibinfo {author} {\bibfnamefont {T.~D.}\ \bibnamefont {{Abbott}}}, \bibinfo
  {author} {\bibfnamefont {F.}~\bibnamefont {{Acernese}}}, \bibinfo {author}
  {\bibfnamefont {K.}~\bibnamefont {{Ackley}}}, \bibinfo {author}
  {\bibfnamefont {C.}~\bibnamefont {{Adams}}}, \bibinfo {author} {\bibfnamefont
  {T.}~\bibnamefont {{Adams}}}, \bibinfo {author} {\bibfnamefont
  {P.}~\bibnamefont {{Addesso}}}, \bibinfo {author} {\bibfnamefont {R.~X.}\
  \bibnamefont {{Adhikari}}}, \bibinfo {author} {\bibfnamefont {V.~B.}\
  \bibnamefont {{Adya}}},  \emph {et~al.},\ }\bibfield  {title} {\enquote
  {\bibinfo {title} {{GW170817: Observation of Gravitational Waves from a
  Binary Neutron Star Inspiral}},}\ }\href {\doibase
  10.1103/PhysRevLett.119.161101} {\bibfield  {journal} {\bibinfo  {journal}
  {Phys. Rev. Lett.}\ }\textbf {\bibinfo {volume} {119}},\ \bibinfo {eid}
  {161101} (\bibinfo {year} {2017}{\natexlab{a}})},\ \Eprint
  {http://arxiv.org/abs/1710.05832} {arXiv:1710.05832} \BibitemShut {NoStop}%
\bibitem [{\citenamefont {{Abbott}}\ \emph {et~al.}(2010)\citenamefont
  {{Abbott}}, \citenamefont {{Abbott}}, \citenamefont {{Acernese}},
  \citenamefont {{Adhikari}}, \citenamefont {{Ajith}}, \citenamefont {{Allen}},
  \citenamefont {{Allen}}, \citenamefont {{Alshourbagy}}, \citenamefont
  {{Amin}}, \citenamefont {{Anderson}} \emph {et~al.}}]{2010ApJ...713..671A}%
  \BibitemOpen
  \bibfield  {author} {\bibinfo {author} {\bibfnamefont {B.~P.}\ \bibnamefont
  {{Abbott}}}, \bibinfo {author} {\bibfnamefont {R.}~\bibnamefont {{Abbott}}},
  \bibinfo {author} {\bibfnamefont {F.}~\bibnamefont {{Acernese}}}, \bibinfo
  {author} {\bibfnamefont {R.}~\bibnamefont {{Adhikari}}}, \bibinfo {author}
  {\bibfnamefont {P.}~\bibnamefont {{Ajith}}}, \bibinfo {author} {\bibfnamefont
  {B.}~\bibnamefont {{Allen}}}, \bibinfo {author} {\bibfnamefont
  {G.}~\bibnamefont {{Allen}}}, \bibinfo {author} {\bibfnamefont
  {M.}~\bibnamefont {{Alshourbagy}}}, \bibinfo {author} {\bibfnamefont {R.~S.}\
  \bibnamefont {{Amin}}}, \bibinfo {author} {\bibfnamefont {S.~B.}\
  \bibnamefont {{Anderson}}},  \emph {et~al.},\ }\bibfield  {title} {\enquote
  {\bibinfo {title} {{Searches for Gravitational Waves from Known Pulsars with
  Science Run 5 LIGO Data}},}\ }\href {\doibase 10.1088/0004-637X/713/1/671}
  {\bibfield  {journal} {\bibinfo  {journal} {\apj}\ }\textbf {\bibinfo
  {volume} {713}},\ \bibinfo {pages} {671--685} (\bibinfo {year} {2010})},\
  \Eprint {http://arxiv.org/abs/0909.3583} {arXiv:0909.3583} \BibitemShut
  {NoStop}%
\bibitem [{\citenamefont {{Johnson-McDaniel}}\ and\ \citenamefont
  {{Owen}}(2013)}]{2013PhRvD..88d4004J}%
  \BibitemOpen
  \bibfield  {author} {\bibinfo {author} {\bibfnamefont {N.~K.}\ \bibnamefont
  {{Johnson-McDaniel}}}\ and\ \bibinfo {author} {\bibfnamefont {B.~J.}\
  \bibnamefont {{Owen}}},\ }\bibfield  {title} {\enquote {\bibinfo {title}
  {{Maximum elastic deformations of relativistic stars}},}\ }\href {\doibase
  10.1103/PhysRevD.88.044004} {\bibfield  {journal} {\bibinfo  {journal}
  {\prd}\ }\textbf {\bibinfo {volume} {88}},\ \bibinfo {eid} {044004} (\bibinfo
  {year} {2013})},\ \Eprint {http://arxiv.org/abs/1208.5227} {arXiv:1208.5227}
  \BibitemShut {NoStop}%
\bibitem [{\citenamefont {{Ushomirsky}}\ \emph {et~al.}(2000)\citenamefont
  {{Ushomirsky}}, \citenamefont {{Cutler}},\ and\ \citenamefont
  {{Bildsten}}}]{2000MNRAS.319..902U}%
  \BibitemOpen
  \bibfield  {author} {\bibinfo {author} {\bibfnamefont {G.}~\bibnamefont
  {{Ushomirsky}}}, \bibinfo {author} {\bibfnamefont {C.}~\bibnamefont
  {{Cutler}}}, \ and\ \bibinfo {author} {\bibfnamefont {L.}~\bibnamefont
  {{Bildsten}}},\ }\bibfield  {title} {\enquote {\bibinfo {title}
  {{Deformations of accreting neutron star crusts and gravitational wave
  emission}},}\ }\href {\doibase 10.1046/j.1365-8711.2000.03938.x} {\bibfield
  {journal} {\bibinfo  {journal} {Mon. Not. R. Astron. Soc.}\ }\textbf
  {\bibinfo {volume} {319}},\ \bibinfo {pages} {902--932} (\bibinfo {year}
  {2000})},\ \Eprint {http://arxiv.org/abs/astro-ph/0001136} {astro-ph/0001136}
  \BibitemShut {NoStop}%
\bibitem [{\citenamefont {{Abbott}}\ \emph {et~al.}(2004)\citenamefont
  {{Abbott}}, \citenamefont {{Abbott}}, \citenamefont {{Adhikari}},
  \citenamefont {{Ageev}}, \citenamefont {{Allen}}, \citenamefont {{Amin}},
  \citenamefont {{Anderson}}, \citenamefont {{Anderson}}, \citenamefont
  {{Araya}}, \citenamefont {{Armandula}} \emph {et~al.}}]{2004PhRvD..69h2004A}%
  \BibitemOpen
  \bibfield  {author} {\bibinfo {author} {\bibfnamefont {B.}~\bibnamefont
  {{Abbott}}}, \bibinfo {author} {\bibfnamefont {R.}~\bibnamefont {{Abbott}}},
  \bibinfo {author} {\bibfnamefont {R.}~\bibnamefont {{Adhikari}}}, \bibinfo
  {author} {\bibfnamefont {A.}~\bibnamefont {{Ageev}}}, \bibinfo {author}
  {\bibfnamefont {B.}~\bibnamefont {{Allen}}}, \bibinfo {author} {\bibfnamefont
  {R.}~\bibnamefont {{Amin}}}, \bibinfo {author} {\bibfnamefont {S.~B.}\
  \bibnamefont {{Anderson}}}, \bibinfo {author} {\bibfnamefont {W.~G.}\
  \bibnamefont {{Anderson}}}, \bibinfo {author} {\bibfnamefont
  {M.}~\bibnamefont {{Araya}}}, \bibinfo {author} {\bibfnamefont
  {H.}~\bibnamefont {{Armandula}}},  \emph {et~al.},\ }\bibfield  {title}
  {\enquote {\bibinfo {title} {{Setting upper limits on the strength of
  periodic gravitational waves from PSR J1939+2134 using the first science data
  from the GEO 600 and LIGO detectors}},}\ }\href {\doibase
  10.1103/PhysRevD.69.082004} {\bibfield  {journal} {\bibinfo  {journal}
  {\prd}\ }\textbf {\bibinfo {volume} {69}},\ \bibinfo {eid} {082004} (\bibinfo
  {year} {2004})},\ \Eprint {http://arxiv.org/abs/gr-qc/0308050}
  {gr-qc/0308050} \BibitemShut {NoStop}%
\bibitem [{\citenamefont {{Abbott}}\ \emph {et~al.}(2005)\citenamefont
  {{Abbott}}, \citenamefont {{Abbott}}, \citenamefont {{Adhikari}},
  \citenamefont {{Ageev}}, \citenamefont {{Allen}}, \citenamefont {{Amin}},
  \citenamefont {{Anderson}}, \citenamefont {{Anderson}}, \citenamefont
  {{Araya}}, \citenamefont {{Armandula}} \emph {et~al.}}]{2005PhRvL..94r1103A}%
  \BibitemOpen
  \bibfield  {author} {\bibinfo {author} {\bibfnamefont {B.}~\bibnamefont
  {{Abbott}}}, \bibinfo {author} {\bibfnamefont {R.}~\bibnamefont {{Abbott}}},
  \bibinfo {author} {\bibfnamefont {R.}~\bibnamefont {{Adhikari}}}, \bibinfo
  {author} {\bibfnamefont {A.}~\bibnamefont {{Ageev}}}, \bibinfo {author}
  {\bibfnamefont {B.}~\bibnamefont {{Allen}}}, \bibinfo {author} {\bibfnamefont
  {R.}~\bibnamefont {{Amin}}}, \bibinfo {author} {\bibfnamefont {S.~B.}\
  \bibnamefont {{Anderson}}}, \bibinfo {author} {\bibfnamefont {W.~G.}\
  \bibnamefont {{Anderson}}}, \bibinfo {author} {\bibfnamefont
  {M.}~\bibnamefont {{Araya}}}, \bibinfo {author} {\bibfnamefont
  {H.}~\bibnamefont {{Armandula}}},  \emph {et~al.},\ }\bibfield  {title}
  {\enquote {\bibinfo {title} {{Limits on Gravitational-Wave Emission from
  Selected Pulsars Using LIGO Data}},}\ }\href {\doibase
  10.1103/PhysRevLett.94.181103} {\bibfield  {journal} {\bibinfo  {journal}
  {\prl}\ }\textbf {\bibinfo {volume} {94}},\ \bibinfo {eid} {181103} (\bibinfo
  {year} {2005})},\ \Eprint {http://arxiv.org/abs/gr-qc/0410007}
  {gr-qc/0410007} \BibitemShut {NoStop}%
\bibitem [{\citenamefont {{Abbott}}\ \emph {et~al.}(2007)\citenamefont
  {{Abbott}}, \citenamefont {{Abbott}}, \citenamefont {{Adhikari}},
  \citenamefont {{Agresti}}, \citenamefont {{Ajith}}, \citenamefont {{Allen}},
  \citenamefont {{Amin}}, \citenamefont {{Anderson}}, \citenamefont
  {{Anderson}}, \citenamefont {{Arain}} \emph {et~al.}}]{2007PhRvD..76d2001A}%
  \BibitemOpen
  \bibfield  {author} {\bibinfo {author} {\bibfnamefont {B.}~\bibnamefont
  {{Abbott}}}, \bibinfo {author} {\bibfnamefont {R.}~\bibnamefont {{Abbott}}},
  \bibinfo {author} {\bibfnamefont {R.}~\bibnamefont {{Adhikari}}}, \bibinfo
  {author} {\bibfnamefont {J.}~\bibnamefont {{Agresti}}}, \bibinfo {author}
  {\bibfnamefont {P.}~\bibnamefont {{Ajith}}}, \bibinfo {author} {\bibfnamefont
  {B.}~\bibnamefont {{Allen}}}, \bibinfo {author} {\bibfnamefont
  {R.}~\bibnamefont {{Amin}}}, \bibinfo {author} {\bibfnamefont {S.~B.}\
  \bibnamefont {{Anderson}}}, \bibinfo {author} {\bibfnamefont {W.~G.}\
  \bibnamefont {{Anderson}}}, \bibinfo {author} {\bibfnamefont
  {M.}~\bibnamefont {{Arain}}},  \emph {et~al.},\ }\bibfield  {title} {\enquote
  {\bibinfo {title} {{Upper limits on gravitational wave emission from 78 radio
  pulsars}},}\ }\href {\doibase 10.1103/PhysRevD.76.042001} {\bibfield
  {journal} {\bibinfo  {journal} {\prd}\ }\textbf {\bibinfo {volume} {76}},\
  \bibinfo {eid} {042001} (\bibinfo {year} {2007})},\ \Eprint
  {http://arxiv.org/abs/gr-qc/0702039} {gr-qc/0702039} \BibitemShut {NoStop}%
\bibitem [{\citenamefont {{Aasi}}\ \emph {et~al.}(2014)\citenamefont {{Aasi}},
  \citenamefont {{Abadie}}, \citenamefont {{Abbott}}, \citenamefont {{Abbott}},
  \citenamefont {{Abbott}}, \citenamefont {{Abernathy}}, \citenamefont
  {{Accadia}}, \citenamefont {{Acernese}}, \citenamefont {{Adams}},
  \citenamefont {{Adams}} \emph {et~al.}}]{2014ApJ...785..119A}%
  \BibitemOpen
  \bibfield  {author} {\bibinfo {author} {\bibfnamefont {J.}~\bibnamefont
  {{Aasi}}}, \bibinfo {author} {\bibfnamefont {J.}~\bibnamefont {{Abadie}}},
  \bibinfo {author} {\bibfnamefont {B.~P.}\ \bibnamefont {{Abbott}}}, \bibinfo
  {author} {\bibfnamefont {R.}~\bibnamefont {{Abbott}}}, \bibinfo {author}
  {\bibfnamefont {T.}~\bibnamefont {{Abbott}}}, \bibinfo {author}
  {\bibfnamefont {M.~R.}\ \bibnamefont {{Abernathy}}}, \bibinfo {author}
  {\bibfnamefont {T.}~\bibnamefont {{Accadia}}}, \bibinfo {author}
  {\bibfnamefont {F.}~\bibnamefont {{Acernese}}}, \bibinfo {author}
  {\bibfnamefont {C.}~\bibnamefont {{Adams}}}, \bibinfo {author} {\bibfnamefont
  {T.}~\bibnamefont {{Adams}}},  \emph {et~al.},\ }\bibfield  {title} {\enquote
  {\bibinfo {title} {{Gravitational Waves from Known Pulsars: Results from the
  Initial Detector Era}},}\ }\href {\doibase 10.1088/0004-637X/785/2/119}
  {\bibfield  {journal} {\bibinfo  {journal} {\apj}\ }\textbf {\bibinfo
  {volume} {785}},\ \bibinfo {eid} {119} (\bibinfo {year} {2014})},\ \Eprint
  {http://arxiv.org/abs/1309.4027} {arXiv:1309.4027} \BibitemShut {NoStop}%
\bibitem [{\citenamefont {{Abbott}}\ \emph
  {et~al.}(2017{\natexlab{b}})\citenamefont {{Abbott}}, \citenamefont
  {{Abbott}}, \citenamefont {{Abbott}}, \citenamefont {{Abernathy}},
  \citenamefont {{Acernese}}, \citenamefont {{Ackley}}, \citenamefont
  {{Adams}}, \citenamefont {{Adams}}, \citenamefont {{Addesso}}, \citenamefont
  {{Adhikari}} \emph {et~al.}}]{2017ApJ...839...12A}%
  \BibitemOpen
  \bibfield  {author} {\bibinfo {author} {\bibfnamefont {B.~P.}\ \bibnamefont
  {{Abbott}}}, \bibinfo {author} {\bibfnamefont {R.}~\bibnamefont {{Abbott}}},
  \bibinfo {author} {\bibfnamefont {T.~D.}\ \bibnamefont {{Abbott}}}, \bibinfo
  {author} {\bibfnamefont {M.~R.}\ \bibnamefont {{Abernathy}}}, \bibinfo
  {author} {\bibfnamefont {F.}~\bibnamefont {{Acernese}}}, \bibinfo {author}
  {\bibfnamefont {K.}~\bibnamefont {{Ackley}}}, \bibinfo {author}
  {\bibfnamefont {C.}~\bibnamefont {{Adams}}}, \bibinfo {author} {\bibfnamefont
  {T.}~\bibnamefont {{Adams}}}, \bibinfo {author} {\bibfnamefont
  {P.}~\bibnamefont {{Addesso}}}, \bibinfo {author} {\bibfnamefont {R.~X.}\
  \bibnamefont {{Adhikari}}},  \emph {et~al.},\ }\bibfield  {title} {\enquote
  {\bibinfo {title} {{First Search for Gravitational Waves from Known Pulsars
  with Advanced LIGO}},}\ }\href {\doibase 10.3847/1538-4357/aa677f} {\bibfield
   {journal} {\bibinfo  {journal} {\apj}\ }\textbf {\bibinfo {volume} {839}},\
  \bibinfo {eid} {12} (\bibinfo {year} {2017}{\natexlab{b}})},\ \Eprint
  {http://arxiv.org/abs/1701.07709} {arXiv:1701.07709} \BibitemShut {NoStop}%
\bibitem [{\citenamefont {{Dupuis}}\ and\ \citenamefont
  {{Woan}}(2005)}]{2005PhRvD..72j2002D}%
  \BibitemOpen
  \bibfield  {author} {\bibinfo {author} {\bibfnamefont {R.~J.}\ \bibnamefont
  {{Dupuis}}}\ and\ \bibinfo {author} {\bibfnamefont {G.}~\bibnamefont
  {{Woan}}},\ }\bibfield  {title} {\enquote {\bibinfo {title} {{Bayesian
  estimation of pulsar parameters from gravitational wave data}},}\ }\href
  {\doibase 10.1103/PhysRevD.72.102002} {\bibfield  {journal} {\bibinfo
  {journal} {\prd}\ }\textbf {\bibinfo {volume} {72}},\ \bibinfo {eid} {102002}
  (\bibinfo {year} {2005})},\ \Eprint {http://arxiv.org/abs/gr-qc/0508096}
  {gr-qc/0508096} \BibitemShut {NoStop}%
\bibitem [{\citenamefont {{Abbott}}\ \emph
  {et~al.}(2016{\natexlab{c}})\citenamefont {{Abbott}}, \citenamefont
  {{Abbott}}, \citenamefont {{Abbott}}, \citenamefont {{Abernathy}},
  \citenamefont {{Acernese}}, \citenamefont {{Ackley}}, \citenamefont
  {{Adams}}, \citenamefont {{Adams}}, \citenamefont {{Addesso}}, \citenamefont
  {{Adhikari}} \emph {et~al.}}]{2016PhRvL.116m1103A}%
  \BibitemOpen
  \bibfield  {author} {\bibinfo {author} {\bibfnamefont {B.~P.}\ \bibnamefont
  {{Abbott}}}, \bibinfo {author} {\bibfnamefont {R.}~\bibnamefont {{Abbott}}},
  \bibinfo {author} {\bibfnamefont {T.~D.}\ \bibnamefont {{Abbott}}}, \bibinfo
  {author} {\bibfnamefont {M.~R.}\ \bibnamefont {{Abernathy}}}, \bibinfo
  {author} {\bibfnamefont {F.}~\bibnamefont {{Acernese}}}, \bibinfo {author}
  {\bibfnamefont {K.}~\bibnamefont {{Ackley}}}, \bibinfo {author}
  {\bibfnamefont {C.}~\bibnamefont {{Adams}}}, \bibinfo {author} {\bibfnamefont
  {T.}~\bibnamefont {{Adams}}}, \bibinfo {author} {\bibfnamefont
  {P.}~\bibnamefont {{Addesso}}}, \bibinfo {author} {\bibfnamefont {R.~X.}\
  \bibnamefont {{Adhikari}}},  \emph {et~al.},\ }\bibfield  {title} {\enquote
  {\bibinfo {title} {{GW150914: The Advanced LIGO Detectors in the Era of First
  Discoveries}},}\ }\href {\doibase 10.1103/PhysRevLett.116.131103} {\bibfield
  {journal} {\bibinfo  {journal} {Phys. Rev. Lett.}\ }\textbf {\bibinfo
  {volume} {116}},\ \bibinfo {eid} {131103} (\bibinfo {year}
  {2016}{\natexlab{c}})},\ \Eprint {http://arxiv.org/abs/1602.03838}
  {arXiv:1602.03838} \BibitemShut {NoStop}%
\bibitem [{\citenamefont {{Manchester}}\ \emph {et~al.}(2005)\citenamefont
  {{Manchester}}, \citenamefont {{Hobbs}}, \citenamefont {{Teoh}},\ and\
  \citenamefont {{Hobbs}}}]{2005AJ....129.1993M}%
  \BibitemOpen
  \bibfield  {author} {\bibinfo {author} {\bibfnamefont {R.~N.}\ \bibnamefont
  {{Manchester}}}, \bibinfo {author} {\bibfnamefont {G.~B.}\ \bibnamefont
  {{Hobbs}}}, \bibinfo {author} {\bibfnamefont {A.}~\bibnamefont {{Teoh}}}, \
  and\ \bibinfo {author} {\bibfnamefont {M.}~\bibnamefont {{Hobbs}}},\
  }\bibfield  {title} {\enquote {\bibinfo {title} {{The Australia Telescope
  National Facility Pulsar Catalogue}},}\ }\href {\doibase 10.1086/428488}
  {\bibfield  {journal} {\bibinfo  {journal} {Astronomical Journal}\ }\textbf
  {\bibinfo {volume} {129}},\ \bibinfo {pages} {1993--2006} (\bibinfo {year}
  {2005})},\ \bibinfo {note}
  {\url{http://www.atnf.csiro.au/people/pulsar/psrcat/}},\ \Eprint
  {http://arxiv.org/abs/astro-ph/0412641} {astro-ph/0412641} \BibitemShut
  {NoStop}%
\bibitem [{\citenamefont {{Cutler}}(2002)}]{2002PhRvD..66h4025C}%
  \BibitemOpen
  \bibfield  {author} {\bibinfo {author} {\bibfnamefont {C.}~\bibnamefont
  {{Cutler}}},\ }\bibfield  {title} {\enquote {\bibinfo {title} {{Gravitational
  waves from neutron stars with large toroidal B fields}},}\ }\href {\doibase
  10.1103/PhysRevD.66.084025} {\bibfield  {journal} {\bibinfo  {journal}
  {\prd}\ }\textbf {\bibinfo {volume} {66}},\ \bibinfo {eid} {084025} (\bibinfo
  {year} {2002})}\BibitemShut {NoStop}%
\bibitem [{\citenamefont {{Woan}}\ \emph {et~al.}(2018)\citenamefont {{Woan}},
  \citenamefont {{Pitkin}}, \citenamefont {{Haskell}}, \citenamefont
  {{Jones}},\ and\ \citenamefont {{Lasky}}}]{2018arXiv180602822W}%
  \BibitemOpen
  \bibfield  {author} {\bibinfo {author} {\bibfnamefont {G.}~\bibnamefont
  {{Woan}}}, \bibinfo {author} {\bibfnamefont {M.~D.}\ \bibnamefont
  {{Pitkin}}}, \bibinfo {author} {\bibfnamefont {B.}~\bibnamefont {{Haskell}}},
  \bibinfo {author} {\bibfnamefont {D.~I.}\ \bibnamefont {{Jones}}}, \ and\
  \bibinfo {author} {\bibfnamefont {P.~D.}\ \bibnamefont {{Lasky}}},\
  }\bibfield  {title} {\enquote {\bibinfo {title} {{Evidence for a Minimum
  Ellipticity in Millisecond Pulsars}},}\ }\href {\doibase
  10.3847/2041-8213/aad86a} {\bibfield  {journal} {\bibinfo  {journal}
  {Astrophys. J. Lett.}\ }\textbf {\bibinfo {volume} {863}},\ \bibinfo {eid}
  {L40} (\bibinfo {year} {2018})},\ \Eprint {http://arxiv.org/abs/1806.02822}
  {arXiv:1806.02822} \BibitemShut {NoStop}%
\bibitem [{\citenamefont {{Cutler}}\ and\ \citenamefont
  {{Schutz}}(2005)}]{2005PhRvD..72f3006C}%
  \BibitemOpen
  \bibfield  {author} {\bibinfo {author} {\bibfnamefont {C.}~\bibnamefont
  {{Cutler}}}\ and\ \bibinfo {author} {\bibfnamefont {B.~F.}\ \bibnamefont
  {{Schutz}}},\ }\bibfield  {title} {\enquote {\bibinfo {title} {{Generalized
  F-statistic: Multiple detectors and multiple gravitational wave pulsars}},}\
  }\href {\doibase 10.1103/PhysRevD.72.063006} {\bibfield  {journal} {\bibinfo
  {journal} {\prd}\ }\textbf {\bibinfo {volume} {72}},\ \bibinfo {eid} {063006}
  (\bibinfo {year} {2005})},\ \Eprint {http://arxiv.org/abs/gr-qc/0504011}
  {gr-qc/0504011} \BibitemShut {NoStop}%
\bibitem [{\citenamefont {{Jaranowski}}\ \emph {et~al.}(1998)\citenamefont
  {{Jaranowski}}, \citenamefont {{Kr{\'o}lak}},\ and\ \citenamefont
  {{Schutz}}}]{1998PhRvD..58f3001J}%
  \BibitemOpen
  \bibfield  {author} {\bibinfo {author} {\bibfnamefont {P.}~\bibnamefont
  {{Jaranowski}}}, \bibinfo {author} {\bibfnamefont {A.}~\bibnamefont
  {{Kr{\'o}lak}}}, \ and\ \bibinfo {author} {\bibfnamefont {B.~F.}\
  \bibnamefont {{Schutz}}},\ }\bibfield  {title} {\enquote {\bibinfo {title}
  {{Data analysis of gravitational-wave signals from spinning neutron stars:
  The signal and its detection}},}\ }\href {\doibase
  10.1103/PhysRevD.58.063001} {\bibfield  {journal} {\bibinfo  {journal}
  {\prd}\ }\textbf {\bibinfo {volume} {58}},\ \bibinfo {eid} {063001} (\bibinfo
  {year} {1998})},\ \Eprint {http://arxiv.org/abs/gr-qc/9804014}
  {gr-qc/9804014} \BibitemShut {NoStop}%
\bibitem [{\citenamefont {{Fan}}\ \emph {et~al.}(2016)\citenamefont {{Fan}},
  \citenamefont {{Chen}},\ and\ \citenamefont
  {{Messenger}}}]{2016PhRvD..94h4029F}%
  \BibitemOpen
  \bibfield  {author} {\bibinfo {author} {\bibfnamefont {X.}~\bibnamefont
  {{Fan}}}, \bibinfo {author} {\bibfnamefont {Y.}~\bibnamefont {{Chen}}}, \
  and\ \bibinfo {author} {\bibfnamefont {C.}~\bibnamefont {{Messenger}}},\
  }\bibfield  {title} {\enquote {\bibinfo {title} {{Method to detect
  gravitational waves from an ensemble of known pulsars}},}\ }\href {\doibase
  10.1103/PhysRevD.94.084029} {\bibfield  {journal} {\bibinfo  {journal}
  {\prd}\ }\textbf {\bibinfo {volume} {94}},\ \bibinfo {eid} {084029} (\bibinfo
  {year} {2016})},\ \Eprint {http://arxiv.org/abs/1607.06735}
  {arXiv:1607.06735} \BibitemShut {NoStop}%
\bibitem [{\citenamefont {{Smith}}\ and\ \citenamefont
  {{Thrane}}(2018)}]{2018PhRvX...8b1019S}%
  \BibitemOpen
  \bibfield  {author} {\bibinfo {author} {\bibfnamefont {Rory}\ \bibnamefont
  {{Smith}}}\ and\ \bibinfo {author} {\bibfnamefont {Eric}\ \bibnamefont
  {{Thrane}}},\ }\bibfield  {title} {\enquote {\bibinfo {title} {{Optimal
  Search for an Astrophysical Gravitational-Wave Background}},}\ }\href
  {\doibase 10.1103/PhysRevX.8.021019} {\bibfield  {journal} {\bibinfo
  {journal} {Phys. Rev. X}\ }\textbf {\bibinfo {volume} {8}} (\bibinfo {year}
  {2018}),\ 10.1103/PhysRevX.8.021019}\BibitemShut {NoStop}%
\bibitem [{\citenamefont {{Pitkin}}\ \emph {et~al.}(2015)\citenamefont
  {{Pitkin}}, \citenamefont {{Gill}}, \citenamefont {{Jones}}, \citenamefont
  {{Woan}},\ and\ \citenamefont {{Davies}}}]{2015MNRAS.453.4399P}%
  \BibitemOpen
  \bibfield  {author} {\bibinfo {author} {\bibfnamefont {M.}~\bibnamefont
  {{Pitkin}}}, \bibinfo {author} {\bibfnamefont {C.}~\bibnamefont {{Gill}}},
  \bibinfo {author} {\bibfnamefont {D.~I.}\ \bibnamefont {{Jones}}}, \bibinfo
  {author} {\bibfnamefont {G.}~\bibnamefont {{Woan}}}, \ and\ \bibinfo {author}
  {\bibfnamefont {G.~S.}\ \bibnamefont {{Davies}}},\ }\bibfield  {title}
  {\enquote {\bibinfo {title} {{First results and future prospects for
  dual-harmonic searches for gravitational waves from spinning neutron
  stars}},}\ }\href {\doibase 10.1093/mnras/stv1931} {\bibfield  {journal}
  {\bibinfo  {journal} {Mon. Not. R. Astron. Soc.}\ }\textbf {\bibinfo {volume}
  {453}},\ \bibinfo {pages} {4399--4420} (\bibinfo {year} {2015})},\ \Eprint
  {http://arxiv.org/abs/1508.00416} {arXiv:1508.00416} \BibitemShut {NoStop}%
\bibitem [{\citenamefont {{Pitkin}}\ \emph {et~al.}()\citenamefont {{Pitkin}},
  \citenamefont {{Isi}}, \citenamefont {{Veitch}},\ and\ \citenamefont
  {{Woan}}}]{2017arXiv170508978P}%
  \BibitemOpen
  \bibfield  {author} {\bibinfo {author} {\bibfnamefont {M.}~\bibnamefont
  {{Pitkin}}}, \bibinfo {author} {\bibfnamefont {M.}~\bibnamefont {{Isi}}},
  \bibinfo {author} {\bibfnamefont {J.}~\bibnamefont {{Veitch}}}, \ and\
  \bibinfo {author} {\bibfnamefont {G.}~\bibnamefont {{Woan}}},\ }\bibfield
  {title} {\enquote {\bibinfo {title} {{A nested sampling code for targeted
  searches for continuous gravitational waves from pulsars}},}\ }\href@noop {}
  {\ }\Eprint {http://arxiv.org/abs/1705.08978} {arXiv:1705.08978} \BibitemShut
  {NoStop}%
\bibitem [{\citenamefont {{Yao}}\ \emph {et~al.}(2017)\citenamefont {{Yao}},
  \citenamefont {{Manchester}},\ and\ \citenamefont
  {{Wang}}}]{2017ApJ...835...29Y}%
  \BibitemOpen
  \bibfield  {author} {\bibinfo {author} {\bibfnamefont {J.~M.}\ \bibnamefont
  {{Yao}}}, \bibinfo {author} {\bibfnamefont {R.~N.}\ \bibnamefont
  {{Manchester}}}, \ and\ \bibinfo {author} {\bibfnamefont {N.}~\bibnamefont
  {{Wang}}},\ }\bibfield  {title} {\enquote {\bibinfo {title} {{A New
  Electron-density Model for Estimation of Pulsar and FRB Distances}},}\ }\href
  {\doibase 10.3847/1538-4357/835/1/29} {\bibfield  {journal} {\bibinfo
  {journal} {\apj}\ }\textbf {\bibinfo {volume} {835}},\ \bibinfo {eid} {29}
  (\bibinfo {year} {2017})},\ \Eprint {http://arxiv.org/abs/1610.09448}
  {arXiv:1610.09448} \BibitemShut {NoStop}%
\bibitem [{\citenamefont {{Cordes}}\ and\ \citenamefont
  {{Lazio}}()}]{2002astro.ph..7156C}%
  \BibitemOpen
  \bibfield  {author} {\bibinfo {author} {\bibfnamefont {J.~M.}\ \bibnamefont
  {{Cordes}}}\ and\ \bibinfo {author} {\bibfnamefont {T.~J.~W.}\ \bibnamefont
  {{Lazio}}},\ }\bibfield  {title} {\enquote {\bibinfo {title} {{NE2001.I. A
  New Model for the Galactic Distribution of Free Electrons and its
  Fluctuations}},}\ }\href@noop {} {\ }\Eprint
  {http://arxiv.org/abs/astro-ph/0207156} {astro-ph/0207156} \BibitemShut
  {NoStop}%
\bibitem [{\citenamefont {{Lorimer}}(2008)}]{2008LRR....11....8L}%
  \BibitemOpen
  \bibfield  {author} {\bibinfo {author} {\bibfnamefont {D.~R.}\ \bibnamefont
  {{Lorimer}}},\ }\bibfield  {title} {\enquote {\bibinfo {title} {{Binary and
  Millisecond Pulsars}},}\ }\href {\doibase 10.12942/lrr-2008-8} {\bibfield
  {journal} {\bibinfo  {journal} {Living Rev. Relativity}\ }\textbf {\bibinfo
  {volume} {11}} (\bibinfo {year} {2008}),\ 10.12942/lrr-2008-8},\ \Eprint
  {http://arxiv.org/abs/0811.0762} {arXiv:0811.0762} \BibitemShut {NoStop}%
\bibitem [{\citenamefont {{Hogg}}\ \emph {et~al.}(2010)\citenamefont {{Hogg}},
  \citenamefont {{Myers}},\ and\ \citenamefont {{Bovy}}}]{2010ApJ...725.2166H}%
  \BibitemOpen
  \bibfield  {author} {\bibinfo {author} {\bibfnamefont {D.~W.}\ \bibnamefont
  {{Hogg}}}, \bibinfo {author} {\bibfnamefont {A.~D.}\ \bibnamefont {{Myers}}},
  \ and\ \bibinfo {author} {\bibfnamefont {J.}~\bibnamefont {{Bovy}}},\
  }\bibfield  {title} {\enquote {\bibinfo {title} {{Inferring the Eccentricity
  Distribution}},}\ }\href {\doibase 10.1088/0004-637X/725/2/2166} {\bibfield
  {journal} {\bibinfo  {journal} {\apj}\ }\textbf {\bibinfo {volume} {725}},\
  \bibinfo {pages} {2166--2175} (\bibinfo {year} {2010})},\ \Eprint
  {http://arxiv.org/abs/1008.4146} {arXiv:1008.4146} \BibitemShut {NoStop}%
\bibitem [{\citenamefont {{Del Pozzo}}\ \emph {et~al.}(2018)\citenamefont {{Del
  Pozzo}}, \citenamefont {{Berry}}, \citenamefont {{Ghosh}}, \citenamefont
  {{Haines}}, \citenamefont {{Singer}},\ and\ \citenamefont
  {{Vecchio}}}]{2018arXiv180108009D}%
  \BibitemOpen
  \bibfield  {author} {\bibinfo {author} {\bibfnamefont {W.}~\bibnamefont {{Del
  Pozzo}}}, \bibinfo {author} {\bibfnamefont {C.~P.~L.}\ \bibnamefont
  {{Berry}}}, \bibinfo {author} {\bibfnamefont {A.}~\bibnamefont {{Ghosh}}},
  \bibinfo {author} {\bibfnamefont {T.~S.~F.}\ \bibnamefont {{Haines}}},
  \bibinfo {author} {\bibfnamefont {L.~P.}\ \bibnamefont {{Singer}}}, \ and\
  \bibinfo {author} {\bibfnamefont {A.}~\bibnamefont {{Vecchio}}},\ }\bibfield
  {title} {\enquote {\bibinfo {title} {{Dirichlet process Gaussian-mixture
  model: An application to localizing coalescing binary neutron stars with
  gravitational-wave observations}},}\ }\href {\doibase 10.1093/mnras/sty1485}
  {\bibfield  {journal} {\bibinfo  {journal} {Mon. Not. R. Astron. Soc.}\
  }\textbf {\bibinfo {volume} {479}},\ \bibinfo {pages} {601--614} (\bibinfo
  {year} {2018})},\ \Eprint {http://arxiv.org/abs/1801.08009}
  {arXiv:1801.08009} \BibitemShut {NoStop}%
\bibitem [{\citenamefont {Skilling}(2006)}]{Skilling:2006}%
  \BibitemOpen
  \bibfield  {author} {\bibinfo {author} {\bibfnamefont {J.}~\bibnamefont
  {Skilling}},\ }\bibfield  {title} {\enquote {\bibinfo {title} {Nested
  sampling for general bayesian computation},}\ }\href {\doibase
  10.1214/06-BA127} {\bibfield  {journal} {\bibinfo  {journal} {Bayesian
  Anal.}\ }\textbf {\bibinfo {volume} {1}},\ \bibinfo {pages} {833--860}
  (\bibinfo {year} {2006})}\BibitemShut {NoStop}%
\bibitem [{\citenamefont {{Shklovskii}}(1970)}]{1970SvA....13..562S}%
  \BibitemOpen
  \bibfield  {author} {\bibinfo {author} {\bibfnamefont {I.~S.}\ \bibnamefont
  {{Shklovskii}}},\ }\bibfield  {title} {\enquote {\bibinfo {title} {{Possible
  Causes of the Secular Increase in Pulsar Periods.}}}\ }\href@noop {}
  {\bibfield  {journal} {\bibinfo  {journal} {Sov. Astron.}\ }\textbf {\bibinfo
  {volume} {13}},\ \bibinfo {pages} {562} (\bibinfo {year} {1970})}\BibitemShut
  {NoStop}%
\bibitem [{\citenamefont {{Damour}}\ and\ \citenamefont
  {{Taylor}}(1991)}]{1991ApJ...366..501D}%
  \BibitemOpen
  \bibfield  {author} {\bibinfo {author} {\bibfnamefont {T.}~\bibnamefont
  {{Damour}}}\ and\ \bibinfo {author} {\bibfnamefont {J.~H.}\ \bibnamefont
  {{Taylor}}},\ }\bibfield  {title} {\enquote {\bibinfo {title} {{On the
  orbital period change of the binary pulsar PSR 1913 + 16}},}\ }\href
  {\doibase 10.1086/169585} {\bibfield  {journal} {\bibinfo  {journal} {\apj}\
  }\textbf {\bibinfo {volume} {366}},\ \bibinfo {pages} {501--511} (\bibinfo
  {year} {1991})}\BibitemShut {NoStop}%
\bibitem [{\citenamefont {{Keitel}}\ \emph {et~al.}(2014)\citenamefont
  {{Keitel}}, \citenamefont {{Prix}}, \citenamefont {{Papa}}, \citenamefont
  {{Leaci}},\ and\ \citenamefont {{Siddiqi}}}]{2014PhRvD..89f4023K}%
  \BibitemOpen
  \bibfield  {author} {\bibinfo {author} {\bibfnamefont {D.}~\bibnamefont
  {{Keitel}}}, \bibinfo {author} {\bibfnamefont {R.}~\bibnamefont {{Prix}}},
  \bibinfo {author} {\bibfnamefont {M.~A.}\ \bibnamefont {{Papa}}}, \bibinfo
  {author} {\bibfnamefont {P.}~\bibnamefont {{Leaci}}}, \ and\ \bibinfo
  {author} {\bibfnamefont {M.}~\bibnamefont {{Siddiqi}}},\ }\bibfield  {title}
  {\enquote {\bibinfo {title} {{Search for continuous gravitational waves:
  Improving robustness versus instrumental artifacts}},}\ }\href {\doibase
  10.1103/PhysRevD.89.064023} {\bibfield  {journal} {\bibinfo  {journal}
  {\prd}\ }\textbf {\bibinfo {volume} {89}},\ \bibinfo {eid} {064023} (\bibinfo
  {year} {2014})},\ \Eprint {http://arxiv.org/abs/1311.5738} {arXiv:1311.5738}
  \BibitemShut {NoStop}%
\bibitem [{\citenamefont {{Isi}}\ \emph {et~al.}(2017)\citenamefont {{Isi}},
  \citenamefont {{Pitkin}},\ and\ \citenamefont
  {{Weinstein}}}]{2017PhRvD..96d2001I}%
  \BibitemOpen
  \bibfield  {author} {\bibinfo {author} {\bibfnamefont {M.}~\bibnamefont
  {{Isi}}}, \bibinfo {author} {\bibfnamefont {M.}~\bibnamefont {{Pitkin}}}, \
  and\ \bibinfo {author} {\bibfnamefont {A.~J.}\ \bibnamefont {{Weinstein}}},\
  }\bibfield  {title} {\enquote {\bibinfo {title} {{Probing dynamical gravity
  with the polarization of continuous gravitational waves}},}\ }\href {\doibase
  10.1103/PhysRevD.96.042001} {\bibfield  {journal} {\bibinfo  {journal}
  {\prd}\ }\textbf {\bibinfo {volume} {96}},\ \bibinfo {eid} {042001} (\bibinfo
  {year} {2017})},\ \Eprint {http://arxiv.org/abs/1703.07530}
  {arXiv:1703.07530} \BibitemShut {NoStop}%
\bibitem [{\citenamefont {{Prix}}\ and\ \citenamefont
  {{Krishnan}}(2009)}]{2009CQGra..26t4013P}%
  \BibitemOpen
  \bibfield  {author} {\bibinfo {author} {\bibfnamefont {R.}~\bibnamefont
  {{Prix}}}\ and\ \bibinfo {author} {\bibfnamefont {B.}~\bibnamefont
  {{Krishnan}}},\ }\bibfield  {title} {\enquote {\bibinfo {title} {{Targeted
  search for continuous gravitational waves: Bayesian versus maximum-likelihood
  statistics}},}\ }\href {\doibase 10.1088/0264-9381/26/20/204013} {\bibfield
  {journal} {\bibinfo  {journal} {Classical Quantum Gravity}\ }\textbf
  {\bibinfo {volume} {26}},\ \bibinfo {eid} {204013} (\bibinfo {year}
  {2009})},\ \Eprint {http://arxiv.org/abs/0907.2569} {arXiv:0907.2569}
  \BibitemShut {NoStop}%
\bibitem [{\citenamefont {{{LIGO} {S}cientific
  {C}ollaboration}}(2010)}]{aLIGOdesign}%
  \BibitemOpen
  \bibfield  {author} {\bibinfo {author} {\bibnamefont {{{LIGO} {S}cientific
  {C}ollaboration}}},\ }\href@noop {} {\emph {\bibinfo {title} {{Advanced LIGO
  anticipated sensitivity curves}}}},\ \bibinfo {type} {Tech. Rep.}\ \bibinfo
  {number} {LIGO-T0900288-v3}\ (\bibinfo {year} {2010})\ \bibinfo {note}
  {\url{https://dcc.ligo.org/LIGO-T0900288/public}}\BibitemShut {NoStop}%
\bibitem [{\citenamefont {{Abbott}}\ \emph
  {et~al.}(2016{\natexlab{d}})\citenamefont {{Abbott}}, \citenamefont
  {{Abbott}}, \citenamefont {{Abbott}}, \citenamefont {{Abernathy}},
  \citenamefont {{Acernese}}, \citenamefont {{Ackley}}, \citenamefont
  {{Adams}}, \citenamefont {{Adams}}, \citenamefont {{Addesso}}, \citenamefont
  {{Adhikari}} \emph {et~al.}}]{2016LRR....19....1A}%
  \BibitemOpen
  \bibfield  {author} {\bibinfo {author} {\bibfnamefont {B.~P.}\ \bibnamefont
  {{Abbott}}}, \bibinfo {author} {\bibfnamefont {R.}~\bibnamefont {{Abbott}}},
  \bibinfo {author} {\bibfnamefont {T.~D.}\ \bibnamefont {{Abbott}}}, \bibinfo
  {author} {\bibfnamefont {M.~R.}\ \bibnamefont {{Abernathy}}}, \bibinfo
  {author} {\bibfnamefont {F.}~\bibnamefont {{Acernese}}}, \bibinfo {author}
  {\bibfnamefont {K.}~\bibnamefont {{Ackley}}}, \bibinfo {author}
  {\bibfnamefont {C.}~\bibnamefont {{Adams}}}, \bibinfo {author} {\bibfnamefont
  {T.}~\bibnamefont {{Adams}}}, \bibinfo {author} {\bibfnamefont
  {P.}~\bibnamefont {{Addesso}}}, \bibinfo {author} {\bibfnamefont {R.~X.}\
  \bibnamefont {{Adhikari}}},  \emph {et~al.},\ }\bibfield  {title} {\enquote
  {\bibinfo {title} {{Prospects for Observing and Localizing Gravitational-Wave
  Transients with Advanced LIGO and Advanced Virgo}},}\ }\href {\doibase
  10.1007/lrr-2016-1} {\bibfield  {journal} {\bibinfo  {journal} {Living Rev.
  Relativity}\ }\textbf {\bibinfo {volume} {19}},\ \bibinfo {eid} {1} (\bibinfo
  {year} {2016}{\natexlab{d}})}\BibitemShut {NoStop}%
\bibitem [{\citenamefont {{Pitkin}}(2011)}]{2011MNRAS.415.1849P}%
  \BibitemOpen
  \bibfield  {author} {\bibinfo {author} {\bibfnamefont {M.}~\bibnamefont
  {{Pitkin}}},\ }\bibfield  {title} {\enquote {\bibinfo {title} {{Prospects of
  observing continuous gravitational waves from known pulsars}},}\ }\href
  {\doibase 10.1111/j.1365-2966.2011.18818.x} {\bibfield  {journal} {\bibinfo
  {journal} {Mon. Not. R. Astron. Soc.}\ }\textbf {\bibinfo {volume} {415}},\
  \bibinfo {pages} {1849--1863} (\bibinfo {year} {2011})},\ \Eprint
  {http://arxiv.org/abs/1103.5867} {arXiv:1103.5867} \BibitemShut {NoStop}%
\bibitem [{\citenamefont {{Veitch}}\ and\ \citenamefont
  {{Vecchio}}(2010)}]{Veitch:2010}%
  \BibitemOpen
  \bibfield  {author} {\bibinfo {author} {\bibfnamefont {J.}~\bibnamefont
  {{Veitch}}}\ and\ \bibinfo {author} {\bibfnamefont {A.}~\bibnamefont
  {{Vecchio}}},\ }\bibfield  {title} {\enquote {\bibinfo {title} {{Bayesian
  coherent analysis of in-spiral gravitational wave signals with a detector
  network}},}\ }\href {\doibase 10.1103/PhysRevD.81.062003} {\bibfield
  {journal} {\bibinfo  {journal} {Phys. Rev. D}\ }\textbf {\bibinfo {volume}
  {81}},\ \bibinfo {pages} {062003} (\bibinfo {year} {2010})},\ \Eprint
  {http://arxiv.org/abs/0911.3820} {arXiv:0911.3820} \BibitemShut {NoStop}%
\bibitem [{\citenamefont {{Veitch}}\ \emph {et~al.}(2015)\citenamefont
  {{Veitch}}, \citenamefont {{Raymond}}, \citenamefont {{Farr}}, \citenamefont
  {{Farr}}, \citenamefont {{Graff}}, \citenamefont {{Vitale}}, \citenamefont
  {{Aylott}}, \citenamefont {{Blackburn}}, \citenamefont {{Christensen}},
  \citenamefont {{Coughlin}}, \citenamefont {{Del Pozzo}}, \citenamefont
  {{Feroz}}, \citenamefont {{Gair}}, \citenamefont {{Haster}}, \citenamefont
  {{Kalogera}}, \citenamefont {{Littenberg}}, \citenamefont {{Mandel}},
  \citenamefont {{O'Shaughnessy}}, \citenamefont {{Pitkin}}, \citenamefont
  {{Rodriguez}} \emph {et~al.}}]{2015PhRvD..91d2003V}%
  \BibitemOpen
  \bibfield  {author} {\bibinfo {author} {\bibfnamefont {J.}~\bibnamefont
  {{Veitch}}}, \bibinfo {author} {\bibfnamefont {V.}~\bibnamefont {{Raymond}}},
  \bibinfo {author} {\bibfnamefont {B.}~\bibnamefont {{Farr}}}, \bibinfo
  {author} {\bibfnamefont {W.}~\bibnamefont {{Farr}}}, \bibinfo {author}
  {\bibfnamefont {P.}~\bibnamefont {{Graff}}}, \bibinfo {author} {\bibfnamefont
  {S.}~\bibnamefont {{Vitale}}}, \bibinfo {author} {\bibfnamefont
  {B.}~\bibnamefont {{Aylott}}}, \bibinfo {author} {\bibfnamefont
  {K.}~\bibnamefont {{Blackburn}}}, \bibinfo {author} {\bibfnamefont
  {N.}~\bibnamefont {{Christensen}}}, \bibinfo {author} {\bibfnamefont
  {M.}~\bibnamefont {{Coughlin}}}, \bibinfo {author} {\bibfnamefont
  {W.}~\bibnamefont {{Del Pozzo}}}, \bibinfo {author} {\bibfnamefont
  {F.}~\bibnamefont {{Feroz}}}, \bibinfo {author} {\bibfnamefont
  {J.}~\bibnamefont {{Gair}}}, \bibinfo {author} {\bibfnamefont {C.-J.}\
  \bibnamefont {{Haster}}}, \bibinfo {author} {\bibfnamefont {V.}~\bibnamefont
  {{Kalogera}}}, \bibinfo {author} {\bibfnamefont {T.}~\bibnamefont
  {{Littenberg}}}, \bibinfo {author} {\bibfnamefont {I.}~\bibnamefont
  {{Mandel}}}, \bibinfo {author} {\bibfnamefont {R.}~\bibnamefont
  {{O'Shaughnessy}}}, \bibinfo {author} {\bibfnamefont {M.}~\bibnamefont
  {{Pitkin}}}, \bibinfo {author} {\bibfnamefont {C.}~\bibnamefont
  {{Rodriguez}}},  \emph {et~al.},\ }\bibfield  {title} {\enquote {\bibinfo
  {title} {{Parameter estimation for compact binaries with ground-based
  gravitational-wave observations using the LALInference software library}},}\
  }\href {\doibase 10.1103/PhysRevD.91.042003} {\bibfield  {journal} {\bibinfo
  {journal} {\prd}\ }\textbf {\bibinfo {volume} {91}},\ \bibinfo {eid} {042003}
  (\bibinfo {year} {2015})},\ \Eprint {http://arxiv.org/abs/1409.7215}
  {arXiv:1409.7215} \BibitemShut {NoStop}%
\bibitem [{\citenamefont {Pedregosa}\ \emph {et~al.}(2011)\citenamefont
  {Pedregosa}, \citenamefont {Varoquaux}, \citenamefont {Gramfort},
  \citenamefont {Michel}, \citenamefont {Thirion} \emph
  {et~al.}}]{scikit-learn}%
  \BibitemOpen
  \bibfield  {author} {\bibinfo {author} {\bibfnamefont {F.}~\bibnamefont
  {Pedregosa}}, \bibinfo {author} {\bibfnamefont {G.}~\bibnamefont
  {Varoquaux}}, \bibinfo {author} {\bibfnamefont {A.}~\bibnamefont {Gramfort}},
  \bibinfo {author} {\bibfnamefont {V.}~\bibnamefont {Michel}}, \bibinfo
  {author} {\bibfnamefont {B.}~\bibnamefont {Thirion}},  \emph {et~al.},\
  }\bibfield  {title} {\enquote {\bibinfo {title} {Scikit-learn: Machine
  learning in {P}ython},}\ }\href@noop {} {\bibfield  {journal} {\bibinfo
  {journal} {J. Mach. Learn. Res.}\ }\textbf {\bibinfo {volume} {12}},\
  \bibinfo {pages} {2825} (\bibinfo {year} {2011})},\ \Eprint
  {http://arxiv.org/abs/1201.0490} {arXiv:1201.0490} \BibitemShut {NoStop}%
\bibitem [{\citenamefont {{Scott}}(1992)}]{scott}%
  \BibitemOpen
  \bibfield  {author} {\bibinfo {author} {\bibfnamefont {D.~W.}\ \bibnamefont
  {{Scott}}},\ }\href@noop {} {\emph {\bibinfo {title} {{Multivariate Density
  Estimation: Theory, Practice, and Visualization}}}},\ {Wiley Series in
  Probability and Statistics}\ (\bibinfo  {publisher} {Wiley},\ \bibinfo
  {address} {New York},\ \bibinfo {year} {1992})\BibitemShut {NoStop}%
\bibitem [{\citenamefont {{Abbott}}\ \emph
  {et~al.}(2017{\natexlab{c}})\citenamefont {{Abbott}}, \citenamefont
  {{Abbott}}, \citenamefont {{Abbott}}, \citenamefont {{Acernese}},
  \citenamefont {{Ackley}}, \citenamefont {{Adams}}, \citenamefont {{Adams}},
  \citenamefont {{Addesso}}, \citenamefont {{Adhikari}}, \citenamefont {{Adya}}
  \emph {et~al.}}]{2017PhRvD..96l2006A}%
  \BibitemOpen
  \bibfield  {author} {\bibinfo {author} {\bibfnamefont {B.~P.}\ \bibnamefont
  {{Abbott}}}, \bibinfo {author} {\bibfnamefont {R.}~\bibnamefont {{Abbott}}},
  \bibinfo {author} {\bibfnamefont {T.~D.}\ \bibnamefont {{Abbott}}}, \bibinfo
  {author} {\bibfnamefont {F.}~\bibnamefont {{Acernese}}}, \bibinfo {author}
  {\bibfnamefont {K.}~\bibnamefont {{Ackley}}}, \bibinfo {author}
  {\bibfnamefont {C.}~\bibnamefont {{Adams}}}, \bibinfo {author} {\bibfnamefont
  {T.}~\bibnamefont {{Adams}}}, \bibinfo {author} {\bibfnamefont
  {P.}~\bibnamefont {{Addesso}}}, \bibinfo {author} {\bibfnamefont {R.~X.}\
  \bibnamefont {{Adhikari}}}, \bibinfo {author} {\bibfnamefont {V.~B.}\
  \bibnamefont {{Adya}}},  \emph {et~al.},\ }\bibfield  {title} {\enquote
  {\bibinfo {title} {{First narrow-band search for continuous gravitational
  waves from known pulsars in advanced detector data}},}\ }\href {\doibase
  10.1103/PhysRevD.96.122006} {\bibfield  {journal} {\bibinfo  {journal}
  {Phys.~Rev.~D}\ }\textbf {\bibinfo {volume} {96}},\ \bibinfo {eid} {122006}
  (\bibinfo {year} {2017}{\natexlab{c}})},\ \Eprint
  {http://arxiv.org/abs/1710.02327} {arXiv:1710.02327} \BibitemShut {NoStop}%
\bibitem [{\citenamefont {{Abbott}}\ \emph {et~al.}(2009)\citenamefont
  {{Abbott}}, \citenamefont {{Abbott}}, \citenamefont {{Adhikari}},
  \citenamefont {{Ajith}}, \citenamefont {{Allen}}, \citenamefont {{Allen}},
  \citenamefont {{Amin}}, \citenamefont {{Anderson}}, \citenamefont
  {{Anderson}}, \citenamefont {{Arain}} \emph {et~al.}}]{2009RPPh...72g6901A}%
  \BibitemOpen
  \bibfield  {author} {\bibinfo {author} {\bibfnamefont {B.~P.}\ \bibnamefont
  {{Abbott}}}, \bibinfo {author} {\bibfnamefont {R.}~\bibnamefont {{Abbott}}},
  \bibinfo {author} {\bibfnamefont {R.}~\bibnamefont {{Adhikari}}}, \bibinfo
  {author} {\bibfnamefont {P.}~\bibnamefont {{Ajith}}}, \bibinfo {author}
  {\bibfnamefont {B.}~\bibnamefont {{Allen}}}, \bibinfo {author} {\bibfnamefont
  {G.}~\bibnamefont {{Allen}}}, \bibinfo {author} {\bibfnamefont {R.~S.}\
  \bibnamefont {{Amin}}}, \bibinfo {author} {\bibfnamefont {S.~B.}\
  \bibnamefont {{Anderson}}}, \bibinfo {author} {\bibfnamefont {W.~G.}\
  \bibnamefont {{Anderson}}}, \bibinfo {author} {\bibfnamefont {M.~A.}\
  \bibnamefont {{Arain}}},  \emph {et~al.},\ }\bibfield  {title} {\enquote
  {\bibinfo {title} {{LIGO: the Laser Interferometer Gravitational-Wave
  Observatory}},}\ }\href {\doibase 10.1088/0034-4885/72/7/076901} {\bibfield
  {journal} {\bibinfo  {journal} {Rep. Prog. Phys.}\ }\textbf {\bibinfo
  {volume} {72}},\ \bibinfo {eid} {076901} (\bibinfo {year} {2009})},\ \Eprint
  {http://arxiv.org/abs/0711.3041} {arXiv:0711.3041} \BibitemShut {NoStop}%
\bibitem [{\citenamefont {{The LIGO Scientific Collaboration}}\ and\
  \citenamefont {{The Virgo Collaboration}}()}]{2012arXiv1203.2674T}%
  \BibitemOpen
  \bibfield  {author} {\bibinfo {author} {\bibnamefont {{The LIGO Scientific
  Collaboration}}}\ and\ \bibinfo {author} {\bibnamefont {{The Virgo
  Collaboration}}},\ }\bibfield  {title} {\enquote {\bibinfo {title}
  {{Sensitivity Achieved by the LIGO and Virgo Gravitational Wave Detectors
  during LIGO's Sixth and Virgo's Second and Third Science Runs}},}\
  }\href@noop {} {\ }\Eprint {http://arxiv.org/abs/1203.2674} {arXiv:1203.2674}
  \BibitemShut {NoStop}%
\bibitem [{\citenamefont {{Vallisneri}}\ \emph {et~al.}(2015)\citenamefont
  {{Vallisneri}}, \citenamefont {{Kanner}}, \citenamefont {{Williams}},
  \citenamefont {{Weinstein}},\ and\ \citenamefont
  {{Stephens}}}]{2015JPhCS.610a2021V}%
  \BibitemOpen
  \bibfield  {author} {\bibinfo {author} {\bibfnamefont {M.}~\bibnamefont
  {{Vallisneri}}}, \bibinfo {author} {\bibfnamefont {J.}~\bibnamefont
  {{Kanner}}}, \bibinfo {author} {\bibfnamefont {R.}~\bibnamefont
  {{Williams}}}, \bibinfo {author} {\bibfnamefont {A.}~\bibnamefont
  {{Weinstein}}}, \ and\ \bibinfo {author} {\bibfnamefont {B.}~\bibnamefont
  {{Stephens}}},\ }\bibfield  {title} {\enquote {\bibinfo {title} {{The LIGO
  Open Science Center}},}\ }\href {\doibase 10.1088/1742-6596/610/1/012021}
  {\bibfield  {journal} {\bibinfo  {journal} {J. Phys. Conf. Ser.}\ }\textbf
  {\bibinfo {volume} {610}},\ \bibinfo {eid} {012021} (\bibinfo {year}
  {2015})},\ \Eprint {http://arxiv.org/abs/1410.4839} {arXiv:1410.4839}
  \BibitemShut {NoStop}%
\bibitem [{\citenamefont {{Davies}}\ \emph {et~al.}(2017)\citenamefont
  {{Davies}}, \citenamefont {{Pitkin}},\ and\ \citenamefont
  {{Woan}}}]{2017CQGra..34a5010D}%
  \BibitemOpen
  \bibfield  {author} {\bibinfo {author} {\bibfnamefont {G.~S.}\ \bibnamefont
  {{Davies}}}, \bibinfo {author} {\bibfnamefont {M.}~\bibnamefont {{Pitkin}}},
  \ and\ \bibinfo {author} {\bibfnamefont {G.}~\bibnamefont {{Woan}}},\
  }\bibfield  {title} {\enquote {\bibinfo {title} {{A targeted spectral
  interpolation algorithm for the detection of continuous gravitational
  waves}},}\ }\href {\doibase 10.1088/1361-6382/34/1/015010} {\bibfield
  {journal} {\bibinfo  {journal} {Classical Quantum Gravity}\ }\textbf
  {\bibinfo {volume} {34}},\ \bibinfo {eid} {015010} (\bibinfo {year}
  {2017})},\ \Eprint {http://arxiv.org/abs/1603.00412} {arXiv:1603.00412}
  \BibitemShut {NoStop}%
\bibitem [{\citenamefont {{Aasi}}\ \emph
  {et~al.}(2015{\natexlab{b}})\citenamefont {{Aasi}}, \citenamefont {{Abbott}},
  \citenamefont {{Abbott}}, \citenamefont {{Abbott}}, \citenamefont
  {{Abernathy}}, \citenamefont {{Acernese}}, \citenamefont {{Ackley}},
  \citenamefont {{Adams}}, \citenamefont {{Adams}}, \citenamefont {{Addesso}}
  \emph {et~al.}}]{2015ApJ...813...39A}%
  \BibitemOpen
  \bibfield  {author} {\bibinfo {author} {\bibfnamefont {J.}~\bibnamefont
  {{Aasi}}}, \bibinfo {author} {\bibfnamefont {B.~P.}\ \bibnamefont
  {{Abbott}}}, \bibinfo {author} {\bibfnamefont {R.}~\bibnamefont {{Abbott}}},
  \bibinfo {author} {\bibfnamefont {T.}~\bibnamefont {{Abbott}}}, \bibinfo
  {author} {\bibfnamefont {M.~R.}\ \bibnamefont {{Abernathy}}}, \bibinfo
  {author} {\bibfnamefont {F.}~\bibnamefont {{Acernese}}}, \bibinfo {author}
  {\bibfnamefont {K.}~\bibnamefont {{Ackley}}}, \bibinfo {author}
  {\bibfnamefont {C.}~\bibnamefont {{Adams}}}, \bibinfo {author} {\bibfnamefont
  {T.}~\bibnamefont {{Adams}}}, \bibinfo {author} {\bibfnamefont
  {P.}~\bibnamefont {{Addesso}}},  \emph {et~al.},\ }\bibfield  {title}
  {\enquote {\bibinfo {title} {{Searches for Continuous Gravitational Waves
  from Nine Young Supernova Remnants}},}\ }\href {\doibase
  10.1088/0004-637X/813/1/39} {\bibfield  {journal} {\bibinfo  {journal}
  {\apj}\ }\textbf {\bibinfo {volume} {813}},\ \bibinfo {eid} {39} (\bibinfo
  {year} {2015}{\natexlab{b}})},\ \Eprint {http://arxiv.org/abs/1412.5942}
  {arXiv:1412.5942} \BibitemShut {NoStop}%
\bibitem [{\citenamefont {{Abbott}}\ \emph
  {et~al.}(2016{\natexlab{e}})\citenamefont {{Abbott}}, \citenamefont
  {{Abbott}}, \citenamefont {{Abbott}}, \citenamefont {{Abernathy}},
  \citenamefont {{Acernese}}, \citenamefont {{Ackley}}, \citenamefont
  {{Adams}}, \citenamefont {{Adams}}, \citenamefont {{Addesso}}, \citenamefont
  {{Adhikari}} \emph {et~al.}}]{2016PhRvD..94j2002A}%
  \BibitemOpen
  \bibfield  {author} {\bibinfo {author} {\bibfnamefont {B.~P.}\ \bibnamefont
  {{Abbott}}}, \bibinfo {author} {\bibfnamefont {R.}~\bibnamefont {{Abbott}}},
  \bibinfo {author} {\bibfnamefont {T.~D.}\ \bibnamefont {{Abbott}}}, \bibinfo
  {author} {\bibfnamefont {M.~R.}\ \bibnamefont {{Abernathy}}}, \bibinfo
  {author} {\bibfnamefont {F.}~\bibnamefont {{Acernese}}}, \bibinfo {author}
  {\bibfnamefont {K.}~\bibnamefont {{Ackley}}}, \bibinfo {author}
  {\bibfnamefont {C.}~\bibnamefont {{Adams}}}, \bibinfo {author} {\bibfnamefont
  {T.}~\bibnamefont {{Adams}}}, \bibinfo {author} {\bibfnamefont
  {P.}~\bibnamefont {{Addesso}}}, \bibinfo {author} {\bibfnamefont {R.~X.}\
  \bibnamefont {{Adhikari}}},  \emph {et~al.},\ }\bibfield  {title} {\enquote
  {\bibinfo {title} {{Results of the deepest all-sky survey for continuous
  gravitational waves on LIGO S6 data running on the Einstein@Home volunteer
  distributed computing project}},}\ }\href {\doibase
  10.1103/PhysRevD.94.102002} {\bibfield  {journal} {\bibinfo  {journal}
  {\prd}\ }\textbf {\bibinfo {volume} {94}},\ \bibinfo {eid} {102002} (\bibinfo
  {year} {2016}{\natexlab{e}})},\ \Eprint {http://arxiv.org/abs/1606.09619}
  {arXiv:1606.09619} \BibitemShut {NoStop}%
\bibitem [{\citenamefont {{Isi}}\ \emph {et~al.}()\citenamefont {{Isi}},
  \citenamefont {{Mastrogiovanni}}, \citenamefont {{Pitkin}},\ and\
  \citenamefont {{Piccinni}}}]{maxpaper}%
  \BibitemOpen
  \bibfield  {author} {\bibinfo {author} {\bibfnamefont {M.}~\bibnamefont
  {{Isi}}}, \bibinfo {author} {\bibfnamefont {S.}~\bibnamefont
  {{Mastrogiovanni}}}, \bibinfo {author} {\bibfnamefont {M.}~\bibnamefont
  {{Pitkin}}}, \ and\ \bibinfo {author} {\bibfnamefont {O.~J.}\ \bibnamefont
  {{Piccinni}}},\ }\href@noop {} {}\bibinfo {note} {{(to be
  published)}}\BibitemShut {NoStop}%
\bibitem [{\citenamefont {{LIGO Scientific Collaboration}}(2017)}]{LALSuite}%
  \BibitemOpen
  \bibfield  {author} {\bibinfo {author} {\bibnamefont {{LIGO Scientific
  Collaboration}}},\ }\href@noop {} {\enquote {\bibinfo {title} {{LALS}uite},}\
  }\bibinfo {howpublished} {\url{https://wiki.ligo.org/DASWG/LALSuite}}
  (\bibinfo {year} {2017})\BibitemShut {NoStop}%
\bibitem [{\citenamefont {Behnel}\ \emph {et~al.}(2011)\citenamefont {Behnel},
  \citenamefont {Bradshaw}, \citenamefont {Citro}, \citenamefont {Dalcin},
  \citenamefont {Seljebotn},\ and\ \citenamefont {Smith}}]{behnel2010cython}%
  \BibitemOpen
  \bibfield  {author} {\bibinfo {author} {\bibfnamefont {S.}~\bibnamefont
  {Behnel}}, \bibinfo {author} {\bibfnamefont {R.}~\bibnamefont {Bradshaw}},
  \bibinfo {author} {\bibfnamefont {C.}~\bibnamefont {Citro}}, \bibinfo
  {author} {\bibfnamefont {L.}~\bibnamefont {Dalcin}}, \bibinfo {author}
  {\bibfnamefont {D.S.}\ \bibnamefont {Seljebotn}}, \ and\ \bibinfo {author}
  {\bibfnamefont {K.}~\bibnamefont {Smith}},\ }\bibfield  {title} {\enquote
  {\bibinfo {title} {Cython: The best of both worlds},}\ }\href {\doibase
  10.1109/MCSE.2010.118} {\bibfield  {journal} {\bibinfo  {journal} {Comput.
  Sci. Eng.}\ }\textbf {\bibinfo {volume} {13}},\ \bibinfo {pages} {31 --39}
  (\bibinfo {year} {2011})},\ \bibinfo {note}
  {\url{http://cython.org}}\BibitemShut {NoStop}%
\bibitem [{\citenamefont {Kluyver}\ \emph {et~al.}(2016)\citenamefont
  {Kluyver}, \citenamefont {Ragan-Kelley}, \citenamefont {P{\'e}rez},
  \citenamefont {Granger}, \citenamefont {Bussonnier}, \citenamefont
  {Frederic}, \citenamefont {Kelley}, \citenamefont {Hamrick}, \citenamefont
  {Grout}, \citenamefont {Corlay} \emph {et~al.}}]{kluyver2016jupyter}%
  \BibitemOpen
  \bibfield  {author} {\bibinfo {author} {\bibfnamefont {Thomas}\ \bibnamefont
  {Kluyver}}, \bibinfo {author} {\bibfnamefont {Benjamin}\ \bibnamefont
  {Ragan-Kelley}}, \bibinfo {author} {\bibfnamefont {Fernando}\ \bibnamefont
  {P{\'e}rez}}, \bibinfo {author} {\bibfnamefont {Brian}\ \bibnamefont
  {Granger}}, \bibinfo {author} {\bibfnamefont {Matthias}\ \bibnamefont
  {Bussonnier}}, \bibinfo {author} {\bibfnamefont {Jonathan}\ \bibnamefont
  {Frederic}}, \bibinfo {author} {\bibfnamefont {Kyle}\ \bibnamefont {Kelley}},
  \bibinfo {author} {\bibfnamefont {Jessica}\ \bibnamefont {Hamrick}}, \bibinfo
  {author} {\bibfnamefont {Jason}\ \bibnamefont {Grout}}, \bibinfo {author}
  {\bibfnamefont {Sylvain}\ \bibnamefont {Corlay}},  \emph {et~al.},\
  }\bibfield  {title} {\enquote {\bibinfo {title} {Jupyter notebooks---a
  publishing format for reproducible computational workflows},}\ }in\
  \href@noop {} {\emph {\bibinfo {booktitle} {Positioning and Power in Academic
  Publishing: Players, Agents and Agendas: Proceedings of the 20th
  International Conference on Electronic Publishing}}}\ (\bibinfo
  {organization} {IOS Press},\ \bibinfo {address} {Amsterdam},\ \bibinfo {year}
  {2016})\ p.~\bibinfo {pages} {87},\ \bibinfo {note}
  {\url{http://jupyter.org/}}\BibitemShut {NoStop}%
\bibitem [{\citenamefont {Hunter}(2007)}]{Hunter:2007}%
  \BibitemOpen
  \bibfield  {author} {\bibinfo {author} {\bibfnamefont {J.~D.}\ \bibnamefont
  {Hunter}},\ }\bibfield  {title} {\enquote {\bibinfo {title} {Matplotlib: A 2d
  graphics environment},}\ }\href {\doibase 10.1109/MCSE.2007.55} {\bibfield
  {journal} {\bibinfo  {journal} {Computing In Science \& Engineering}\
  }\textbf {\bibinfo {volume} {9}},\ \bibinfo {pages} {90--95} (\bibinfo {year}
  {2007})}\BibitemShut {NoStop}%
\bibitem [{\citenamefont {{Droettboom}}\ \emph {et~al.}(2017)\citenamefont
  {{Droettboom}}, \citenamefont {{Caswell}}, \citenamefont {{Hunter}} \emph
  {et~al.}}]{michael_droettboom_2017_248351}%
  \BibitemOpen
  \bibfield  {author} {\bibinfo {author} {\bibfnamefont {M.}~\bibnamefont
  {{Droettboom}}}, \bibinfo {author} {\bibfnamefont {T.~A.}\ \bibnamefont
  {{Caswell}}}, \bibinfo {author} {\bibfnamefont {J.}~\bibnamefont {{Hunter}}},
   \emph {et~al.},\ }\href {\doibase 10.5281/zenodo.248351} {\enquote {\bibinfo
  {title} {matplotlib/matplotlib: v2.0.0},}\ } (\bibinfo {year} {2017}),\
  \bibinfo {note}
  {\href{https://doi.org/10.5281/zenodo.248351}{10.5281/zenodo.248351}}\BibitemShut
  {NoStop}%
\bibitem [{\citenamefont {{Pitkin}}(2018)}]{psrqpy}%
  \BibitemOpen
  \bibfield  {author} {\bibinfo {author} {\bibfnamefont {M.}~\bibnamefont
  {{Pitkin}}},\ }\bibfield  {title} {\enquote {\bibinfo {title} {{psrqpy: a
  python interface for querying the ATNF pulsar catalogue}},}\ }\href {\doibase
  10.21105/joss.00538} {\bibfield  {journal} {\bibinfo  {journal} {{JOSS}}\
  }\textbf {\bibinfo {volume} {3}},\ \bibinfo {pages} {538} (\bibinfo {year}
  {2018})},\ \Eprint {http://arxiv.org/abs/1806.07809} {arXiv:1806.07809}
  \BibitemShut {NoStop}%
\bibitem [{\citenamefont {{Buchner}}(2016)}]{Buchner2016}%
  \BibitemOpen
  \bibfield  {author} {\bibinfo {author} {\bibfnamefont {J.}~\bibnamefont
  {{Buchner}}},\ }\bibfield  {title} {\enquote {\bibinfo {title} {{A
  statistical test for Nested Sampling algorithms}},}\ }\href {\doibase
  10.1007/s11222-014-9512-y} {\bibfield  {journal} {\bibinfo  {journal} {{J.
  Stat. Comput.}}\ }\textbf {\bibinfo {volume} {26}},\ \bibinfo {pages} {383}
  (\bibinfo {year} {2016})},\ \Eprint {http://arxiv.org/abs/1407.5459}
  {arXiv:1407.5459} \BibitemShut {NoStop}%
\end{thebibliography}%

\end{document}